\documentclass[doc,floatsintext,11pt,biblatex]{apa7}\usepackage[]{graphicx}\usepackage[]{xcolor}
\makeatletter
\def\maxwidth{ %
  \ifdim\Gin@nat@width>\linewidth
    \linewidth
  \else
    \Gin@nat@width
  \fi
}
\makeatother

\definecolor{fgcolor}{rgb}{0.345, 0.345, 0.345}

\usepackage{framed}
\makeatletter
 {\par\unskip\endMakeFramed%
 \at@end@of@kframe}
\makeatother

\definecolor{shadecolor}{rgb}{.97, .97, .97}
\definecolor{messagecolor}{rgb}{0, 0, 0}
\definecolor{warningcolor}{rgb}{1, 0, 1}
\definecolor{errorcolor}{rgb}{1, 0, 0}
\newenvironment{knitrout}{}{} 

\usepackage{alltt}
\usepackage{lipsum}
\usepackage{amsmath,amssymb,amstext, mathrsfs} 
\usepackage{siunitx} 
\usepackage{array}
\usepackage{booktabs}
\usepackage{tikz}
\newcolumntype{M}[1]{>{\centering\arraybackslash}m{#1}}
\usepackage{pgfplots}
\graphicspath{ { } }
\usepackage{amsthm}
\usepackage{pgfplots}
\usepackage{wrapfig}
\usepackage{mathtools}

\theoremstyle{remark}

\theoremstyle{definition}

\pgfplotsset{compat=1.17}
\usetikzlibrary{arrows.meta}
\usepgfplotslibrary{groupplots}
\usepgfplotslibrary{patchplots}
\usepackage{float}
\usepackage{caption}
\usepackage{afterpage}
\renewcommand{\vec}[1]{\overset{\text{\tiny$\rightarrow$}\hspace{-1mm}}{#1}}

\title{What I Disbelieve: Quantifying Belief With Negative Ranking Functions}
\shorttitle{Quantifying Belief}
\authorsnames{Hanbin Go,Britt Anderson}
\authorsaffiliations{Dept. of Psychology, University of Waterloo, CANADA}
\journal{in submission}
\usepackage{fontspec}
\usepackage{microtype}
\abstract{
A challenge in exploring the psychology of human uncertainty is its measurement and quantification. Prior work has focused almost exclusively on probability, while the formal theories developed in philosophy are rarely tested. We tested Wolfgang Spohn’s ranking theory, a normative account in which belief is graded by ordinal ranks of disbelief that combine by minimization rather than addition, as a descriptive model of human belief. Across six experiments, participants assigned ranks of disbelief to propositions involving personal experience, chance devices, sequential learning, and an Ellsberg urn. Ranks were reliable and, through the theory’s two-sided rank function, distinguished belief, disbelief, and suspension of judgment. Participants respected the Law of Negation, never disbelieving both a proposition and its complement, in every task except a lottery. Disjunctions did not follow the theory's Law of Disjunction. Instead of taking the minimum rank of the disjuncts, participants averaged them. In a learning task, ranks aligned more closely than did subjective probabilities with objective probabilities. In an Ellsberg Urn scenario, the paradoxical behavior seen with betting was markedly reduced by ranks. We conclude that human belief has the bipolar, negation-respecting structure that ranking theory proposes, but combines disjuncts by averaging rather than by minimization. As a clear account of human belief dynamics with metrics that can be reliably measured, Ranking Theory provides a valuable framework for investigating and contrasting probabilistic and non-probabilistic accounts of human belief.
}
\IfFileExists{upquote.sty}{\usepackage{upquote}}{}
\begin{document}
\raggedbottom
\maketitle

\section{Introduction}
What I believe \autocite{forster1938what, Russell_2013, einstein2015} is more than a declaration of credence.
It is that collection of statements, ideas, expectations that I hold to be true about the world.
Thus connected to truth, it would seem to be binary.
Things are true or false; therefore, I either believe or disbelieve.
But, as observed by \textcite{spohn2012laws} there is a vagueness about our usage.
I may tell you that I \emph{believe} a particular politician lied.
You ask me if I am sure.
I say ``No.'' Then you tell me that I must mean only that it is likely they lied. ``No,'' I say, ``I really believe it.'' Clearly, my belief is not maximum certainty nor some likelihood below maximal.
How do we come to grips with this vagueness? We will return to Spohn’s theory shortly, but the practical answer we will justify by experiment is that if you want to know what someone believes and how strongly they believe it: ask them.

Thinking of belief as probability (or possibly confidence \autocite{jackson_2024}) is not just folk psychology, and the assumption that belief is some function of probability has formed the basis for much research.
An advantage of this approach is that probability places strong constraints on how beliefs may be combined and updated, making this proposal testable.
For instance, the same subjective probability must always lead to the same degree of belief regardless of context.
Additionally, since probabilities of mutually exclusive events are additive the probability of a disjunction (A or B) is the \emph{sum} of their individual probabilities \autocite{kolmogorov2018foundations}.
When considering sequences of independent events the joint probability is the \emph{product} of their individual probabilities \autocite{feller1991introduction}.
Practically, the probabilistic perspective of belief is investigated by having people make choices in environments with known probabilities (at least known to the experimenter) while observing participant choices and possibly brain activity \autocite{Fiore_2022,Walker_2023}.
This approach to belief estimation assigns a probability to each possible state of the world and uses probability laws to guide decision-making \autocite{cosmides1996humans, sanborn2016bayesian}.
Degrees of belief become real numbers in the range [0,1] \autocite{wyer2005belief}, with higher values indicating greater belief in a proposition’s truth \autocite{huber_ed_2009}.

While offering mathematical rigor, the equating of belief and subjective probability does not seem to capture the phenomenology of human belief systems.
People are bad probabilists \autocite{Nickerson_2002}, so bad that recently some have come to doubt the value of its emphasis in cognitive theories of decision making under uncertainty \autocite{Szollosi_2023}.
Our struggles to form and justify our beliefs seems at odds with a posited capacity for robust statistical estimation that can efficiently guide behavior.
Humans often struggle to determine what to believe \autocite{bland2012different}.
And the mathematics are a double-edged sword.
No Bayesian model can update belief about a proposition held to be true with probability 1 (all other propositions must have probability zero).
This seems at odds with commonsense reasoning about belief revision.
Dropping the threshold of belief from probability 1 to some value smaller than one falls prey to other paradoxes, such as the lottery paradox \autocite{Kyburg1961}.
Imagine a lottery with a great many tickets, say one million.
It seems rational to believe that ticket one will not win, and also that ticket two will not win, and so on.
Since it is considered rational to believe the conjunction of our beliefs, we should not believe that any ticket holder will win the lottery, but we do.
The famous Linda the bank teller paradox of \textcite{tversky1983extensional} is another demonstration of how what we believe does not obey the probability calculus.
The assignment of a single value to each proposition under consideration, which represents its overall plausibility or likelihood, seems insufficient to capture the different levels of belief people might have about a fixed likelihood estimate; it cannot account for differences in a framing of acceptance versus rejection, or a sensitivity to the quality of evidence \autocite{spohn2012laws}.
As \textcite{keynes2013treatise} argued in his Principle of Indifference, when an agent lacks sufficient information, such as being indoors all day without knowledge of the weather, the probability account leads to an erroneous assumption that the absence of evidence implies equal probabilities.
Ignorance or lack of external relevant evidence should not automatically lead to such equal probability assignments.
Paradoxes and contradictions may result \autocite{keynes2013treatise}.
Another distinction often made in philosophical circles is the notion of (non-)monotonicity.
We revise and retract beliefs.
We do not revise probability the same way.
Knowing that Polly is a bird we may hold a belief that Polly flies and a high probability that birds fly.
Learning that Polly is an ostrich, we revise our belief that Polly flies, but not our probability estimate that birds fly. 

Belief is not just an isolated ``state'', it is a ``force''; it influences how people think, learn, and make decisions.
Belief representation is not monolithic: beliefs are formed, stored, remembered, updated, and used.
Restricting accounts of belief to subjective probability estimates disconnects belief from factors like logic, evidence, memory, attention, emotion, motivation, and social context all of which play a role in belief representation \autocite{uso2016belief, grayling2011psychology, halpern1992two, wyer2005belief}.
Beliefs are not merely passive reflections but active engagements with truth, guided by normative standards of correctness \autocite{wedgwood2002aim}.

The nature of belief, whether it can be partial, and if so, how it is updated has been an active area of philosophical discussion \autocite{huber_ed_2009}.
This paper concerns itself with one particular approach to this issue, the Negative Ranking Theory (NRT) of Wolfgang Spohn, a summary of which we provide shortly.
Spohn’s theory is a normative theory and as such argues for what people should do, and not necessarily what they do do.
The research presented here shares our initial effort to address this question.
It is motivated by the intuition that evolution can be an effective teacher, and that optimal methods attract.
For a start, we wondered can people even reliably do a task that asks them to specify negative ranks, and if so, do the results seem sensible.
Then we explore whether there is an advantage over the more common position that human belief is just a synonym for subjective probability or confidence \autocite{Fleming_2024}.
Does NRT simply offer an alternative vocabulary, albeit with possible advantages for explanation or illustration?
Or does NRT actually have predictive and explanatory advantages over the contemporary probability based account?

\subsection{Formalizing Belief}

One of the main challenges in belief representation is modeling it formally and rigorously.
Cognitive science and artificial intelligence employ various models for this purpose, including probability theory, Dempster-Shafer theory, fuzzy reasoning, Bayesian reasoning, possibility theory, default reasoning, nonmonotonic reasoning, and ranking theory \autocite{pearl1988probabilistic, kern20215}.
Each model has distinct logical assumptions, methods, applications, strengths, and weaknesses.
For instance, probability theory (about which more shortly) is more rigorous in its mathematical foundation, while Dempster-Shafer theory allows for a more flexible representation of uncertainty (particularly when evidence is incomplete) \autocite{pearl1988probabilistic, shafer1976mathematical}.
Bayesian reasoning, commonly applied in probabilistic predictions by integrating prior knowledge with new evidence, and NRT can be more suitable for representing degrees of disbelief in nonmonotonic contexts \autocite{griffiths2006optimal, spohn2012laws}.
These models also have unique relationships and comparisons; some are more flexible or have strict sets of rules, while others are more practical or theoretical.

Probability theory has traditionally been regarded as \emph{the} framework for measuring uncertainty, linking evidence to hypotheses in order to promote rational beliefs \autocite{galavotti_interpretation_2017}.
However, this approach often fails to account for the subjective aspects of human belief, as it assumes that a complete and objective assessment of evidence will lead to a singular, rational probability assignment \autocite{galavotti_interpretation_2017}.
This interpretation overlooks the personal and contextual nature of belief.
Subjective interpretations, such as Bruno de Finetti’s, emphasize that beliefs are shaped by individual perspectives and the specific information available to a person \autocite{de_finetti_theory_1975, galavotti_interpretation_2017}.
De Finetti highlighted that evaluating probability involves both objective components, such as data and facts, and subjective components, such as opinions based on this evidence \autocite{galavotti_interpretation_2017}.
These views challenge the notion that \emph{objective} probability can serve as a universally applicable measure of belief \autocite{galavotti_interpretation_2017}.

While few doubt that environmental evidence plays a role in belief formation, the absence of reliable statistics in many situations means it cannot be a general explanation for belief representation \autocite{galavotti_interpretation_2017}.
Frequentist interpretations of probability apply to much of the natural sciences, but even here they fail to provide convincing answers when applied to single events, which includes many everyday scenarios that are distinct in time, place, persons, and context \autocite{galavotti_interpretation_2017, von_mises_probability_1957}.
Determining the appropriate reference class for probability calculations is often difficult; sometimes impossible \autocite{galavotti_interpretation_2017}.
Bayesian methods which are prominent in our field offer a pluralistic approach that combines objective and subjective elements \autocite{galavotti_interpretation_2017}.
But it seems unlikely that humans can realistically perform the complex calculations required by the theory in all but the most trivial of scenarios \autocite{henckel2022belief}.

While empirically minded psychologists seem to gravitate to probability accounts, perhaps because of our familiarity with these methods brought about by our statistical training, other professions with an interest in models of belief have chosen differently. 
Many philosophers prefer epistemic theories, like logicism \autocite{galavotti_interpretation_2017, carnap1962logical}.
The complexity of the formalisms of such theories seems to have deterred many outside of philosophy.
In economics, and the social sciences more broadly, subjectivism is prevalent.
Beliefs are interpreted as expectations that shape forecasts.
The subjectivity of subjectivism often leads to it being viewed skeptically by natural scientists.
It is often misconceived as a mere expression of personal opinions without empirical grounding \autocite{galavotti_interpretation_2017}. 

Despite these challenges there is no doubt we do believe some things and not others and believe in some things more strongly than others.
An interpretative and formal theory is still required.
While logical interpretations of probability are insightful, they do not seem to fully capture the complexities of human belief.
There is still room for alternative models \autocite{spohn2012laws}.

NRT, introduced by Wolfgang Spohn in 1983, provides just such an alternative.
Rather than assigning probabilities to events, it employs a system based on natural numbers and infinity to quantify degrees of disbelief or belief in propositions.
This framework is rooted in formal epistemology and principles of rationality \autocite{skovgaard2016ranking, spohn2012laws, spohn2013ranking}.
NRT expresses logical relations and supporting evidence by assigning ranks of disbelief or belief to propositions.
This method effectively addresses paradoxes like the lottery paradox previously discussed.
By assigning a high rank of disbelief to each ticket, ranking theory acknowledges the low chance of any single ticket winning while maintaining the conditional belief that one ticket must win.
This approach allows varying degrees of belief to coexist without conflict, avoiding the need for constant probabilistic recalculations \autocite{spohn2012laws}.
Ranking theory provides a more adaptable framework for managing beliefs, streamlining belief adjustments based on new evidence without the complexity involved in probability-based models.
These practical advantages make ranking theory a valuable tool for representing human beliefs and uncertainties in a more practical manner \autocite{spohn2012laws}.

Despite its potential, ranking theory faces many of the challenges just outlined above for extant theories.
It can be complex to understand and apply due to its substantial conceptual shift from traditional probabilistic models.
This shift requires a deeper understanding of formal logic and epistemology.
Additionally, ranking theory involves managing a broad scale that includes natural numbers plus infinity, which presents unique challenges in both learning and practical applications.
Most importantly for psychology, empirical testing is needed \autocite{kern20215, skovgaard2016ranking}.

\subsection{The Basics of Ranking Theory}

\textcite{spohn2012laws} is the best source for a detailed explanation of his theory and for his philosophical critique of the alternatives currently debated in the philosophical community.
Here we try to first provide a minimalist version of NRT sufficient to follow along with the experiments.
Anyone wanting the full details should refer to Spohn or perhaps \textcite{huber2013belief} for a shorter, but still comprehensive explication. 

The key conception for NRT is the assignment of a non-negative integer (0, 1, 2, ...) or $+ ~ \infty$ as an answer to the question ``How much do I disbelieve this proposition?''.
If you do not disbelieve it all your answer is zero.
If you find it impossible, then your disbelief is maximal, i.e. $+ ~ \infty$.
All other values represent partial disbelief and are ordered.
Thus, a disbelief rank of 3 means you disbelieve the proposition more than a disbelief of rank 2.
The term \emph{negative} can be confusing on a first reading.
What is negative is not the number of the rank you assign, but that you are ranking your DISbelief.
It has been known for a century that we can reliably assign numbers to sensory magnitudes \autocite{cross2013judgments}, and numerical ranks can work well for more abstract attributes, e.g.
document relevance \autocite{maddalena2015judging}.
The assignment of negative ranks for disbelief would be plausibly another example of magnitude estimation. 

The degree of disbelief is usually designated with a $\kappa$, and the negative ranking \emph{function} comes from the adoption of a possible world semantics \autocite{stalnaker1976possible}, though one in which the sets of possible worlds are assigned ranks rather than the more common probabilistic quantities.
The NRT literature uses $W$ for the set of all possible worlds that are mutually exclusive and jointly exhaustive \autocite{spohn2012laws}.

Informally, one can think of an algebra specifying subsets of our space of all possible worlds, and the rules for how those subsets can be combined and quantified.
For example, when flipping a coin, the space of all possible worlds might include heads, tails, and the coin landing on its side.
The algebra provides the operational details for how these subsets can be combined and measured.
With this logic, we consider an algebra, \(\mathcal{A}\) for subsets of $W$, which we refer to as propositions \autocite{spohn2012laws}.

Some consequences of this algebra of disbelief are the ones we test here.
First, we highlight the \emph{Law of Negation}.
The Law of Negation states that one cannot simultaneously disbelieve both \(\mathcal{A}\) and \(\bar{\mathcal{A}}\).
In words \(\bar{\mathcal{A}}\) means not \(\mathcal{A}\) or the complement of \(\mathcal{A}\). \(\mathcal{A}\) and \(\bar{\mathcal{A}}\) together make up the whole universe of possible worlds.
To disbelieve both (that is to assign a \(\kappa > 0\) for both) is to disbelieve in existence.
Note that Spohn is describing a normative system, a system for how one ought to believe.
It is an empirical question if people do or do not believe thusly. 

As an aside, one can assign both \(\mathcal{A}\) and \(\bar{\mathcal{A}}\) a rank 0.
That means neither is disbelieved.
This is a state of equipoise regarding the proposition that defined the partition of \(W\) into \(\mathcal{A}\) and \(\bar{\mathcal{A}}\). 

The second law of NRT that we evaluate is the \emph{Law of Disjunction}.
As a formula this reads: \( \kappa(A \cup B) = \min\{\kappa(A), \kappa(B)\} \).
In words, the law of disjunction states that the negative rank of the disjunction (the logical OR) of two propositions \(A\) and \(B\) is equal to the \emph{minimum} of the ranks of the individual propositions.
This means that the disbelief in \(A \cup B\) is determined by the lesser of the disbeliefs in \(A\) and \(B\).
If you have minimal skepticism about either event \(A\) or \(B\) occurring, then your skepticism about the combined possibility aligns with the lesser skepticism towards the individual events.
In contrast to classical probability, where the likelihood of \(A \cup B\) is calculated by adding the probabilities of \(A\) and \(B\) subtracting the overlap (if any), ranking theory simplifies to adopting the minimal disbelief between \(A\) and \(B\).
This approach focuses on the most plausible scenario rather than accumulating probabilities.

The math for Spohn’s theory works out nicer with disbelief as the principal concept \autocite{kern20215,spohn2009survey,spohn2012laws,spohn2012laws}.
However, earlier philosophical development had been from a positive belief perspective \autocite[e.g.]{cohen1977probable,dubois2023possibility}.
This, combined with an intuition that speaking about degrees of belief rather than degrees of disbelief ``feels'' more natural to many makes a comparison of negative and positive belief an important empirical issue.

Positive ranking functions are often denoted by \(\beta\), and the relation to negative ranks is direct.
I believe in a proposition \(\mathcal{A}\) to the extent I disbelieve \(\bar{\mathcal{A}}\).
As a formula this is: \( \beta(A) = \kappa(\bar{A}) \).

Two-sided ranking functions spanning the full range from certain conviction to unshakable disbelief are possible by combining these two ranking functions. $\tau$ is the customary symbol for a two-sided rank.
As a formula, we write \(\tau(A) = \kappa(\bar{A}) - \kappa(A) = \beta(A) - \kappa(A) \).
With this formula a $\tau$ greater than zero indicates belief, less than zero disbelief, and zero indicates that the proposition is neither believed nor disbelieved. 

\subsection{Probability-Ranking Translation}

As noted above, there has been a tradition of equating belief with subjective probability, and there is a way, perhaps a bit tortured, to try and unify the two.
The negative rank of the conjunction of two propositions (how much do I disbelieve this and that) is the sum of their individual ranks.
While the probability of a conjunction is the product of their probabilities.
Those who remember logarithms will recognize this.
The log of a product is the sum of the individual logs.
The challenge of viewing ranks as the logarithms of probabilities comes from the disjunction, the ``or'' operation; it is a minimum for ranks and it is a sum for probabilities.
There are ways to get this equivalence if instead of the usual logarithm bases we are familiar with (base 10, base 2, the natural logarithm) we consider logarithms with infinitesimally small bases \autocite{raidl2017}.
This translation would preserve some of the logical properties, but might strain our credulity a bit.
Nevertheless, since such a translation is in principle possible we can try to fit a series of ranks with a logarithmic function where the base is free to vary and be very small.
Despite this parallel, and our investigation of it, we share the resistance of \textcite{spohn2013ranking} to reducing ranks to purely probabilistic terms.
Belief cannot always be consistently or deductively represented probabilistically, which supports the independent development of ranking theory as a distinct approach to belief representation \autocite{kern20215}.


\section{Experiment 1 and 2: Expressing Disbelief through Negative Ranking Functions}
\label{chap:E1}

Experiments 1 and 2 investigated whether humans could express degrees of disbelief using negative ranking functions and explored the relationship between these ranks and subjective probabilities.
They can be largely seen as a replication pair with Experiment 1 done on-line and Experiment 2 an in-lab replication.
The experiments examined participants’ degrees of disbelief for propositions where it was likely participants would only have subjective probability estimates. 
The scenarios were the language demographics of Quebec, September weather in an American city, and a co-worker’s music preferences at a karaoke party. 
We chose such scenarios with a goal of not priming the participants to think probabilistically.

\subsection{Methods}

Undergraduate students from the University of Waterloo participated in both experiments. 
The study received ethics clearance from the University of Waterloo Research Ethics Board (REB 44180) before it was conducted. 
All participants were informed about the purpose of the study and their right to withdraw at any time without facing any penalty. 
To protect participants’ privacy, data was de-identified before being securely stored. 
All participants provided their informed consent before participating.

In Experiment 1, 168 students completed the on-line study for course credit, with a median age of 20 years. 
The participants included 101 females, 66 males, and 1 who did not disclose their gender.

Experiment 2, conducted in a subsequent semester, involved 63 students in an in-lab setting, with a median age of 19 years, including 48 females, 13 males, and 2 who preferred not to answer.

The study utilized a custom-built on-line platform created with JavaScript to collect participant responses \autocite{de2023jspsych}. 
The platform featured an interface that enabled participants to enter their responses for negative ranks, as well as a 10 by 10 interactive grid where they could click on boxes to indicate possible event cards.
This provided a post-hoc report of subjective probability. 
Participants finished the study in under 30 minutes.

\textbf{Grading of Disbelief — Negative Ranking Functions}

Participants began by completing a series of questionnaires designed to assess their degrees of disbelief toward various propositions. 
Each proposition was assigned a non-negative integer ranging from zero (not surprising) to infinity (impossible), representing varying levels of disbelief or surprise. 
Participants were to grade their degrees of disbelief: 0 for not surprising, a value greater than 0 for surprising, and infinity for impossible. 
The topics of these questionnaires included:

\begin{enumerate}
    \item \textbf{Language Demographics in Quebec:} Participants graded their disbelief about meeting a Quebec native (born and raised) who speaks only one language.
The languages included were French, Spanish, English, Greek, Polish, Korean, Urdu, Japanese, or the disjunction Spanish or Urdu.
    
    \item \textbf{September Weather in a Designated Visiting City:} Participants graded their disbelief regarding daytime high temperatures for a hypothetical day trip to Boston, Massachusetts, United States in September.
The temperature ranges included -29 to -20°C, -19 to -10°C, -9 to 0°C, 1 to 10°C, 11 to 20°C, 21 to 30°C, 31 to 40°C, 41 to 50°C, and the disjunction -9 to 0 or 1 to 10°C.
    
    \item \textbf{Co-Workers’ Music Preferences During a Hypothetical Karaoke Party:} Participants imagined being employees at an automotive service centre invited to a work karaoke party.
They were informed that their co-worker Tom, a Caucasian male in his 20s who grew up in a rural area and loves wearing his cowboy hat, would be attending.
Participants then graded their disbelief about Tom enjoying singing along to various music genres such as Pop, Rock, Hip Hop, Heavy Metal, Country, Jazz, Opera, Folk, or the disjunction Jazz or Opera.
\end{enumerate}

The selected scenarios reflected everyday situations with unclear objective probabilities. 
Comprehensive details of the ranking questionnaires are documented in Appendix \ref{AppendixE1}.

\textbf{Open Sampling Format (OSF)}

To measure participants’ representation of probabilities, an OSF was used following each questionnaire. 
Participants engaged with a 10x10 grid where each cell represented a possible outcome.

Initially, the grid was generated using a ranking-to-probability equation with a logarithmic base set between 0 and 1, initially at 0.5.
This arbitrary choice was based on theoretical predictions of a logarithmic relationship. 
Participants then adjusted the grid to reflect their reported degrees of disbelief from the questionnaires.

Each grid cell represented a possible outcome. 
Participants filled the grid according to their disbelief levels, with higher frequencies indicating greater perceived commonality. 
They were informed that some outcomes might be missing or appear more frequently than expected. 
Participants adjusted the number of event cards for each outcome according to their reported disbelief levels. 
They could select an outcome from a drop-down menu to replace a cell or choose not to make any changes if the event cards already reflected their level of disbelief (see Figure \ref{fig:E1_E3_OSF}).

This graphical representation aimed to capture subjective probabilities without directly asking about them. 
Adjusting the grid was intended to reduce the priming effect of thinking in terms of probability and provide a more accurate representation of expectations \autocite{tiede2022revisiting}. 
Literature suggests people make better decisions when probabilities are shown graphically rather than numerically. 
Unlike past studies where OSF probabilities were pre-made, this study allowed participants to create and modify the OSF themselves, introducing a novel, interactive method.

\textbf{End of Task Survey}

\textbf{Subjective Probability Estimates}: Following the experiment, participants completed the same three questionnaires as before, but this time using percentages. 
The questionnaires covered language demographics, weather in a designated city, and co-workers’ music preferences. 
This step aimed to evaluate whether the OSF correlated with subjective probabilities. 
To avoid priming disbelief ranks, subjective probability estimates were gathered at the end of the experiment.

\textbf{Difficulty and Preference Assessment}: Participants were then asked to evaluate the difficulty of quantifying their degrees of disbelief and their preference for belief versus disbelief assignments using a Likert scale. 
The questions assessed were:
\begin{enumerate}
    \item The difficulty of quantifying degrees of disbelief/surprise.
    \item Whether it would have been more natural or easier to assign degrees of belief rather than degrees of disbelief/surprise.
\end{enumerate}
These questions aimed to capture participants’ subjective experiences regarding the ease of the task and their preference for belief versus disbelief assignments.

Data from the questionnaires and OSF tasks were analyzed using R (version 4.6.0) for statistical computing and visualization. 
To examine the relationship between observed negative ranks and subjective probabilities, a non-linear least squares regression was used. 
This regression model estimated the optimal logarithmic base \( b \), optimizing the fit between the ranks and probabilities. 
Descriptive statistics, correlation analyses, and regression models were utilized to further explore the relationships between these and other measured variables. 
The ggplot2 package in R was used to generate figures providing visual representations of the data \autocite{ggplot2016}.

For data analysis involving statistical tests, such as curve fitting (non-linear least squares regression) and linear mixed-effects models (predicting disjunction scores), the inverse hyperbolic sine (arsinh) function was applied to the ranks.
This transformation was used to minimize the impact of outliers, which are prevalent in unrestricted ranks.
While logarithmic transformation is often employed for such cases it cannot handle zero values.
The arsinh transformation can handle negative numbers. 

\subsection{Results}

The negative ranks from three questionnaires: language demographics, temperature ranges, and music preferences, were analyzed using descriptive statistics, focusing on the median and Interquartile Range (IQR) to illustrate response variability.
Lower IQR values indicate greater consensus, while higher IQR values suggest broader disbelief variability.
For each questionnaire, negative ranks for each proposition were visualized with a logarithmic y-axis for better visibility.

\begin{center}
  
\begin{knitrout}
\definecolor{shadecolor}{rgb}{0.969, 0.969, 0.969}\color{fgcolor}
\includegraphics[width=\maxwidth,]{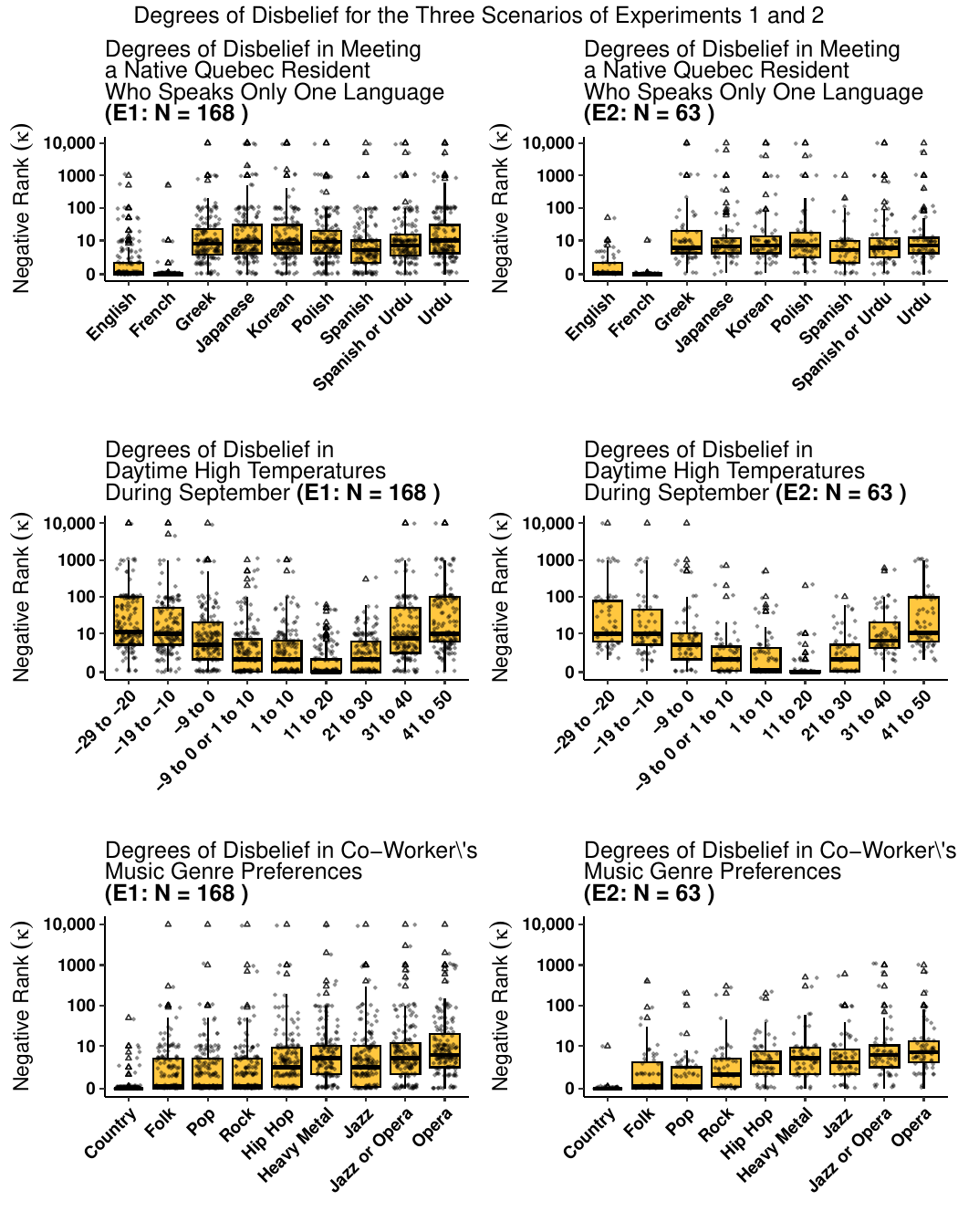} 
\end{knitrout}
\end{center}

\clearpage

\begin{center}
  \captionof{figure}{Boxplots for all three questions of both Experiments 1 and 2 are shown. The left column shows data for Experiment 1, and the right column for Experiment 2. The top row shows the degrees of disbelief (negative rank) for encountering a Quebec resident who speaks only one language. The y-axis is displayed on a logarithmic scale for enhanced visibility. Both Experiments show low disbelief for French and English and higher disbelief for less commonly spoken languages. The second row shows the boxplots for the degrees of disbelief (negative rank) across various daytime high temperature ranges in Boston during September. The 11 to 20°C temperature range exhibits the lowest disbelief. Extreme temperatures, such as -29 to -20°C and 41 to 50°C, show higher disbelief. The bottom row shows boxplots for the degrees of disbelief (negative rank) for various music genres a co-worker might sing along to at a karaoke party. Country music exhibits the lowest disbelief, suggesting participants consider it typical for the co-worker. In contrast, genres such as opera show higher disbelief, reflecting their perceived rarity.}
  \label{fig:E1_E2_All_boxplots}
\end{center}

Figure~\ref{fig:E1_E2_All_boxplots} and Table~\ref{combined_E1_E2_stats} present the degrees of disbelief for language demographics across Experiments 1 and 2, showing consistent patterns of disbelief across both experiments. 
French was widely accepted as the predominant language in Quebec, with a median disbelief of 0 and IQR of 0 in both experiments, indicating no disbelief.

English had a median disbelief of 1 and IQR of 2 in both Experiments 1 and 2, showing it was slightly surprising but maintained the lowest possible median disbelief across both experiments.

In contrast, languages such as Polish, Japanese, and Urdu received higher disbelief scores, with greater variability (IQR) in both experiments, reflecting their perceived rarity as sole languages among Quebec residents.

Overall, the findings in Experiment 2 confirmed the initial results of Experiment 1, showing that disbelief levels for each language remained consistent across both experiments.
This pattern demonstrated the perceived commonality of French and English in Quebec and the higher disbelief associated with less commonly spoken languages, reinforcing the robustness of these results.

\begin{table}[ht]
\centering
\begin{tabular}{rlrrrr}
  \hline
 & Propositions & Median (E1) & IQR (E1) & Median (E2) & IQR (E2) \\ 
  \hline
1 & English & 1.00 & 2.00 & 1.00 & 2.00 \\ 
  2 & French & 0.00 & 0.00 & 0.00 & 0.00 \\ 
  3 & Greek & 8.50 & 26.00 & 6.00 & 16.00 \\ 
  4 & Japanese & 9.00 & 36.00 & 7.00 & 9.00 \\ 
  5 & Korean & 8.50 & 27.25 & 7.00 & 9.50 \\ 
  6 & Polish & 9.00 & 21.75 & 7.00 & 14.50 \\ 
  7 & Spanish & 5.00 & 8.00 & 5.00 & 8.00 \\ 
  8 & Spanish or Urdu & 8.00 & 21.25 & 6.00 & 9.00 \\ 
  9 & Urdu & 10.00 & 45.25 & 7.00 & 8.50 \\ 
   \hline
\end{tabular}
\caption{Descriptive statistics for Language Demographics, showing the median and interquartile range (IQR) of negative rank scores for each language. The IQR represents the middle 50\% of the data, indicating more consensus with smaller values and greater variability in disbelief scores with larger values.} 
\label{combined_E1_E2_stats}
\end{table}

Figure~\ref{fig:E1_E2_All_boxplots} (middle row) and Table~\ref{combined_E1_E2_stats} illustrate participants’ levels of disbelief regarding various daytime high temperatures in September across both experiments.
A consistent U-shaped distribution was observed, with moderate temperatures (11 to 20°C) eliciting low disbelief and extreme temperatures generating higher disbelief scores.

In Experiment 1, moderate temperatures (11 to 20°C) had a median disbelief of 0 and an IQR of 2, reflecting broad consensus on these temperatures as typical and expected for September.
In contrast, extreme temperatures such as -29 to -20°C and 41 to 50°C showed the highest disbelief, with median disbelief scores of 15 and 13.5, respectively, and notable variability in ratings (IQR of 94). 

Experiment 2 replicated these findings, with moderate temperatures similarly perceived as typical for September in Boston, showing low disbelief.
Extreme temperatures continued to produce the highest median disbelief and IQR values, indicating variability in participants’ assessments of rarity.

The broader IQRs, generally observed for rarer outcomes, highlight the diversity in expectations and the degree of surprise, likely influenced by personal experiences and regional climate knowledge.
For detailed statistics on each temperature range, refer to Table~\ref{tbl:E1_E2_Q2_stats}.

\begin{table}[ht]
\centering
\begin{tabular}{rlrrrr}
  \hline
 & Temperature Range & Median (E1) & IQR (E1) & Median (E2) & IQR (E2) \\ 
  \hline
2 & -29 to -20 & 15.00 & 94.00 & 10.00 & 84.00 \\ 
  1 & -19 to -10 & 10.00 & 57.50 & 10.00 & 40.00 \\ 
  3 & -9 to 0 & 6.00 & 18.00 & 5.00 & 8.00 \\ 
  4 & -9 to 0 or 1 to 10 & 2.00 & 7.25 & 2.00 & 3.50 \\ 
  5 & 1 to 10 & 2.00 & 7.00 & 1.00 & 4.00 \\ 
  6 & 11 to 20 & 0.00 & 2.00 & 0.00 & 1.00 \\ 
  7 & 21 to 30 & 2.00 & 6.00 & 2.00 & 4.50 \\ 
  8 & 31 to 40 & 8.00 & 47.00 & 7.00 & 16.00 \\ 
  9 & 41 to 50 & 13.50 & 94.00 & 11.00 & 94.00 \\ 
   \hline
\end{tabular}
\caption{Descriptive statistics for Temperature Ranges, showing the median and interquartile range (IQR) of negative rank scores for each temperature range in Experiments 1 and 2. The IQR represents the middle 50\% of the data, indicating greater consensus with smaller values and increased variability in disbelief scores with larger values.} 
\label{tbl:E1_E2_Q2_stats}
\end{table}

Figure~\ref{fig:E1_E2_All_boxplots} (bottom row) and Table~\ref{tbl:E1_E2_Q3_stats} revealed participants’ levels of disbelief regarding various music genres that Tom, a hypothetical co-worker, might sing along to at a karaoke party.
Across both experiments, country music elicited the least disbelief, with median and IQR values of \(0\) in Experiment 1 and \(0\) in Experiment 2, indicating strong consensus among participants that this genre was expected for Tom.
Similarly, genres such as Pop, Rock, and Folk elicited surprise but to a minimal degree, thus could be considered plausible genres.

Conversely, genres like Heavy Metal, Jazz, and Opera showed higher disbelief across both experiments.
Opera yielded the highest disbelief, with a median of \(6\) in Experiment 1 and \(7\) in Experiment 2, along with substantial variability in ratings (IQR of \(17\) and \(9.5\)).
This variability likely reflects participants’ differing cultural expectations regarding genre preferences for a typical co-worker.

The consistent disbelief patterns across experiments suggest that cultural expectations play a significant role in shaping perceptions of musical preferences.
While some genres were widely accepted as typical, others were perceived as surprising.
These findings illustrate how cultural context, rather than probability alone, influences judgments in the hypothetical scenario.
Detailed statistics for each genre are provided in Table~\ref{tbl:E1_E2_Q3_stats}.

\begin{table}[ht]
\centering
\begin{tabular}{rlrrrr}
  \hline
 & Music Genre & Median (E1) & IQR (E1) & Median (E2) & IQR (E2) \\ 
  \hline
1 & Country & 0.00 & 0.00 & 0.00 & 0.00 \\ 
  2 & Folk & 1.00 & 5.00 & 1.00 & 4.00 \\ 
  3 & Heavy Metal & 5.00 & 8.00 & 5.00 & 7.00 \\ 
  4 & Hip Hop & 3.00 & 9.00 & 4.00 & 5.50 \\ 
  5 & Jazz & 3.00 & 9.00 & 4.00 & 6.50 \\ 
  6 & Jazz or Opera & 5.00 & 10.25 & 6.00 & 7.50 \\ 
  7 & Opera & 6.00 & 17.00 & 7.00 & 9.50 \\ 
  8 & Pop & 1.00 & 5.00 & 1.00 & 3.00 \\ 
  9 & Rock & 1.00 & 5.00 & 2.00 & 4.00 \\ 
   \hline
\end{tabular}
\caption{Descriptive statistics for Music Preferences, showing the median and interquartile range (IQR) of negative rank scores for each genre in Experiments 1 and 2. The IQR represents the middle 50\% of the data, indicating greater consensus with smaller values and increased variability in disbelief scores with larger values.} 
\label{tbl:E1_E2_Q3_stats}
\end{table}

The relationship between negative ranks and subjective probabilities was expected to resemble a logarithmic function, where the negative ranking function is the logarithm of the probability with base \( b \), where \( b \in (0,1) \) .

For propositions where participant responses reflected no disbelief (a negative rank of zero), the range of subjective probabilities spanned from 0 to 1 as can easily be seen in Figure~\ref{fig:e12_subjProb_negRank_withFits} (the dashed blue line represents the threshold where disbelief is greater than 0).

Additionally, Figure~\ref{fig:e12_subjProb_negRank_withFits} shows the nonlinear relationship between negative rank and subjective probability.
The threshold at which a proposition becomes surprising varies and highlights the context-dependent nature of subjective probability thresholds for disbelief.
While it is true that generally a high probability estimate was associated with a lack of surprise or disbelief there was much less consistency for low probability estimates.
Negative ranks offer a greater granularity in assessing degrees of disbelief.

Overall, lower probabilities were associated with greater degrees of disbelief.
However, probabilities ranging from 0 to 1 may not always elicit surprise or disbelief, demonstrating the value of negative rank in explicitly reporting varying degrees of disbelief.

\begin{knitrout}
\definecolor{shadecolor}{rgb}{0.969, 0.969, 0.969}\color{fgcolor}\begin{figure}[H]
\includegraphics[width=\maxwidth,]{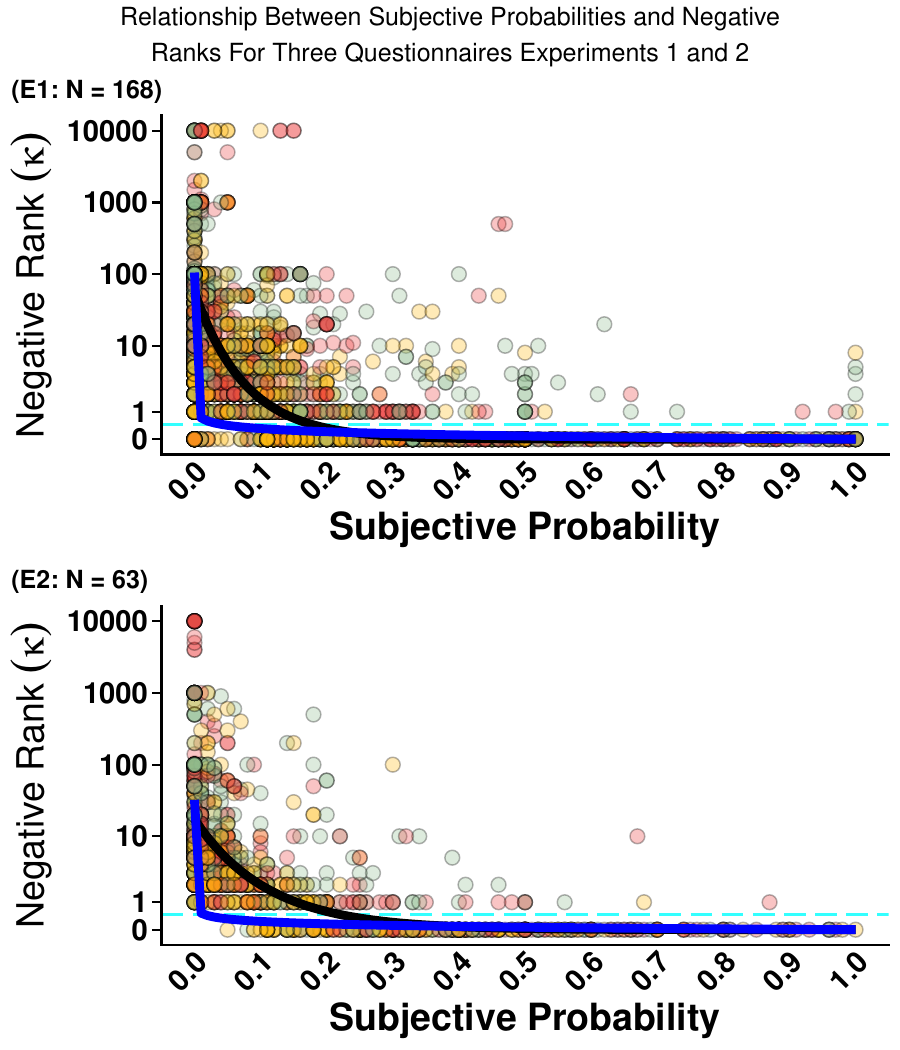} \caption[Scatter plot showing the relationship between subjective probabilities and negative ranks ($\kappa$) across all questionnaires]{Scatter plot showing the relationship between subjective probabilities and negative ranks ($\kappa$) across all questionnaires. The y-axis is scaled to effectively display negative ranks, with labels representing the original rank values to aid interpretation. The top panel is the data for Experiment 1 and the bottom panel for Experiment 2. For Experiment 1 a logarithmic function (black) had a pseudo R squared of 3.90 and exponential decay model (blue) had a pseudo R squared of  8.35. For Experiment 2 the pseudo R squared for the logarithmic fit (black) was -6.38 and for the exponential decay model (blue) 40.50.}\label{fig:e12_subjProb_negRank_withFits}
\end{figure}

\end{knitrout}

Figure~\ref{fig:e12_subjProb_negRank_withFits} confirmed the nonlinear relationship between subjective probabilities and negative ranks.
Unlike in Experiment 1 — where a negative rank of 0 extended across the entire probability scale — Experiment 2 displayed a distinct pattern: subjective probabilities approaching or at 0 consistently corresponded with negative ranks of 1 or higher, indicating disbelief.
This pattern clearly demonstrated that extremely low probability values were reliably associated with higher degrees of disbelief.
For this cohort a negative rank of 0 (which indicates either a lack of disbelief or belief) had subjective probabilities ranging from 0.1 to 1.

Across the three questionnaires, the point at which a subjective probability was associated with disbelief varied.
Specifically, for music preferences, a subjective probability of less than 0.2 was linked to disbelief, while for temperature ranges, a subjective probability of less than 0.35 elicited disbelief for most participants.
This variation highlighted that the subjective probabilities considered surprising or disbelieved depended on the specific context.

There are some theoretical reasons to think that probability and negative ranks for disbelief should be logarithmically related, albeit with a small base.
This relationship is described by the equation:
\[
\kappa(A) = \log_b(P(A)) - \log_b(\max_{S \in W} P(S))
\]
where \( P(A) \) represented the subjective probability of event \( A \) and \( b \) is the logarithmic base, ranging between 0 and 1.
To manage extreme values and enhance the visual representation in plots the asinh was applied to the negative ranks (\(\kappa\)).
This function effectively stabilized extreme values without distorting the underlying data distribution.

A non-linear least squares regression was employed to estimate the logarithmic base \( b \) yielding for Experiment 1:
\[
b = 0.001, \quad \text{SE} = 0.000, \quad t(1) = 4.91, \emph{p}  < .001 
, \] and for Experiment 2 \[
b = 0.0002, \quad \text{SE} = 0.000, \quad t(1) = 4.81, \emph{p}  < .001
. \]

There were significant logarithmic fits.
Pseudo \( R^2 \) was computed to provide an approximation of the variance explained by the model in the non-linear context, and was supplemented by the AIC (Akaike Information Criterion) and BIC (Bayesian Information Criterion).
The model yielded in Experiment 1 a pseudo \( R^2 \) of \(3.90\) for the logarithmic relationship (see Figure~\ref{fig:e12_subjProb_negRank_withFits}) and in Experiment 2 a pseudo \( R^2 \) of \(-6.38\).
A negative pseudo \( R^2 \) suggests that the model does not explain the variance in negative ranks and performs worse than a horizontal line (through the mean value of negative ranks), implying that the logarithmic model is not suitable for this data, (see Figure~\ref{fig:e12_subjProb_negRank_withFits}).

While there was a statistically detectable relationship in Experiment 1 very little variance was explained.
So, as a comparison, we also fit an exponential decay model with the secondary purpose of exploring how disbelief might diminish exponentially as subjective probability increases, reflecting the natural decay of disbelief with increased certainty.
The equation for this model is described as:
\[
\kappa(A) = a \cdot e^{b \cdot P(A)}
\]
where \( a \) and \( b \) were parameters to be estimated.

The model fitting process involved non-linear least squares regression, yielding statistically significant parameters for Experiment 1:
\[
a = 4.73, \quad b = -12.97, \quad \text{SE}_{a} = 0.142, \quad \text{SE}_{b} = 1.318, \quad \textit{p}_{a}  < .001, \quad \textit{p}_{b}  < .001
\] and for Experiment 2 of \[
a = 3.54, \quad b = -9.01, \quad \text{SE}_{a} = 0.062, \quad \text{SE}_{b} = 0.599, \quad \textit{p}_{a}  < .001, \quad \textit{p}_{b}  < .001
. \] 

The fit of the exponential model in Experiment 1 was a pseudo \( R^2 \) value of \(0.084\), indicating that approximately \(8.35\%\) of the variance in negative ranks was explained by this exponential relationship (see Figure~\ref{fig:e12_subjProb_negRank_withFits}), and for Experiment 2 was a pseudo \( R^2 \) value of \(0.405\), indicating that approximately \(40.50\%\) of the variance in negative ranks was explained by this exponential relationship (see Figure~\ref{fig:e12_subjProb_negRank_withFits}).

To determine which model more accurately described the relationship between subjective probability and negative ranks (arsinh transformed), we computed the AIC and BIC values.
These criteria assessed model fit while penalizing the number of parameters to avoid over-fitting.
The logarithmic model had one parameter, as opposed to two parameters estimated in the exponential model.
The results indicated that the exponential model provided a better fit of Experiment 1 data with an AIC of 25393.18 compared to 25582.68 for the logarithmic model, yielding a difference of 189.50.
Similarly, the BIC for the exponential model was 25412.09, whereas it was 25595.28 for the logarithmic model, resulting in a difference of 183.20.
Results were similar in Experiment 2 with the exponential model providing a substantially better fit, with an AIC of 5592.55 compared to 6469.18 for the logarithmic model, yielding a difference of 876.63.
Similarly, the BIC for the exponential model was 5608.51, whereas it was 6479.82 for the logarithmic model, resulting in a difference of 871.31.

\begin{knitrout}
\definecolor{shadecolor}{rgb}{0.969, 0.969, 0.969}\color{fgcolor}\begin{figure}[H]
\includegraphics[width=0.8\linewidth,]{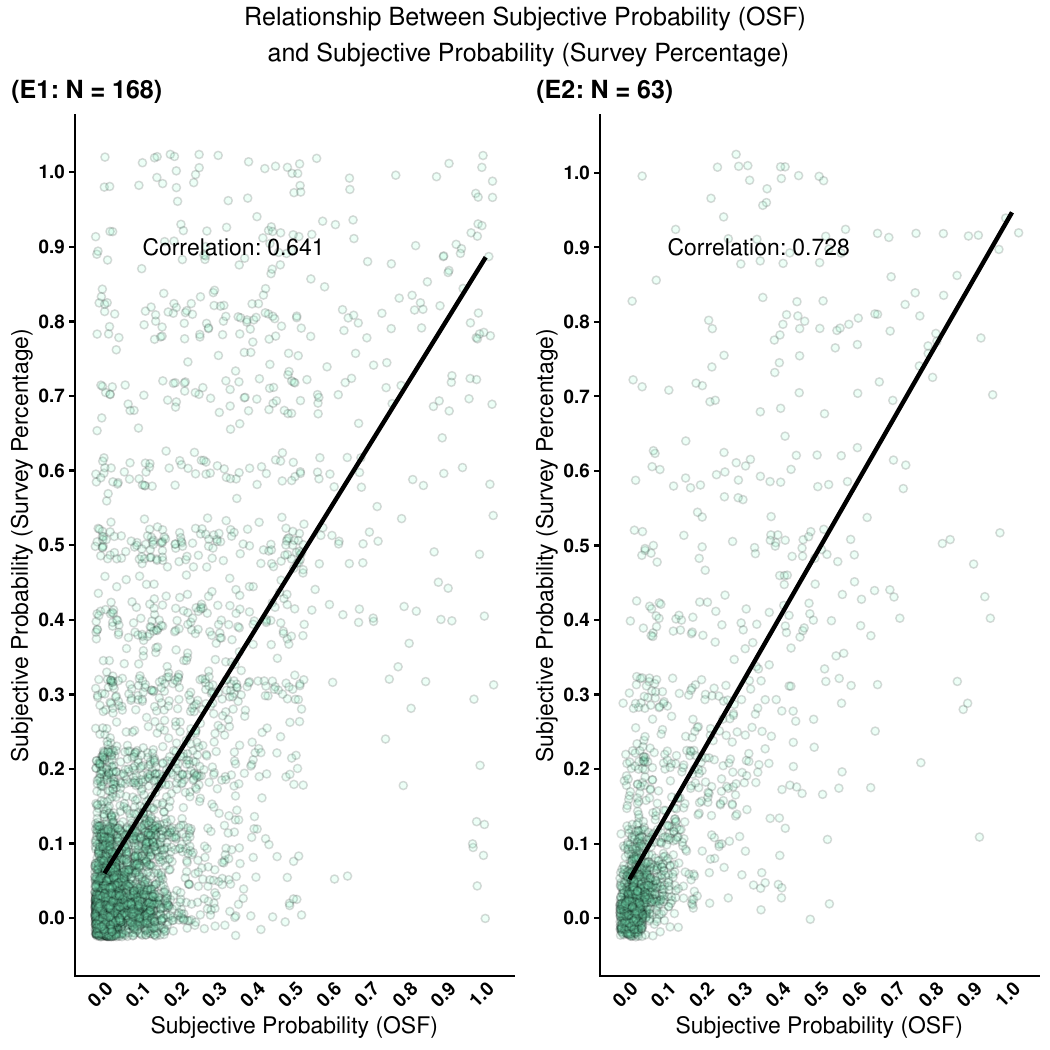} \caption[Scatterplot showing the relationship between subjective probabilities recorded via OSF and subjective probabilities recorded as percentages at the end of the task]{Scatterplot showing the relationship between subjective probabilities recorded via OSF and subjective probabilities recorded as percentages at the end of the task. For Experiment 1 the correlation coefficient was  0.641 , and for Experiment 2  0.728  indicating a strong relationship between the two measures for both experiments.}\label{fig:e12_prob_plots}
\end{figure}

\end{knitrout}

The relationship between subjective probabilities recorded via the OSF immediately after each scenario and the subjective probabilities recorded as percentages at the end of the task showed a correlation in Experiment 1 of 0.641 that was significant (\( \textit{p}  < .001\); Figure \ref{fig:e12_prob_plots} left panel).
Similarly for Experiment 2 there was a strong correlation (r = 0.728, \emph{p}  < .001; Figure \ref{fig:e12_prob_plots} right panel).

Recall that an interesting consequence of negative ranking theory is that when judging an OR statement of two mutually exclusive propositions the negative rank assigned to the \emph{disjunction} should be the minimum of the negative rank assigned to each individually.
We intentionally included a disjunctive question in each scenario.
For example in the language scenario it involved the propositions ``Spanish'' and separately ``Urdu'', as well as ``Spanish or Urdu''.

The disjunction box plot, shown in Figure \ref{fig:e12_disjunction_plots}, illustrates these relationships.
For the language and music scenarios the rank assigned to the disjunction was not the minimum, but an intermediate value.

Lastly, the percentage of participants who had their disjunction equal to the minimum was 38.69\%,
while
35.12\% of participants were equal to the maximum, and 26.79\% of the participant were between minimum and maximum.

\begin{knitrout}
\definecolor{shadecolor}{rgb}{0.969, 0.969, 0.969}\color{fgcolor}\begin{figure}[H]
\includegraphics[width=\linewidth,]{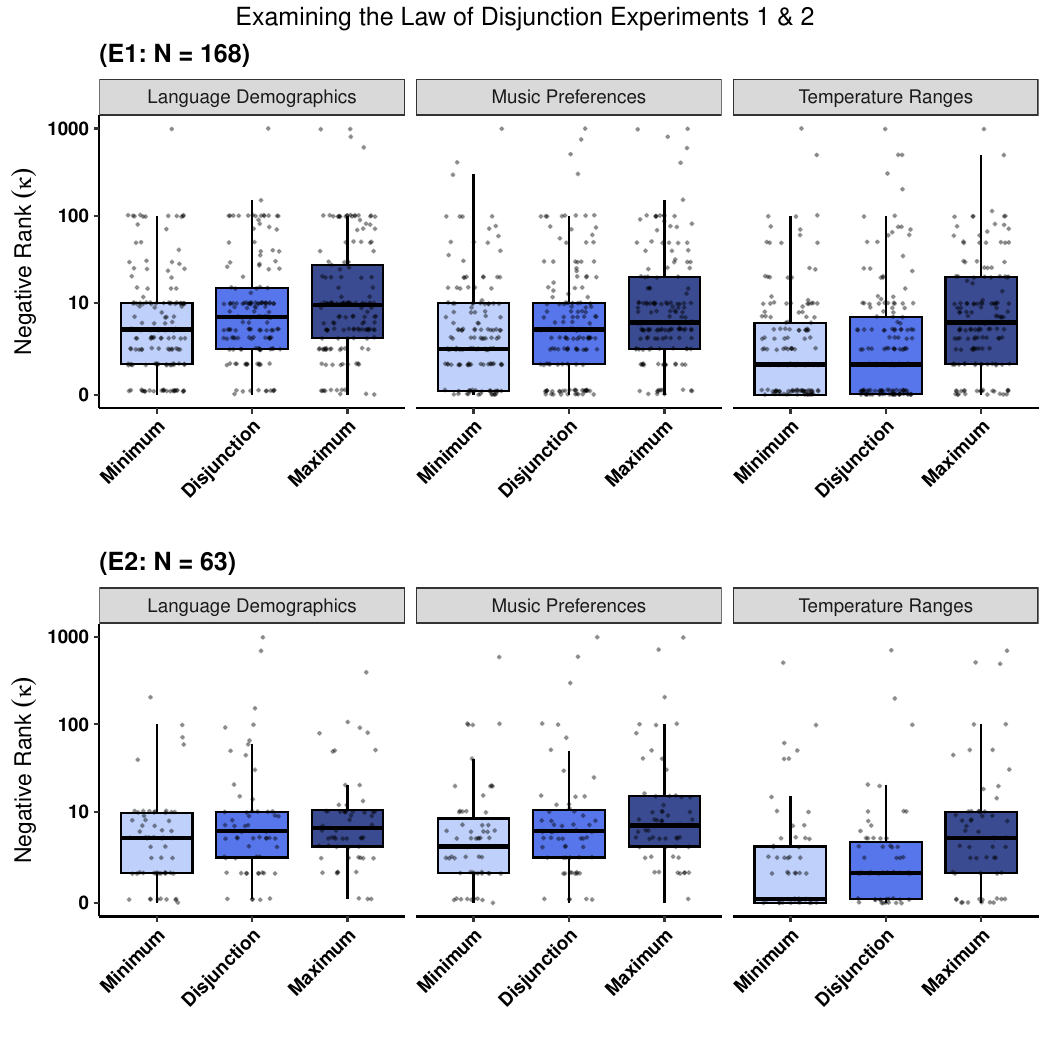} \caption[Examining the law of disjunction]{Examining the law of disjunction. The y-axis is on a logarithmic scale to display a wider range of observations. The disjunction rank corresponds to the logical ``or'' relationship, reflecting the combined disbelief that at least one of the propositions is true. The dots are jittered points representing individual participant responses.}\label{fig:e12_disjunction_plots}
\end{figure}

\end{knitrout}

Results were similar for Experiment 2 (Figure~\ref{fig:e12_disjunction_plots}).
Participant disjunction rankings showed variability: 38.62\% aligned with the minimum, while 39.15\% with the maximum, and 25.93\% were in-between.
These results illustrate the diverse judgments of participants when grading their disbelief for disjunctions.

Further exploring this disjunctive relationship we performed a linear mixed-effects model to predict disjunction ranks in instances where the minimum rank differed from the maximum rank among the propositions.
The number of observations removed due to ties between the minimum and maximum ranks of the two propositions was 26.79\%.

Negative ranks were transformed using the arsinh method to accommodate extreme values, which ranged from 0 to very high numbers.
The Pearson correlation coefficient showed a high multicollinearity of 0.96, which occurred because the two independent variables, minimum rank and maximum rank, were highly correlated, indicating that the two propositions in the disjunction had similar disbelief ratings.
We performed the linear mixed-effects model to try and get some further insight, but readers should note the violation of statistical assumptions.

A null model was conducted to predict disjunction (arsinh-transformed) with a random intercept for subjects.
This model yielded an Intraclass Correlation Coefficient (ICC) of 0.74.
This ICC value indicated that 74.03\% of the total variance in disjunction scores was attributable to differences between subjects.
This suggested that individual differences significantly predicted disjunction, necessitating the inclusion of random intercepts in subsequent models to account for subject variability.

Interactions between questionnaire type and both minimum and maximum ranks were examined.
To account for individual variability, a random intercept for participants was included.
Using temperature ranges as the reference category for questionnaire type, the estimated effect of the minimum rank (\(\beta\) = 1.05, SE = 0.03, t = 33.1, \textit{p}  < .001) reflected the relationship between the minimum rank and disjunction rank within the reference category of temperature ranges and when the maximum rank was 0.
Similarly, the non-significant estimate for the maximum rank (\(\beta\) = 0.01, SE = 0.03, t = 0.48, \textit{p}  = .632) applied to this same condition.
However, there were significant interactions where questionnaire type influenced both the minimum and maximum ranks.
Specifically, the interactions with music preferences and language demographics significantly decreased the estimate for minimum rank (\(\beta\) = -0.92, SE = 0.08, t = -12.3, \textit{p}  < .001); (\(\beta\) = -1.06, SE = 0.09, t = -11.8, \textit{p}  < .001) and increased the estimate for maximum rank (\(\beta\) = 0.84, SE = 0.06, t = 13.81, \textit{p}  < .001); (\(\beta\) = 0.59, SE = 0.07, t = 8.71, \textit{p}  < .001).
These interaction effects indicated that the relationship between minimum or maximum rank and disjunction varied significantly across questionnaire types.
Refer to Table~\ref{tbl:E1_lmer_model} for full details of the predicted model.

Figure~\ref{fig:e12_disjunction_min-max_plots} illustrates that disjunction was effectively predicted by the minimum rank for temperature ranges aligning with the expectations of negative ranking theory.
In contrast, the maximum rank seemed to better predict disjunction for the other scenarios though ranks for disjunctions were still nominally lower than the maximum rank in most cases.
Consequently, the simple effects for the reference category were not generalizable across all contexts.
Significant interaction effects between the minimum rank and questionnaire type, particularly for temperature ranges and music preferences indicates that context significantly influences the predictive power of the minimum and maximum ranks on disjunction.

\begin{table}[ht]
\centering
\begin{tabular}{rrrrrr}
  \hline
 & Est. & S.E. & t val. & d.f. & p \\ 
  \hline
(Intercept) & 0.09 & 0.09 & 0.95 & 356.52 & 0.34 \\ 
  Minimum Rank & 1.05 & 0.03 & 33.10 & 359.99 & 0.00 \\ 
  Maximum Rank & 0.01 & 0.03 & 0.48 & 358.64 & 0.63 \\ 
  Music Preferences & -0.37 & 0.19 & -1.96 & 310.81 & 0.05 \\ 
  Language Demographics & 1.00 & 0.13 & 7.63 & 280.67 & 0.00 \\ 
  Minimum Rank:Music Preferences & -0.92 & 0.08 & -12.30 & 325.89 & 0.00 \\ 
  Minimum Rank:Language Demographics & -1.06 & 0.09 & -11.80 & 346.13 & 0.00 \\ 
  Maximum Rank:Music Preferences & 0.84 & 0.06 & 13.81 & 281.88 & 0.00 \\ 
  Maximum Rank:Language Demographics & 0.59 & 0.07 & 8.71 & 347.12 & 0.00 \\ 
   \hline
\end{tabular}
\caption{Results of the linear mixed-effects model predicting disjunction ranks. This table displays the estimated coefficients, standard errors, t-values, and p-values for the arsinh-transformed minimum and maximum ranks, alongside questionnaire types and their interaction effects.} 
\label{tbl:E1_lmer_model}
\end{table}

\begin{knitrout}
\definecolor{shadecolor}{rgb}{0.969, 0.969, 0.969}\color{fgcolor}\begin{figure}[H]
\includegraphics[width=\maxwidth,]{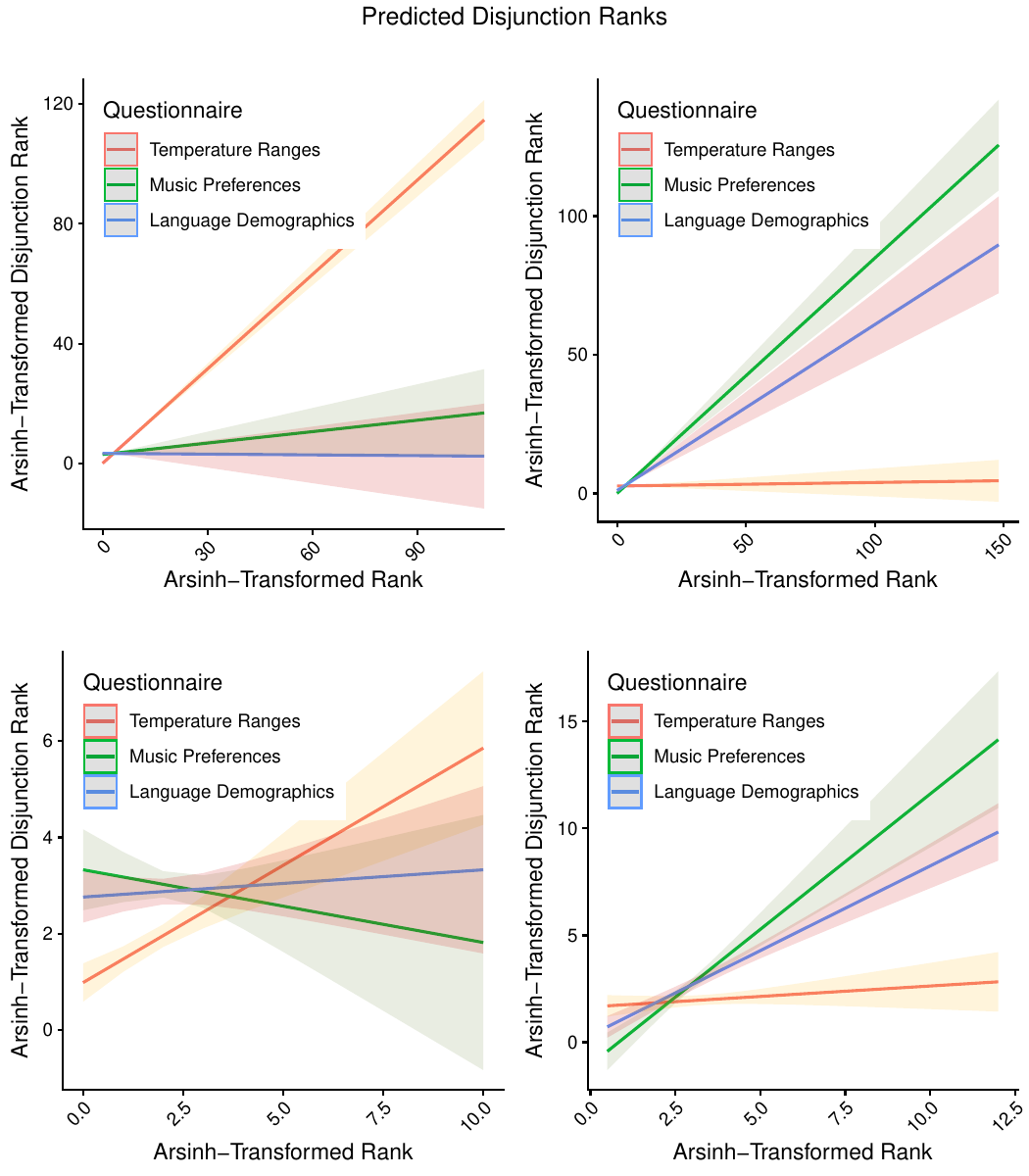} \caption[Predicted disjunction rank values ($\kappa$) for minimum (left) and maximum (right) across different questionnaire types (E1]{Predicted disjunction rank values ($\kappa$) for minimum (left) and maximum (right) across different questionnaire types (E1: top row; E2 bottom row).}\label{fig:e12_disjunction_min-max_plots}
\end{figure}

\end{knitrout}

This analysis was repeated for Experiment 2.
There were 25.4\%, ties in these data, which were excluded from the analysis.
These data too were highly collinear (0.73).

The null mode showed an Intraclass Correlation Coefficient (ICC) of 0.36, indicating that 36.42\% of the variance in disjunction scores was due to differences between subjects.
This significant variation justified the inclusion of random intercepts in further models to account for subject variability.

Subsequent analysis with a linear mixed-effects model included fixed effects for minimum and maximum ranks, and their interactions with questionnaire type.
Using temperature ranges as the reference category for questionnaire type, the estimated effect of the minimum rank (\(\beta\) = 0.49, SE = 0.1, t = 5.01, \textit{p}  < .001) represented the relationship between the minimum rank and disjunction rank within the reference category and when the maximum rank was 0.
With the same reference category, the maximum rank under the same condition did not significantly predict the disjunction ranks (\(\beta\) = 0.1, SE = 0.08, t = 1.21, \textit{p}  = .227).
However, significant interactions indicated that questionnaire type influenced both the minimum and maximum ranks.
Specifically, interactions with music preferences and language demographics significantly decreased the estimate for minimum rank (\(\beta\) = -0.64, SE = 0.2, t = -3.17, \textit{p}  = .002); (\(\beta\) = -0.43, SE = 0.15, t = -2.96, \textit{p}  = .004) and increased the estimate for the maximum rank (\(\beta\) = 1.17, SE = 0.2, t = 5.96, \textit{p}  < .001); (\(\beta\) = 0.7, SE = 0.11, t = 6.39, \textit{p}  < .001).
These interaction effects indicated that the relationships between minimum or maximum rank and disjunction varied significantly across questionnaire types, as observed in Experiment 1.
Refer to Table~\ref{tbl:E2_lmer_model} for full model details.

Figure \ref{fig:e12_disjunction_min-max_plots} again illustrates the significant predictiveness of the minimum rank for temperature ranges of moderate magnitude ( 0.49).
For other scenarios, the maximum rank appeared to be a stronger predictor of disjunction; however, the predicted disjunction values were context-dependent and generally fell below the maximum rank.
These findings emphasized that the main conclusions were based on interaction effects rather than the simple effects of the reference category, which were not generalizable across all contexts.

\begin{table}[ht]
\centering
\begin{tabular}{rrrrrr}
  \hline
 & Est. & S.E. & t val. & d.f. & p \\ 
  \hline
(Intercept) & 0.67 & 0.21 & 3.13 & 131.96 & 0.00 \\ 
  Minimum Rank & 0.49 & 0.10 & 5.01 & 131.74 & 0.00 \\ 
  Maximum Rank & 0.10 & 0.08 & 1.21 & 131.94 & 0.23 \\ 
  Music Preferences & -1.43 & 0.37 & -3.83 & 111.71 & 0.00 \\ 
  Language Demographics & -0.47 & 0.31 & -1.48 & 111.16 & 0.14 \\ 
  Minimum Rank:Music Preferences & -0.64 & 0.20 & -3.17 & 131.97 & 0.00 \\ 
  Minimum Rank:Language Demographics & -0.43 & 0.15 & -2.96 & 118.14 & 0.00 \\ 
  Maximum Rank:Music Preferences & 1.17 & 0.20 & 5.96 & 131.83 & 0.00 \\ 
  Maximum Rank:Language Demographics & 0.70 & 0.11 & 6.39 & 100.53 & 0.00 \\ 
   \hline
\end{tabular}
\caption{Results of the linear mixed-effects model predicting disjunction ranks. This table displays the estimated coefficients, standard errors, t-values, and p-values for the arsinh-transformed minimum and maximum ranks, alongside questionnaire types and their interaction effects.} 
\label{tbl:E2_lmer_model}
\end{table}

At the conclusion of the task, participants were asked to rate the difficulty of grading degrees of disbelief and to hypothetically evaluate whether it would be easier to assign degrees of belief instead.

In Experiment 1 most participants found the task moderately challenging, with the most common rating being a 4 on a scale from 1 (Very Easy) to 7 (Very Difficult).
This rating, chosen by approximately 30\% of participants, suggests a balanced complexity of the task.
The fewer responses at the extremes of the scale, with ratings of 1 and 7, imply that participants neither found the task as very easy or very difficult.
This distribution is depicted in Figure~\ref{fig:e1E2CombPlotAlt}.
When asked hypothetically about the ease of assigning degrees of belief over disbelief, the majority of responses centered around a neutral rating of 4, ``Neither Agree nor Disagree''.
This suggests a neutral stance by most participants, as illustrated.

\begin{knitrout}
\definecolor{shadecolor}{rgb}{0.969, 0.969, 0.969}\color{fgcolor}\begin{figure}[H]
\includegraphics[width=\maxwidth,]{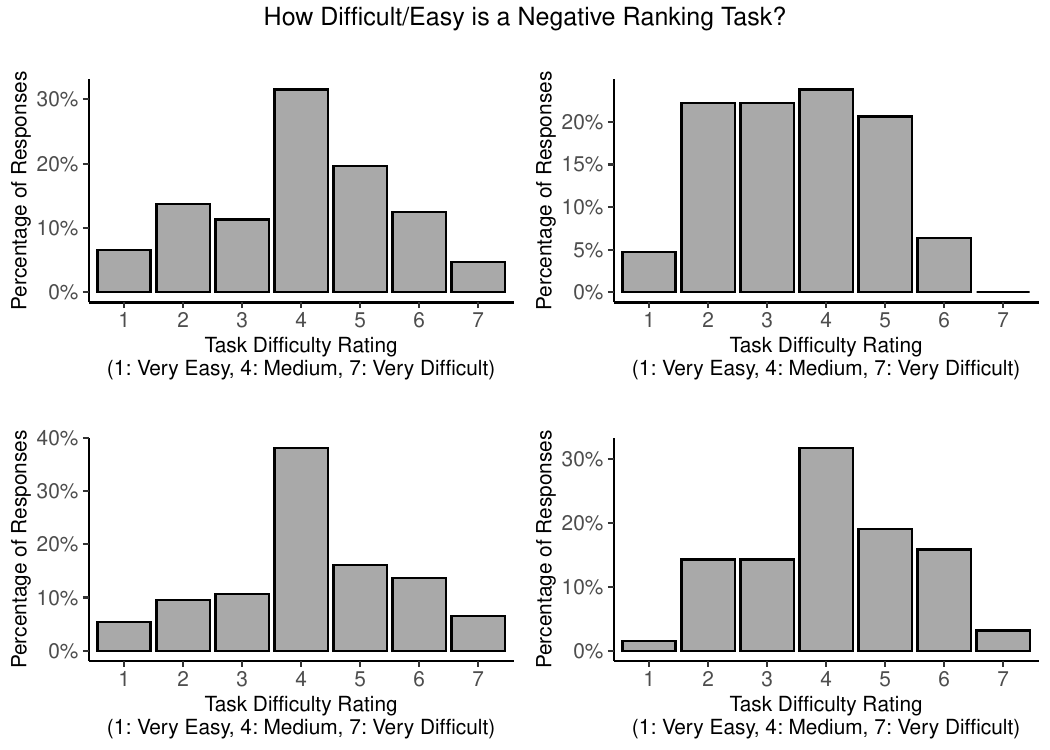} \caption[Bar graphs depicting the distribution of responses to the perceived difficulty (top row) or ease (bottom row) of assigning degrees of belief over degrees of disbelief for Experiment 1 (left column) and Experiment 2 (right column)]{Bar graphs depicting the distribution of responses to the perceived difficulty (top row) or ease (bottom row) of assigning degrees of belief over degrees of disbelief for Experiment 1 (left column) and Experiment 2 (right column). The most frequent ratings were neutral (Rating 4)}\label{fig:e1E2CombPlotAlt}
\end{figure}

\end{knitrout}

Experiment 2 showed broadly similar ratings (Figure~\ref{fig:e1E2CombPlotAlt}).
Difficulty ratings ranged from 1 ``very easy,'' to 7 ``very difficult,'' with the majority clustering between 2 and 5 and a modal rating of 4 as observed in Experiment 1.
Figure~\ref{fig:e1E2CombPlotAlt} illustrates participants’ perceptions regarding the ease of assigning belief levels compared to disbelief levels.

\subsection{Discussion: Experiments 1 \& 2}
These studies evaluated the feasibility of integrating ranking theory into behavioral experiments in both laboratory and on-line settings.
Participants consistently expressed degrees of disbelief through negative ranks with good concordance between the in-laboratory and on-line testing environments.
This was true even though the scenarios chosen had unclear probabilities. 

The use of negative ranks expands the response range from 0 to 1 to 0 to infinity.
Participants use whole numbers, and do not need to track significant digits to express their disbelief orderings on a decimal scale.
Ranks do not require that all options be kept in working memory.
To make sure all probability reports tally to one one must, in principle, keep a tally of all the probabilities for all the other choices and then renormalize if a new option is added or one taken away.
Dealing with whole numbers and not needing to keep track of sums may compensate for any conceptual difficulty in understanding what is meant by a ''degree of disbelief''.
Whatever the reason, participants do not find it that hard.

While the range could vary from 0 to infinity, most participants were content with relatively small numbers.
Some participants used a scale from 0 to 10, while others opted for a broader range, such as 0 to 100.
Rankings were intuitively sensible.
In the aggregate, and speaking practically, these results demonstrate that participants can use ranks to express degrees of disbelief and that these ranks concord with common sense intuition.

The relation between ranks and subjective probability was also sensible, but noisy.
While a rank of 0 means unsurprising or that a proposition is not disbelieved, the subjective probability that was associated with this was variable and context dependent.
If subjective probability \emph{is} the basis of belief then this variability is puzzling.
Again, while smaller subjective probabilities were, in general, predictive of higher degrees of disbelief, even when estimates of subjective probability were small there was a large dispersion of negative ranks.

Our interest in using negative ranks grew from Spohn’s \emph{normative} theory.
That theory suggests that negative ranks might be logarithmically related to probabilities, albeit with a very small base.
We found little support for this idea in our data.
An exponential decay model was often better in fitting the relationship.
A stronger mark against Spohn’s theory as a full and faithful account of human belief quantification is the failure to find obedience to the law of disjunction.
The minimum degree of belief in a disjunct predicted the joint ranking only for the temperature scenario; for language demographics and music preferences it was instead the maximum degree of belief that had the greater effect. In general, it appeared the participants were averaging rather than adopting the minimum as required by the law.

We began our studies with negative ranks, because that is what Spohn emphasizes as the mathematically most tractable and justifiable, but positive ranks can be computed within the theory and are related to negative ranks.
It seems natural to wonder if positive ranks of belief, instead of negative ranks of disbelief, would be easier for participants and might reveal differences.
The next experiment used a full range of ranks from -infinity to + infinity with zero a point of equipoise: neither believed nor disbelieved; to span from absolute disbelief (or impossibility) to absolute belief (certainty).

\section{Experiment 3: Examining the Full Spectrum of Belief with Positive and Negative Ranking Functions}

\subsection{Introduction}

It is considered nonsensical to disbelieve both a proposition and its negation so ranking theory holds that our degree of disbelief for the disjunction of $A$ and $\bar{A}$ must be 0 (at least one of them is not disbelieved).
We can reframe the negative rank of $\bar{A}$ as a positive rank of $A$.
It is customary to use the letter $\beta$ for the positive ranking function.
This reframing gives us the formula $\beta(A) = \kappa(\bar{A})$.
From this we can compute the two sided ranking function ($\tau$) by subtracting negative ranks from positive ranks.
Exploring this two side ranking function was the goal of Experiment 3.
By including both positive and negative ranks we hoped to deepen our understanding of how people might be using these ranks when operating in settings of uncertainty. 

\subsection{Methods}

A total of 200 undergraduate students from the University of Waterloo participated in this on-line experiment in exchange for course credit.
No exclusion criteria were applied (median age: 19 years; 156 females; 44 males).

Much of the methodology was similar to Experiments 1 \& 2, but we expanded the scenarios and augmented the rating procedures.
Details for these changes follow.

Experiment 3 integrated two new scenarios: ``real estate investment'' and ``lottery ticket.''
These new scenarios were added to the pre-existing ``temperature ranges'' and ``co-worker’s music preferences'' scenarios, and the ``language demographics'' was dropped (this was done to keep the overall duration of the study roughly the same).

The categories for the two new scenarios were: 
\begin{enumerate}
    \item \textbf{Real Estate Investment:} Participants graded their disbelief regarding the value of a Toronto property 5 years from today (negative ranks: increased, decreased, remained the same; positive ranks: did not increase, did not decrease, did not remain the same, and the disjunction, increased or remained the same).

    \item \textbf{Lottery Ticket Scenario:} Participants graded their degrees of disbelief if they won the lottery after purchasing 900 out of 1000 tickets (negative ranks: won the lottery; positive ranks: lost the lottery, and the disjunction: won or lost the lottery).
\end{enumerate}

These scenarios were intended to include situations in which a reasonable estimate of probability could be made.
The lottery ticket scenario was explicit (900 out of 1000 tickets), while the other three relied on broader forms of prior knowledge.
Detailed descriptions of the ranking questionnaires are available in the Appendix \ref{AppendixE3}.

To verify the consistency in the use of positive and negative ranks, participants responded to a simple question for two questionnaires:

\begin{enumerate}
    \item \textbf{Real Estate Investment:} Do you believe the value of your Toronto property will increase 5 years from today? Yes or no.
    
    \item \textbf{Lottery Ticket Scenario:} Do you believe you will win the lottery? Yes or no.
\end{enumerate}

These yes/no responses were compared with the computed two-sided ranks from positive and negative ranks.
To reduce collinearity, we formed the disjunct from propositions that were more extreme than in Experiments 1 and 2.

\subsection{Results}

Figure~\ref{fig:E3_All_boxplots} depicts participants’ levels of disbelief and belief (shown as their disbelief in the converse, that is the ``not'' version) for all four scenarios.
The disbelief rankings are shown in green and the negative ranks for their negations in orange.
There is quite good agreement with the two basically being reflections.
This pattern or reflection is expected if participants are following the law of negation.

\begin{knitrout}
\definecolor{shadecolor}{rgb}{0.969, 0.969, 0.969}\color{fgcolor}\begin{figure}[H]
\includegraphics[width=\maxwidth,]{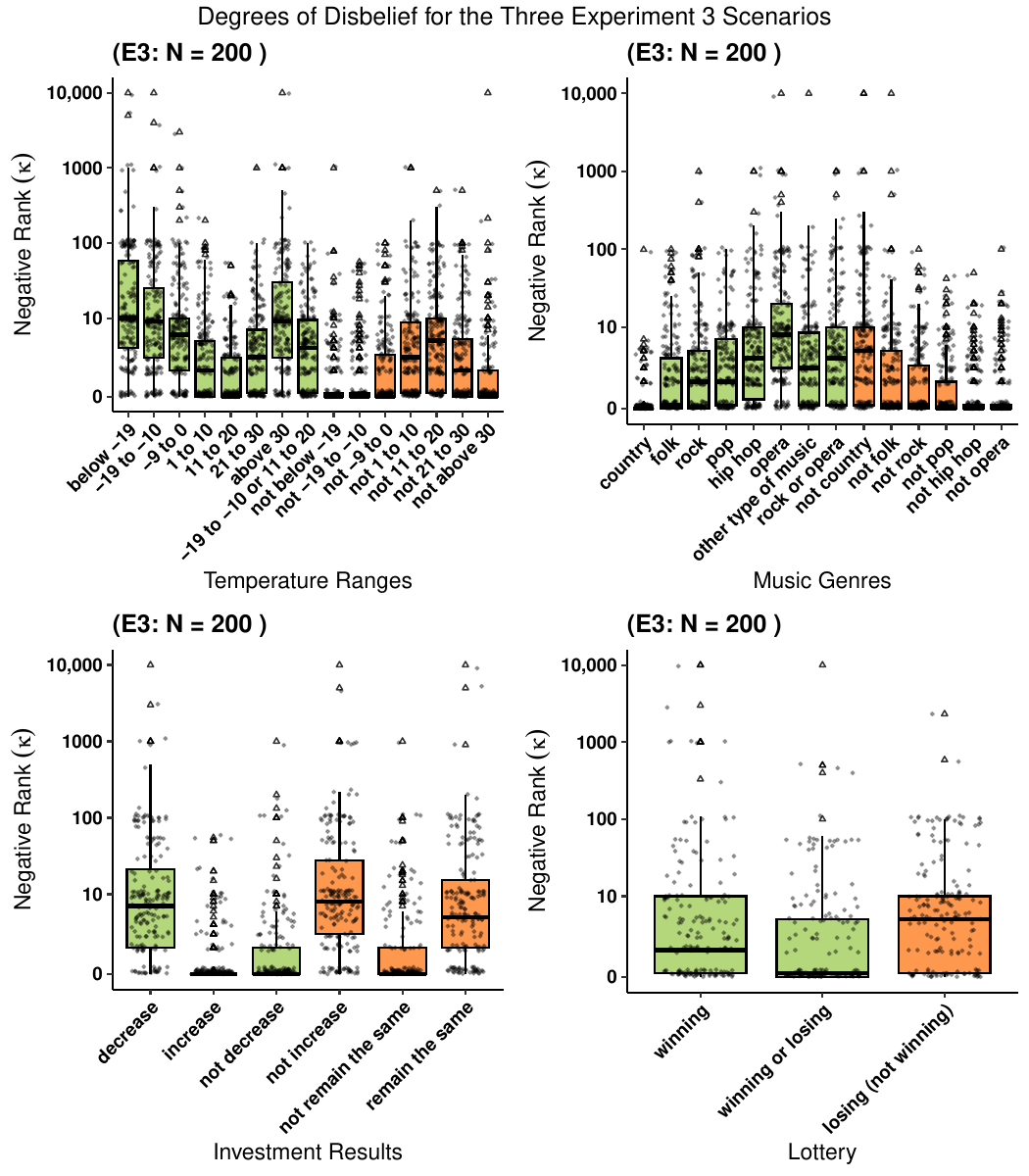} \caption[Boxplots illustrating the degrees of disbelief (green) or belief (orange) as negative ranks for a proposition or its negation for the different questionnaires of Experiment 3]{Boxplots illustrating the degrees of disbelief (green) or belief (orange) as negative ranks for a proposition or its negation for the different questionnaires of Experiment 3. The y-axis is presented on a logarithmic scale to display a broader range of observations. The median and interquartile range (IQR) are indicated for each genre, with lower values signifying lower disbelief.}\label{fig:E3_All_boxplots}
\end{figure}

\end{knitrout}

Figure~\ref{fig:E3_rank_OSF1} illustrates how exponential function better fits the correlation between subjective probability and ranks.
For this graphic we combine the four distinct questionnaires.
Again, a wide range of subjective probabilities for a negative rank of 0 imply that unsurprising events were not limited to high probabilities.

\begin{knitrout}
\definecolor{shadecolor}{rgb}{0.969, 0.969, 0.969}\color{fgcolor}\begin{figure}[H]
\includegraphics[width=\maxwidth,]{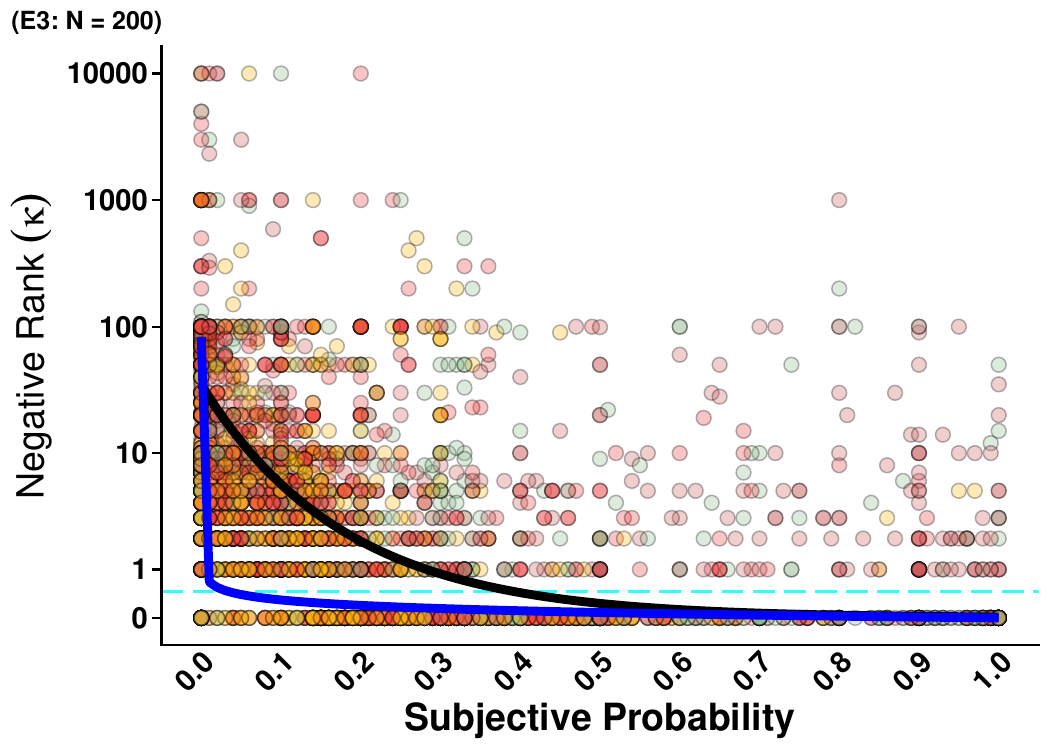} \caption[Scatter plots showing the relationship between subjective probabilities (via OSF) and negative ranks ($\kappa$) across four questionnaires]{Scatter plots showing the relationship between subjective probabilities (via OSF) and negative ranks ($\kappa$) across four questionnaires: Temperature Ranges, Music Preferences, Real-estate Investment, and Lottery. The y-axis is scaled to display negative ranks effectively, while the labels show the original rank values for easier interpretation. These plots highlight the non-linear and context-dependent nature of the relationship between subjective probabilities and negative ranks. A logarithmic fit is shown in blue accounts and an exponential decay model is shown in black.}\label{fig:E3_rank_OSF1}
\end{figure}

\end{knitrout}

\begin{knitrout}
\definecolor{shadecolor}{rgb}{0.969, 0.969, 0.969}\color{fgcolor}\begin{figure}[H]
\includegraphics[width=\maxwidth,]{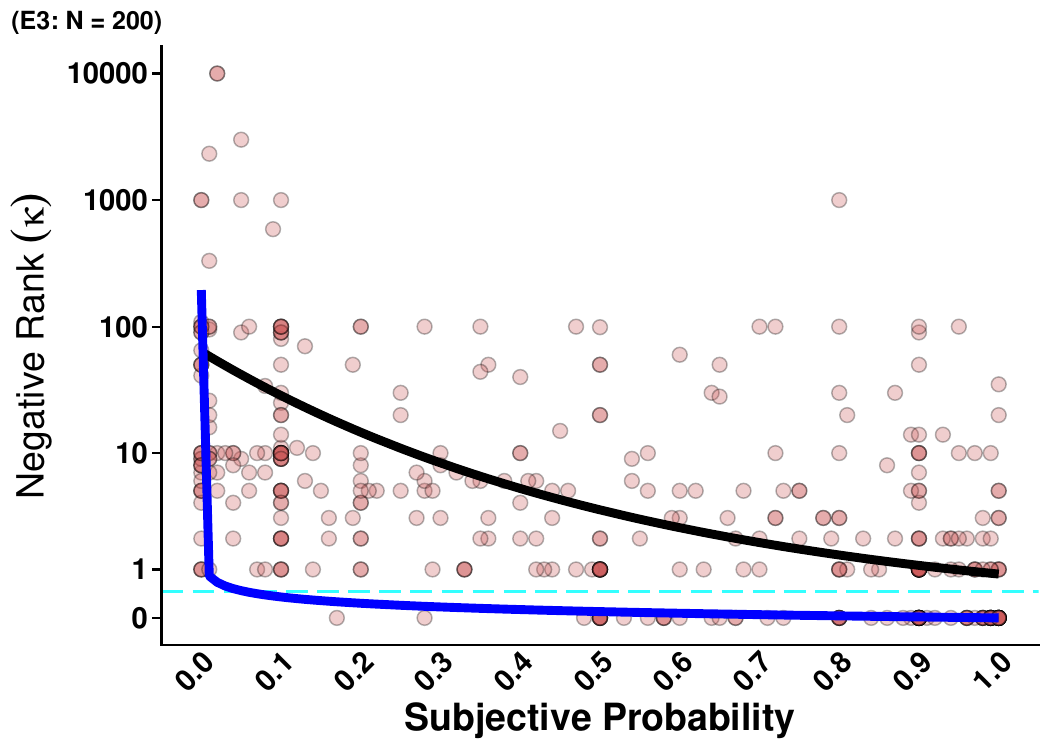} \caption[The scatter plot for the lottery data only are shown, and these are again fit with a logarithmic relation (blue) and an exponential decay model (black)]{The scatter plot for the lottery data only are shown, and these are again fit with a logarithmic relation (blue) and an exponential decay model (black). Even for these ostensible probabilities the data and relationships look the same.}\label{fig:E3_rank_OSF1_lottery}
\end{figure}

\end{knitrout}

The lottery example provides the best opportunity to compare ranks and subjective probability estimates (Figure~\ref{fig:E3_rank_OSF1_lottery}) since in this scenario there is a transparent objective probability that can be inferred from the number of tickets and some reasonable assumptions of fairness.
Despite limiting to these data the same general pattern of an exponential decay with wide variation in the matching of probability to disbelief occurs. 

The visual impressions of a superior fit for the exponential is confirmed statistically. 
For the logarithmic model, the base \( b \) was estimated using non-linear least squares regression:
\[
b = 0.0009, \quad \text{SE} = 0.0003, \quad t(1) = 3.38, \emph{p}  < .001
.\] Again, consistent with Experiments 1 \& 2, the fit was poor with a pseudo \( R^2 \) of -0.02.
For the exponential decay model, parameters \( a \) and \( b \) were also estimated using non-linear least squares regression:
\[
a = 4.24, \quad b = -5.56, \quad \text{SE}_{a} = 0.169, \quad \text{SE}_{b} = 0.593, \quad \textit{p}_{a}  < .001, \quad \textit{p}_{b}  < .001
,\] and yielded a pseudo \( R^2 \) of \(0.050\).
Comparing the two models we computed the AIC and BIC.
The exponential model provided a better fit, with an AIC of 23970.52, compared to 24226.72 for the logarithmic model, yielding a difference of 256.20.
Similarly, the BIC for the exponential model was 23989.25, whereas it was 24239.20 for the logarithmic model, resulting in a difference of 249.96.

There is a relationship between subjective probability and negative ranks.
Improbable events tend to be judged as more surprising (or more disbelieved), but this relationship is not tight, and it is context dependent.
Perhaps this is due to our inability to distinguish no disbelief from actual positive belief.
Figure~\ref{fig:E3_subProb_twosidedRank} plots subjective probabilities (x-axis) and two-sided ranks (arsinh transformed) on the y-axis for all four questionnaires.
While low probabilities align with greater disbelief, and high probabilities with higher belief ranks the relationship remains noisy.
In contrast, the investment and lottery questionnaires, which are arguably closer to objective probability scenarios, showed a more consistent relationship.
These two scenarios are also the ones with more extreme and fewer options.

\begin{knitrout}
\definecolor{shadecolor}{rgb}{0.969, 0.969, 0.969}\color{fgcolor}\begin{figure}[H]
\includegraphics[width=\maxwidth,]{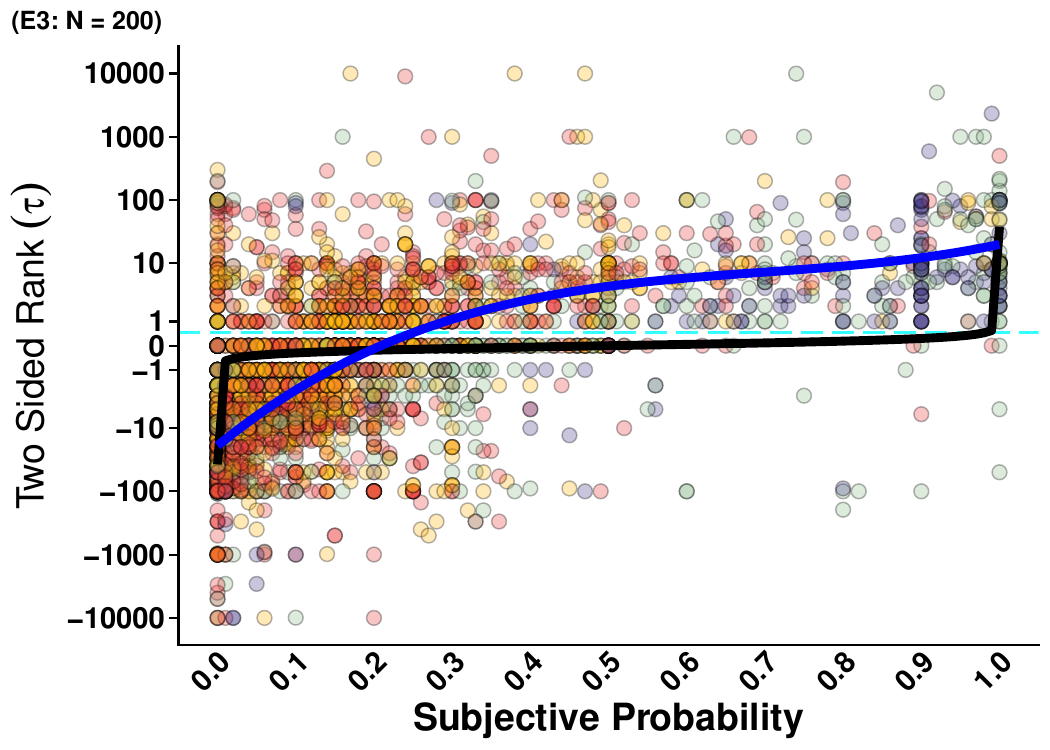} \caption[Scatterplot showing the relationship between subjective probabilities and two-sided ranks ($\tau$) for each distribution configuration]{Scatterplot showing the relationship between subjective probabilities and two-sided ranks ($\tau$) for each distribution configuration. The y-axis is scaled to effectively display both positive and negative ranks, with labels representing the original rank values to aid interpretation. The dotted line represents a two-sided rank of zero, indicating neutral judgment.}\label{fig:E3_subProb_twosidedRank}
\end{figure}

\end{knitrout}

We repeated an assessment of disjunctions for the two sided rank function.
Again, the median disjunction values were usually intermediate, and not the theoretically required minimum.

The percentages of participants whose disjunctions equaled the minimum or maximum did represent a change. 31.25\% of participants ranked the disjunction equal to the minimum. 15.75\% equal to the maximum. 37\% ranked the disjunction between the minimum and maximum.
This was particularly notable for the two scenarios where there were fewer options and actual objective probability may have been easier to derive.

\begin{knitrout}
\definecolor{shadecolor}{rgb}{0.969, 0.969, 0.969}\color{fgcolor}\begin{figure}[H]
\includegraphics[width=\maxwidth,]{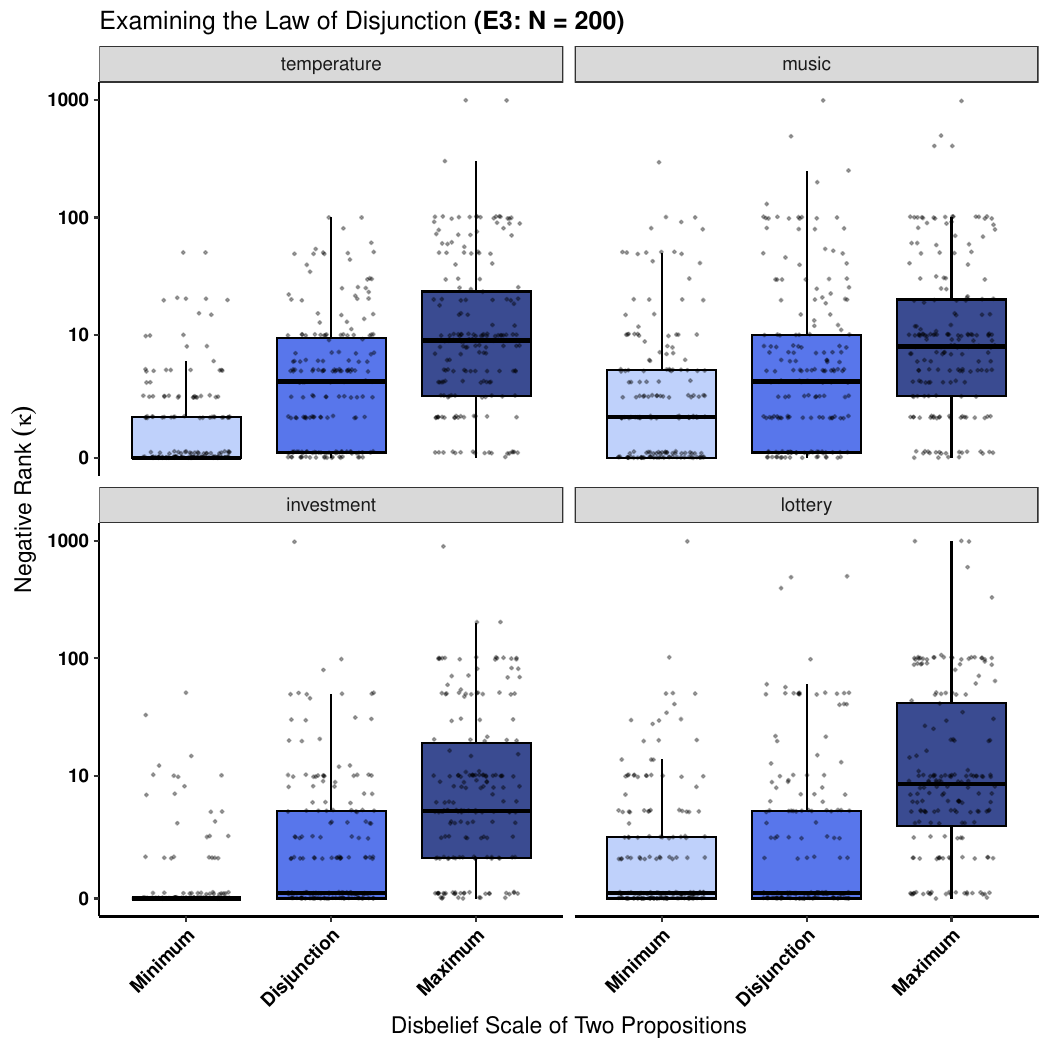} \caption[Boxplots examining the law of disjunction for four questionnaires]{Boxplots examining the law of disjunction for four questionnaires: Temperature Ranges, Music Preferences, Investment Scenarios, and Lottery Outcomes. The y-axis is on a logarithmic scale to display a wider range of observations. The minimum rank represents the least disbelieved of the two propositions involved in the disjunction, while the maximum rank represents the more disbelieved of the two propositions. The disjunction rank corresponds to the logical ``or'' relationship involving the two propositions. The dots are jittered points representing individual participant responses.}\label{fig:E3_disjunction_plot}
\end{figure}

\end{knitrout}

\begin{knitrout}
\definecolor{shadecolor}{rgb}{0.969, 0.969, 0.969}\color{fgcolor}\begin{figure}[H]
\includegraphics[width=\maxwidth,]{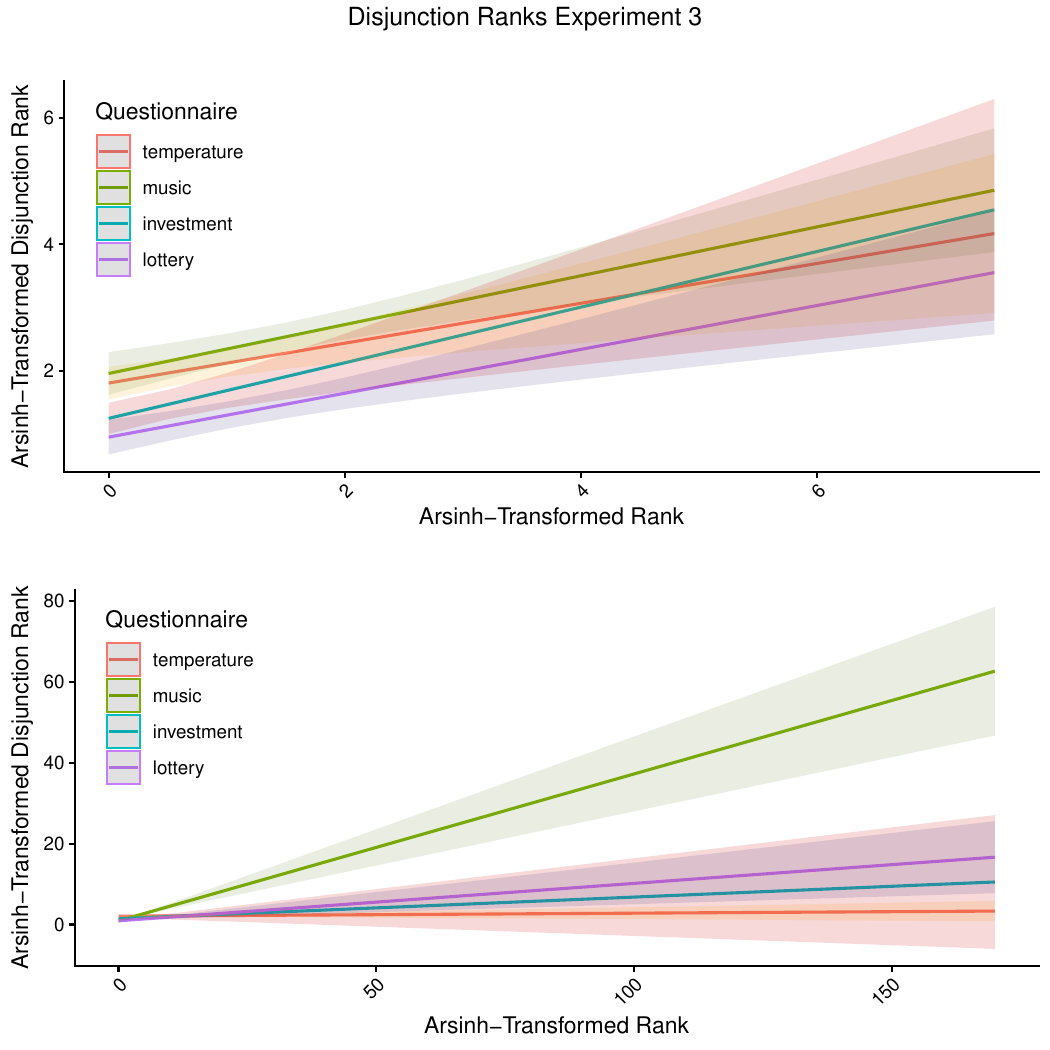} \caption[Predicted disjunction rank values]{Predicted disjunction rank values. The upper plot looks at the relation between the rank reported for the disjunction (y-axis) and the *minimum* rank of the disjuncts. The lower plot has the same from except that we plot the disjunction rank against the *maximum* of the disjuncts. For all four scenarios there is a clear linear relationship between the minimum and the disjunction rank while this is only true for 1 of the four scenarios for the maximum. While many participants did not pick the minimum of the disjuncts, it did appear that this ranking had a greater effect on the composite judgment. Shaded areas denote confidence intervals, demonstrating the significant positive relationship between minimum rank and disjunction rank across varying types of questionnaires.}\label{fig:E3_disjunction_min_max_plot}
\end{figure}

\end{knitrout}

\subsection{Discussion}

Experiment 3 extended our assessment of ranks to include positive and negative ranks.
We found substantially the same pattern as for Experiments 1 \& 2.
Neither negative nor positive ranks seem substantially easier than the other.
The relationship to subjective probability remained highly variable.

While the participants ranks did not strictly adhere with theory there was some alignment.
The law of negation was followed in most of the scenarios by most of the participants.
Participants did not hold contradictory beliefs though they were unaware of the formal principles of ranking theory that both a proposition and its negation could not be ranked greater than zero.
The sole exception was the lottery scenario.
Many participants expressed disbelief in both winning and losing.
We attribute this to a skepticism specific to gambling contexts. 

When evaluating the law of disjunction in Experiments 1 \& 2 we ran into a problem with frequent ties.
In Experiment 3 we redesigned the disjunct propositions to be more discrepant in their degree of disbelief, which made ties less frequent.
This reduced multicollinearity.
Results were more definitive.
Less disbelieved propositions (minimum rank) better predicted the disjunction rank.

We again found that ranks are not a simple transform of probability.
By including positive ranks we could disambiguate negative ranks of zero (which could be either neither believed nor disbelieved or truly believed).
For this practical reason two sided ranking functions may be the better choice in some experimental scenarios.

The influence of context on the threshold for disbelief was again evident.
Context seems to shape the expression of disbelief.
This variation complicates the use of subjective probabilities to infer disbelief, and further emphasizes the importance of considering contextual factors in belief assessments.

The complexity observed for the relationship between subjective probabilities and ranks led us to Experiment 4.
In the prior experiments the probability of an event was mostly vague.
In the one scenario that arguably had a known objective probability (lottery) the results were nominally different.
In addition, all the prior experiments used static scenarios.
The situation was fixed.
Perhaps where objective probabilities were clearer, or where the participant had a chance to learn the probabilistic relationships, we might see more concordance for ranks and probability reports.
We thus adapted the ranking approach to another protocol from our group that was designed to measure probability learning \autocite{diberardino2023probability}.
 
\section{Experiment 4: Evaluating Ranking Functions in a Probability Learning Task}

\subsection{Introduction}
Experiment 4 explored the dynamics of ranking theory within a learning context.
Studies of probability learning often infer that participants have a representation of probability for their choices \autocite{go2022saccadic, yu2012influence}.
A common procedure is to ask participants to make subjective probability judgments about outcome distributions, and it is common in this setting to equate belief with subjective probability.
We had participants engage in a children’s hide-and-seek strategy game against a computer opponent \autocite{diberardino2023probability}.
Each computer opponent had a unique probability distribution that they used to guide their choice of where to hide or where to seek in a simulated three room game of hide-and-seek.
For instance, an opponent might choose Room A to hide 80\% of the time, Room B 15\%, and Room C 5\%.
After playing a series of hides or seeks the participants provided their ranks and subjective probabilities for each room, reflecting their perceptions of the opponents’ hiding and seeking behaviors. 

Our principal objective was to compare ranks and subjective probability estimates to the ground truth probability.
We surmised that if human belief is fundamentally represented through negative ranking functions, these should correspond more closely with objective probabilities than subjective probability assessments.
This rationale rests on the idea that translating rank-based representations into probability measures could introduce errors due to the intrinsic differences between these formats.
Such errors may reflect the potential inaccuracies inherent in converting a rank-based system into a probability-based one.
Assuming that ranks come first derives from the ideas that negative ranks are a more straightforward cognitive expression of disbelief due to being easier to compute \autocite{spohn2013ranking}.

\subsection{Methods}
We recruited 268 undergraduate students from the University of Waterloo to participate in the experiment for course credit, with options for on-line or in-person.
After excluding 67 participants for not adhering to instructions, a total of 201 participants were included in the analysis.
The median age of the participants was 20 years, with 140 females, 56 males, and 5 undisclosed.
Furthermore, we had 188 participants from the on-line sample and 13 participants from the in-person sample for the analysis.

Exclusion criteria were determined based on extreme violations of the instructions.
For the probability condition, participants were instructed that the probabilities for Rooms A, B, and C should sum to 1.
However, the JavaScript program did not enforce this rule to avoid biasing participants’ responses and to accommodate those with difficulties in basic arithmetic.
Participants were excluded if their responses for the probabilities of Rooms A, B, and C summed to exactly 3 in any of the eight probability entries or if their responses summed to a value close to 0 (i.e. less than or equal to \( 1 \times 10^{-10} \)) more than 50\% of the time (i.e., five or more out of the eight total probability questionnaires: four distributions with two entries each for the ``hiding'' and ``seeking'' conditions). This exclusion criterion addressed patterns suggestive of disengagement or attempts to bypass task requirements. Exclusion criteria were applied solely to participants’ extreme errors in subjective probability responses and not ranks. Therefore, the exclusion process may have a bias in favor of a subjective probability account.

The study received ethics clearance from the University of Waterloo Research Ethics Board (REB 45091) before data collection. Participants were informed about the study’s purpose and their right to withdraw at any time without penalty. To ensure privacy, all data were securely stored, and participants provided informed consent prior to participation.

\begin{figure}[H]
\centering
  \includegraphics[width=0.7\textwidth]{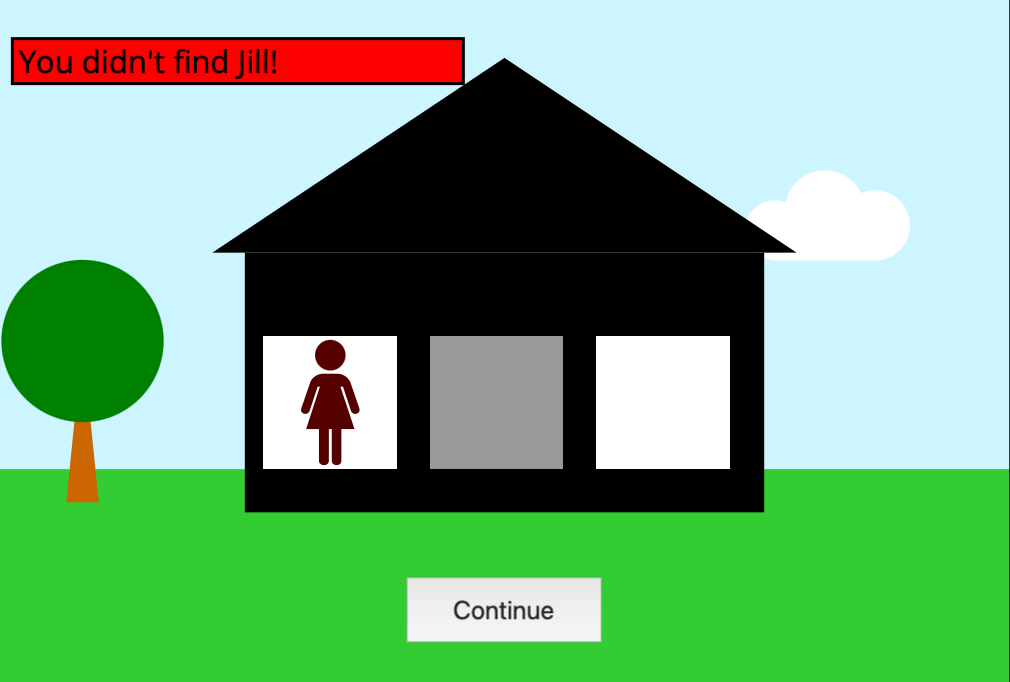}

  \caption{Participant view of the screen after selecting the middle room in a ‘seek’ trial within the three-room dollhouse condition. The selected room turns gray, revealing the child’s location and presenting a notification indicating the outcome. In this example, the participant did not find Jill. The figure illustrates a trial in the hide-and-seek gameplay, where participants select one of three rooms in the dollhouse to seek the hidden child.}
  \label{fig:hideandseek_rooms}
\end{figure}

We adapted a computerized version of the children’s hide-and-seek game, in which participants played against a computer opponent \autocite{diberardino2023probability}. Each computer opponent used a consistent probability distribution to select their hiding and seeking locations, which was unknown to the participant. The computer opponent selected hiding and seeking locations using the same probability distribution. The game interface featured a cartoon-style dollhouse with three visible rooms, each assigned a probability that summed up to 1, shown in Figure \ref{fig:hideandseek_rooms}. The assignment of probability distributions to rooms varied randomly per game, as detailed in Table \ref{tbl:opponents_distribution_long}. A progress bar at the top of the screen displayed the participant’s completion progress. We used the JsPsych JavaScript library to implement and collect on-line data \autocite{de2023jspsych}.

\begin{table}[H] 
    \centering
    \begin{tabular}{cM{25mm}M{60mm}M{30mm}}
       \toprule
        Opponent & Opponent Image & Distribution Identity & Distribution Graph \\
        \midrule
        (Practice) Sally &
        \raisebox{-0.5\height}{\includegraphics[width=6mm]{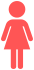}} &
        40-40-20 &
        \resizebox{20mm}{15mm}{%
        \begin{tikzpicture}
          \begin{axis}[ybar, bar width=25pt, ymin=0, ymax=1, xtick=data, symbolic x coords={A,B,C},tick label style={font=\huge}]
            \addplot coordinates {(A,0.4) (B,0.4) (C,0.2)};
          \end{axis}
        \end{tikzpicture}
        } \\
        \midrule
        Bo &
        \raisebox{-0.5\height}{\includegraphics[width=4.4mm]{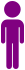}} &
        100-0-0 &
        \resizebox{20mm}{15mm}{%
        \begin{tikzpicture}
          \begin{axis}[ybar, bar width=25pt, ymin=0, ymax=1, xtick=data, symbolic x coords={A,B,C},tick label style={font=\huge}]
            \addplot coordinates {(A,1) (B,0) (C,0)};
          \end{axis}
        \end{tikzpicture}
        } \\
        \midrule
        Jill &
        \raisebox{-0.5\height}{\includegraphics[width=6mm]{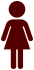}} &
        80-15-5 &
        \resizebox{20mm}{15mm}{%
        \begin{tikzpicture}
          \begin{axis}[ybar, bar width=25pt, ymin=0, ymax=1, xtick=data, symbolic x coords={A,B,C},tick label style={font=\huge}]
            \addplot coordinates {(A,0.8) (B,0.15) (C,0.05)};
          \end{axis}
        \end{tikzpicture}
        } \\
        \midrule
        Kala &
        \raisebox{-0.5\height}{\includegraphics[width=6mm]{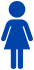}} &
        70-30-0 &
        \resizebox{20mm}{15mm}{%
        \begin{tikzpicture}
          \begin{axis}[ybar, bar width=25pt, ymin=0, ymax=1, xtick=data, symbolic x coords={A,B,C},tick label style={font=\huge}]
            \addplot coordinates {(A,0.7) (B,0.3) (C,0)};
          \end{axis}
        \end{tikzpicture}
        } \\
        \midrule
        Toby &
        \raisebox{-0.5\height}{\includegraphics[width=4.4mm]{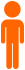}} &
        33.33-33.33-33.33 &
        \resizebox{20mm}{15mm}{%
        \begin{tikzpicture}
          \begin{axis}[ybar, bar width=25pt, ymin=0, ymax=1, xtick=data, symbolic x coords={A,B,C},tick label style={font=\huge}]
            \addplot coordinates {(A,0.333) (B,0.333) (C,0.333)};
          \end{axis}
        \end{tikzpicture}
        } \\
        \bottomrule
    \end{tabular}
  
    \caption{Opponents faced by participants in each condition. Each distribution is presented here in decreasing room probabilities, though probabilities were randomly assigned to rooms during the experiment. Opponent identities were also randomly assigned to each game without replacement, and the order of the distributions was randomized except for the practice, which always came first.}
    \label{tbl:opponents_distribution_long}
\end{table}
The experiment was conducted either on-line or in-person at the University of Waterloo, with identical procedures for both settings.
Participants first completed demographic surveys detailing their biological sex, age, faculty, program name, and current term of study.

The experiment began with practice.
Participants engaged in three seeking turns followed by three hiding turns.
Participants faced one of five simulated opponents – Sally, Kala, Bo, Jill, or Toby (see Table \ref{tbl:opponents_distribution_long}).
As in standard hide-and-seek games, each participant was instructed to find each opponent while seeking and avoid each opponent while hiding.
During the practice round, the probabilities of being assigned to Rooms A, B, and C were distributed randomly across participants following a [40,40,20] distribution.
This meant that all participants saw the same distribution for the practice trials, but that the matching of rooms and probabilities was counter-balanced between participants.

Each individual action during hiding or seeking was considered a ``trial''.
The experiment alternated between ``seeking'' and ``hiding'' rounds, starting with a seeking round.
Each set of 10 trials formed a ``round'', and participants engaged in 7 seeking rounds and 7 hiding rounds per game, totaling 70 seeking and 70 hiding trials per game. 

In each game, a different computer opponent was randomly assigned one of the predefined probability distributions: [100,0,0], [70,30,0], [80,15,5], [$33.\overline{3}$, $33.\overline{3}$, $33.\overline{3}$].
These distributions were also randomly assigned to the rooms, ensuring variability in the specific distribution per game.
For example, if Sally was assigned the [100,0,0] distribution for one game, the remaining characters (Kala, Bo, Jill, Toby) would receive either [70,30,0], [80,15,5], or [$33.\overline{3}$, $33.\overline{3}$, $33.\overline{3}$] for their respective games.
This randomized distribution of probabilities not only dictated the predictable behavior pattern of each character for the duration of the game but also introduced an element of unpredictability in how participants could strategize against different opponents.

The uniform distribution [$33.\overline{3}$, $33.\overline{3}$, $33.\overline{3}$] was specifically designed to induce equal beliefs among participants regarding the likelihood of their opponent hiding or seeking in any of the rooms.
A distribution including a zero probability room aimed to maximize participants’ degrees of disbelief or surprise about the opponent hiding or seeking in that room.
After each game, participants provided their degrees of disbelief and belief and their subjective probability estimates for each room, counterbalanced to reflect their internal model of the computer opponent’s seeking and hiding behavior.

\textbf{Seeking Phase:} When seeking, participants were informed via a dialogue box (``Look for [child]!'') that an opponent, referred to as ``[child]'', was hidden in one of the rooms.
Participants chose a room to search by clicking on it, turning the room light gray.
Success (``You found [child]!'', in green) or failure (``You didn’t find [child]!'', in red) notifications were displayed accordingly, with the opponent’s location revealed post-attempt, as shown in Figure \ref{fig:hideandseek_rooms}.

\textbf{Hiding Phase:} When hiding, participants received a prompt (``Hide from [child]!'') and selected a room to hide in using the mouse.
The outcome was revealed through either a success (``[child] didn’t find you!'', in green) or failure (``[child] found you!'', in red) notification, with the opponent’s icon shown in the selected room. 

In both the seek and hiding phases, we utilized the colors green and red to inform participants if they had successfully hidden or sought their opponent (in green) or had failed to hide or seek their opponent (in red).
This was done to give additional visual feedback in addition to the box dialogue.

\textbf{Experiment Questionnaires:} After completing the hide-and-seek game with each computer opponent, the participants were asked to fill out two questionnaires — the ``Ranking Questionnaire'' and the ``Subjective Probability Questionnaire.'' The order in which they received these questionnaires was randomized.
Each type addressed both “hiding” and “seeking” behaviors of a computer opponent in a simulated hide-and-seek game.
For each item on the questionnaires, participants had to assign a numerical value that represented their level of belief/disbelief or their subjective probability estimates.

\textbf{Ranking Questionnaire}: Participants were asked to grade their degrees of disbelief of the opponent hiding or seeking in each of the three rooms, as well as the disbelief regarding the opponent not being in each of the room — negation of these propositions.
Negation was measured to compute two-sided ranking functions in the data analysis.
Each item required a non-negative integer value, with ``0'' indicating no disbelief or surprise, positive integers indicating increasing degrees of disbelief, and ``infinity'' signifying an impossible event.
Additionally, participants were asked about disjunctions, or the disbelief that the opponent was in any of the rooms.

\textbf{Subjective Probability Questionnaire}: Participants were also asked to estimate the probability of the opponent hiding or seeking in each of the rooms, along with the probability of the opponent not being in each room (negation).
Probabilities for disjunction scenarios, in which the opponent could be hiding or seeking in any combination of two rooms, were additionally collected.
Participants were instructed to give a value of ``0'' for an impossible event, while the probability of a certain event was expressed as ``1.'' Additionally, guidelines required that the probability estimates for Rooms A, B, and C must sum to 1.

Finally, participants completed a final survey assessing the difficulty of grading their subjective probability estimates and disbelief levels.

Inspection revealed two participant strategies while seeking: maximizing and matching \autocite{diberardino2023probability}.
The maximizing strategy involves always seeking at the location with the highest probability of finding the opponent.
The matching strategy seeks proportionally to the probability of where the opponent hides.
Traditionally, the maximizing strategy is considered optimal for seeking \autocite{diberardino2023probability}.

When translated to the hiding rounds these strategies become minimizing and antimatching.
The minimizing strategy always chooses to hide in the room where the opponent is least likely to search.
The antimatching strategy involves selecting a hiding spot in a manner that deliberately diverges from any predictable pattern \autocite{diberardino2023probability}.

\begin{figure}[H]
\centering
  \includegraphics[width=0.5\textwidth]{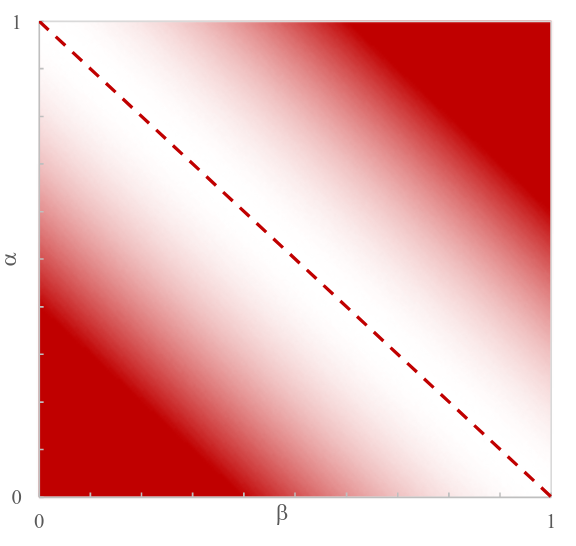}
  \caption{For seeking strategies defined by the equation $\alpha + \beta = 1$, error (indicated by red regions) is minimized when $\vec{\epsilon} = \vec{0}$. Numerous points on or near this line suggest a strong model fit, whereas clusters of points in areas with high error signal a poor fit.}
  \label{fig:visualerror}
\end{figure}

\textbf{Fitting the Line on the Diagonal and Error Minimization}: We used the method of \textcite{diberardino2023probability} to fit strategy plots.
Strategy plots are an optimal linear combination of matching/minimizing and matching/antimatching strategies with an error offset.
The strategic fit is optimal (i.e., error is minimized) when the strategy combination lies along the diagonal line from (1,0) to (0,1) (Figure~\ref{fig:visualerror}).

\subsection{Results}

Participants learn the probabilities and use common policies.
To visualize this we generated strategy plots (Figure~\ref{fig:E4_strategyPlot}).
The most interesting distributions for this purpose are the [70, 30, 0] and the [80, 15, 5] distributions since this is where there was the greatest variation in participants’ choices.
Overall, the strategy plots show that participants learned the distribution and deployed, for the most part, the two strategies (maximizing and matching) commonly observed in the literature (Figure~\ref{fig:E4_strategyPlot}). 

\begin{knitrout}
\definecolor{shadecolor}{rgb}{0.969, 0.969, 0.969}\color{fgcolor}\begin{figure}[H]
\includegraphics[width=\textwidth,]{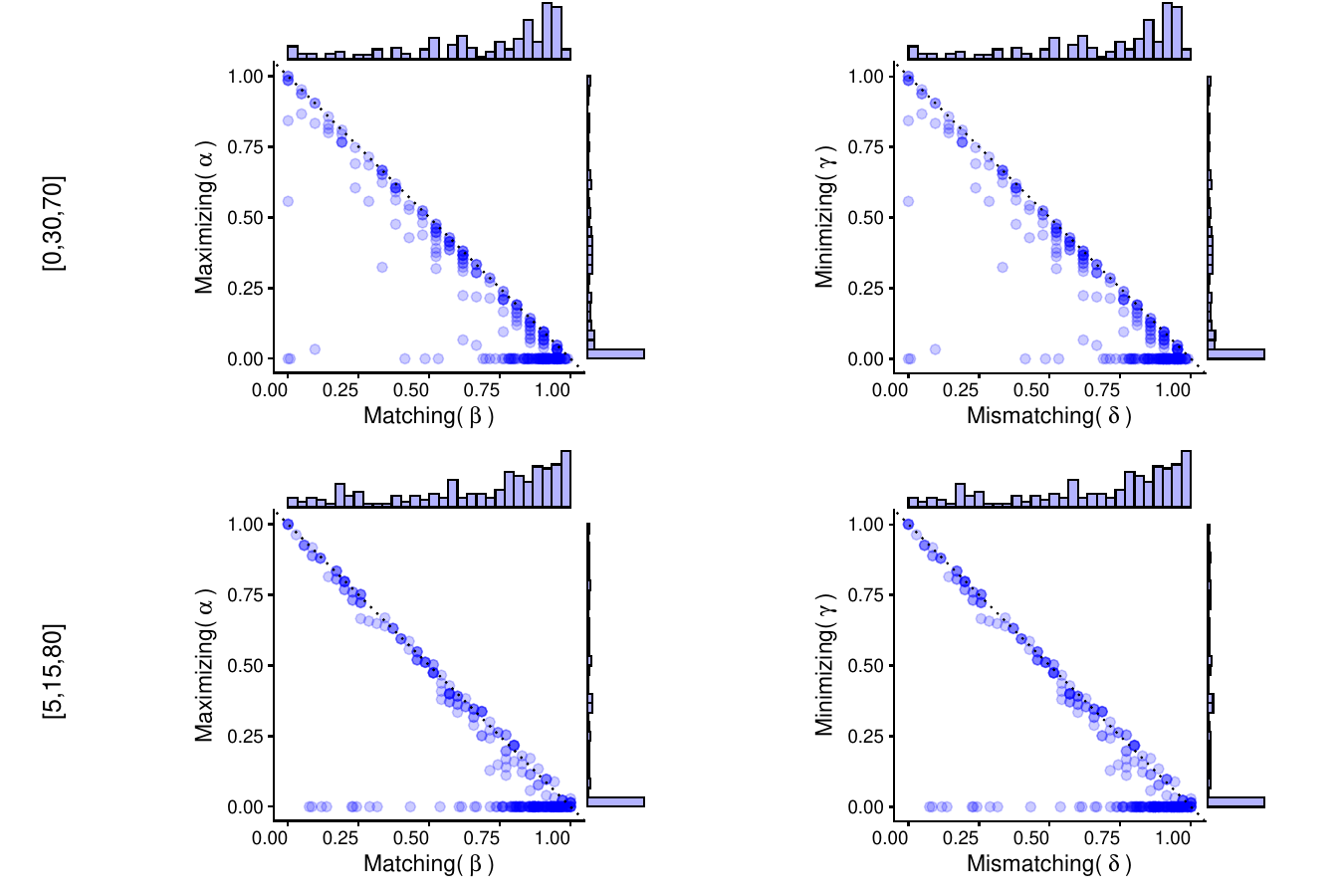} \caption[Participants’ strategy space plots]{Participants’ strategy space plots. Each plot shows the relationship between two strategies (maximizing and matching for seeking; minimizing and antimatching for hiding). The diagonal lines represent a pure mixture of these two strategies. Points represent individual participants’ strategies mixtures. Histograms along the margins show the frequency distribution of each strategy component.}\label{fig:E4_strategyPlot}
\end{figure}

\end{knitrout}

Since participants learned the task we next compared their belief ranks to the objective probability distribution of the various opponents (Figure~\ref{fig:E4_objProb_twosidedRank}).
For the [100-0-0] distribution, participants generally ranked the 100\% probability room with a two-sided rank of infinity.
Conversely, rooms with a 0\% probability were considered impossible, marked by -infinity.

The occurrence of a two-sided rank of 0 was particularly prevalent in the uniform [$33.\overline{3}$, $33.\overline{3}$, $33.\overline{3}$] distribution.
This suggests that participants had a balanced perception regarding the presence or absence of the opponent in any room.
It indicates that no single room was definitively perceived as more or less believable than the others reflecting an equilibrium in participants’ belief systems when faced with equal probabilities.

In the [80-15-5] and [70-30-0] distributions, the rooms with the highest probabilities (80\% and 70\%) elicited strong belief, while intermediate probabilities (30\% and 15\%) showed reduced belief, with 15\% room being associated with some disbelief.
The least probable outcomes (0\% and 5\%) were met with strong disbelief.
These findings suggest good alignment between participants’ ranks and the learned probabilities.
This was confirmed by a correlation between ranks and objective probabilities of 0.77 (\(p\)  < .001).

\begin{knitrout}
\definecolor{shadecolor}{rgb}{0.969, 0.969, 0.969}\color{fgcolor}\begin{figure}[H]
\includegraphics[width=\maxwidth,]{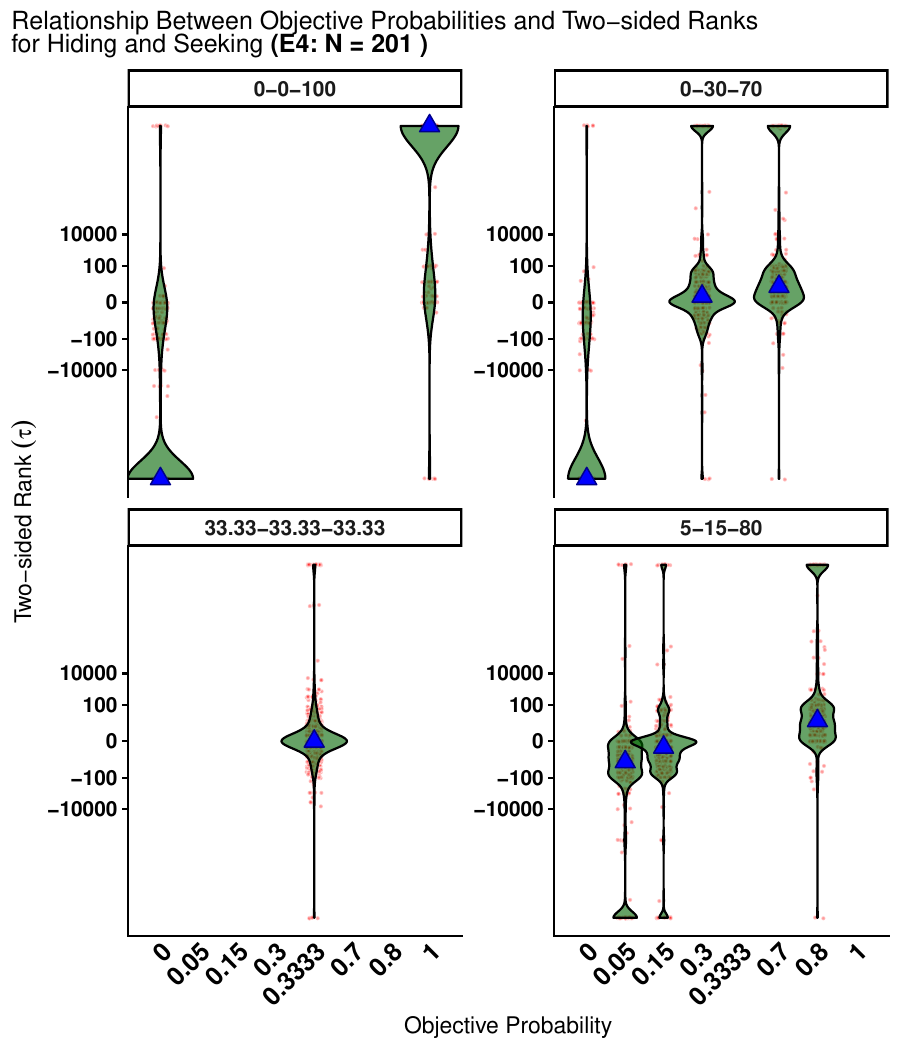} \caption[Violin plot illustrating participants’ degrees of disbelief and belief for each probability distribution, adjusted using the arsinh transformation to scale values from -infinity (representing impossibility) to infinity (representing certainty)]{Violin plot illustrating participants’ degrees of disbelief and belief for each probability distribution, adjusted using the arsinh transformation to scale values from -infinity (representing impossibility) to infinity (representing certainty). This plot compares the relationship between objective probabilities and the two-sided ranks, where negative values indicate disbelief and positive values indicate belief. The triangle within each violin plot represents the median, highlighting how participants’ belief systems adapt to different probability distributions.}\label{fig:E4_objProb_twosidedRank}
\end{figure}

\end{knitrout}

Perhaps more interesting is to compare ranks and subjective probabilities since the latter are so often taken as a proxy for degrees of belief.
In the [100-0-0] distribution, participants typically assigned a subjective probability of 1 to rooms with a guaranteed presence of the opponent, corresponding with a negative rank of 0, signaling complete certainty.
In contrast, rooms with a 0\% likelihood saw subjective probabilities of 0, matched with maximum disbelief (-infinity), indicating perceived impossibility.
However, a few participants assigned non-infinite negative ranks to these rooms, suggesting a perception of unlikelihood rather than impossibility.

Similar patterns emerged in the [70-30-0] distribution, where zero-probability rooms generally received maximum disbelief ratings.
The [70-30-0] and [80-15-5] distributions exhibited a broader range of subjective probability responses than the [100-0-0] distribution, suggesting increased difficulty in making subjective probability estimates.
Disbelief decreased non-linearly as subjective probability increased.

In the uniform distribution [$33.\overline{3}$, $33.\overline{3}$, $33.\overline{3}$], while many probability estimates were near \(\frac{1}{3}\), a wide range of subjective estimates was observed, suggesting that some participants struggled to represent the environmental statistics accurately through probabilities.
Additionally, many participants indicated it would not be surprising to find the opponent in any room, while others expressed varying levels of surprise.

Again, as it is theoretically predicted that subjective probability and ranks could show a logarithmic relationship we again analyzed non-linear fits to these data.
The asinh was applied to the negative ranks to address skewness and extreme values.
Both a logarithmic and exponential fit explained about half of the data variance.
For the logarithmic model we used a non-linear squares regression to estimate the logarithm’s base.


The logarithmic model fit gave a pseudo $R^2$ of 0.55 (Figure~\ref{fig:E4_subjProb_negRank_fitted}).
While the exponential decay model (parameters estimated by non-linear least squares regression) 
showed a pseudo $R^2$ value of 0.548 (Figure~\ref{fig:E4_subjProb_negRank_fitted}).

\begin{knitrout}
\definecolor{shadecolor}{rgb}{0.969, 0.969, 0.969}\color{fgcolor}\begin{figure}[H]
\includegraphics[width=\maxwidth,]{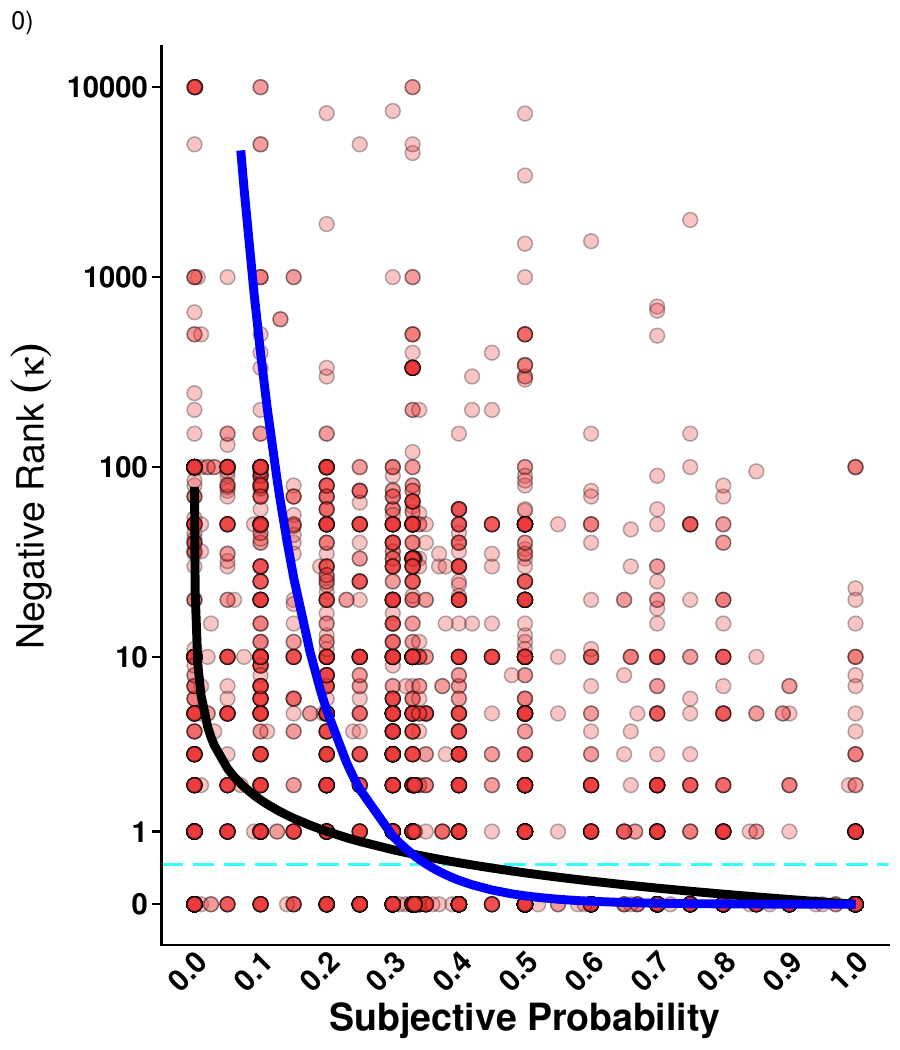} \caption[Scatter plot illustrating the relationship between subjective probabilities and negative ranks ($\kappa$) across all questionnaires]{Scatter plot illustrating the relationship between subjective probabilities and negative ranks ($\kappa$) across all questionnaires. The y-axis is scaled to display negative ranks effectively, while the labels show the original rank values for easier interpretation. A logarithmic model (black) and exponential model (blue) are fitted to the data. The logarithmic function explains 55.37 \% of the variance in negative ranks while the exponential function explains 54.77 \% of the variance.}\label{fig:E4_subjProb_negRank_fitted}
\end{figure}

\end{knitrout}
To compare the two models we calculated the AIC and BIC.
The logarithmic model showed a superior fit, with a lower AIC of 32292.26 compared to 32359.05 for the exponential model — a difference of 66.78.
The BIC results similarly favored the logarithmic model, with a lower BIC of 32305.23 compared to 32378.49 — a difference of 73.27.
These findings deviate from those in Experiments 1 through 3, suggesting that the logarithmic model better explains the relationship between subjective probabilities and negative ranks in a probability context.


In this experiment we were able to compare subjective and objective probability estimates.
This is relevant as subjective probability is often proposed as the proxy for degrees of belief, but if subjective probability correlates poorly with objective probability it gets harder to make this case if one presumes that our beliefs should concord with environmental experience.
The Pearson correlation between the two was 0.87 ( p =  < .001).
Despite this relationship it is worth noting that the median subjective probabilities did not consistently align with the objective probabilities across various distributions (Figure~\ref{fig:E4_objProb_twosidedRank}).

\begin{knitrout}
\definecolor{shadecolor}{rgb}{0.969, 0.969, 0.969}\color{fgcolor}\begin{figure}[H]
\includegraphics[width=\maxwidth,]{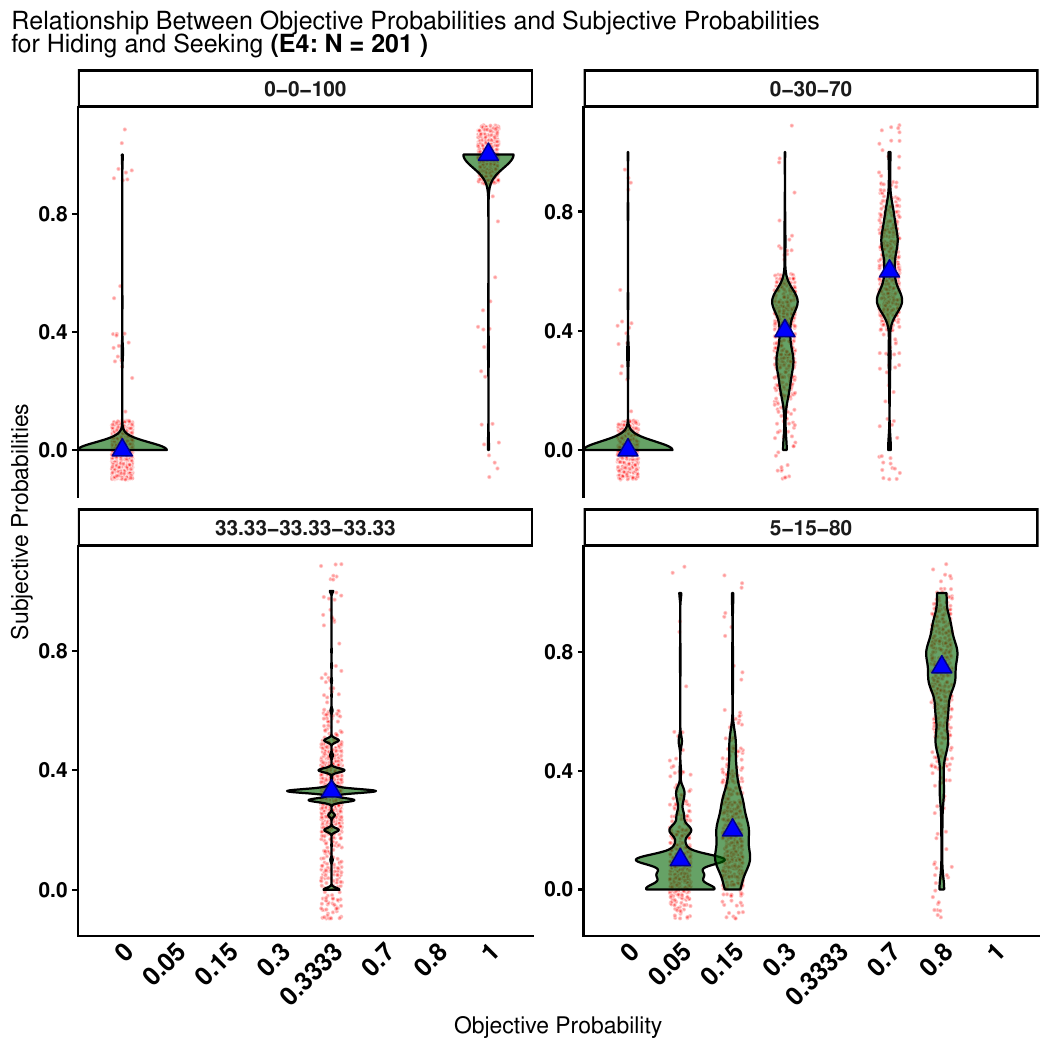} \caption[Violin plot illustrating participants’ subjective probabilities against objective probabilities for each probability distribution]{Violin plot illustrating participants’ subjective probabilities against objective probabilities for each probability distribution. The triangle within each violin plot represents the median. The horizontal red dotted lines indicate the corresponding objective probabilities for each distribution.}\label{fig:E4_objProb_subjProb}
\end{figure}

\end{knitrout}

Ideally one would like to compare all three metrics together: objective probability, subjective probability estimates and rankings of belief.
We found one qualitative method for this for one of the conditions.
We visualized subjective probabilities and negative ranks with objective probabilities on a probability simplex.
This geometric representation plots the participants’ probability distributions as points within a triangle.
Each vertex of the triangle corresponds to the distribution that chooses that room 100\% of the time.
We can place participants on this simplex as well.
For the uniform distribution condition we can make a direct translation into the simplex space for both participants’ subjective probabilities and ranks. 

\begin{figure}[H]
  \centering
  \small\textnormal{Subjective Probabilities and Negative Ranks in a Probability Simplex} \\[0.5em] 
  \includegraphics[width=1\textwidth]{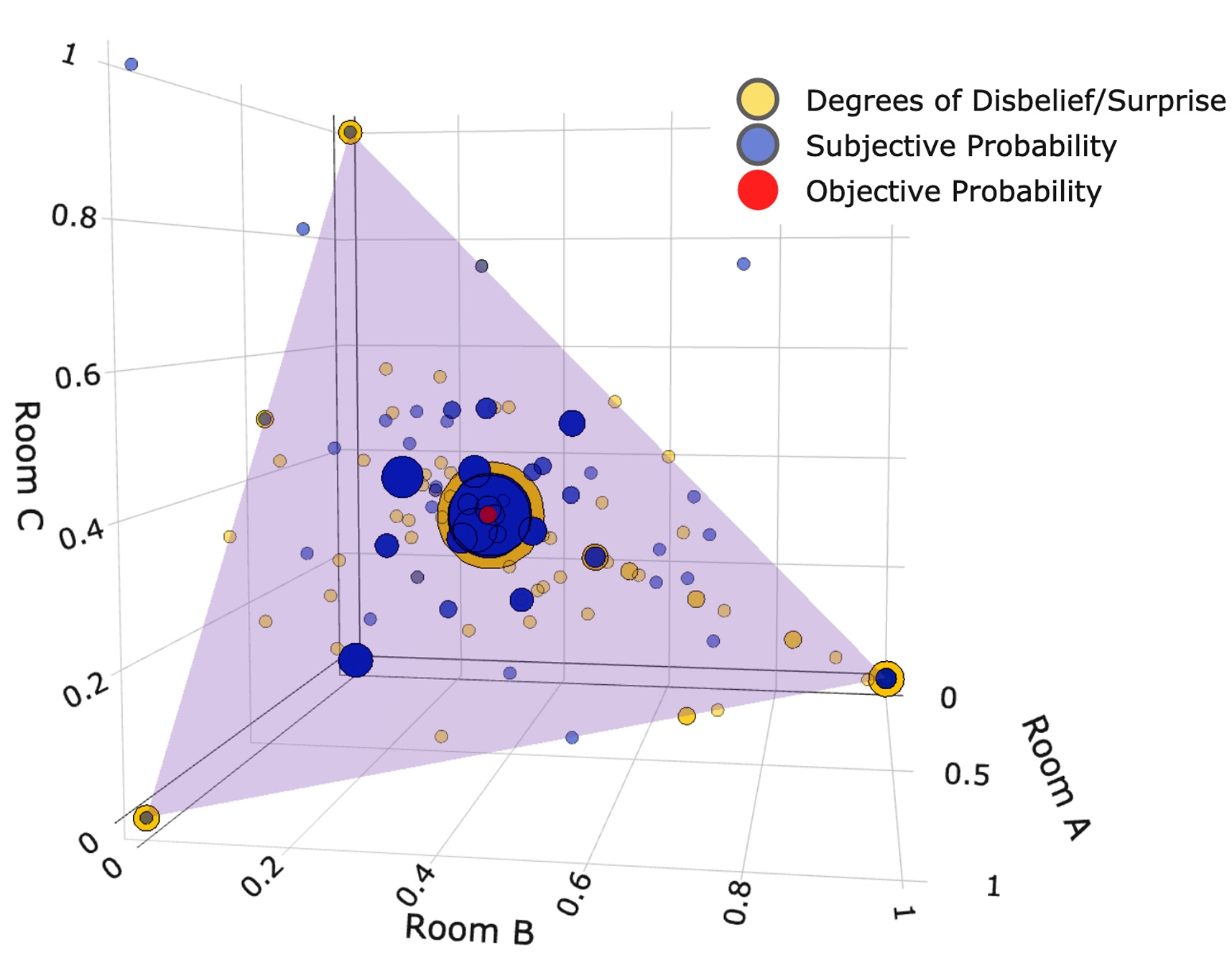}

  \caption{A probability simplex ($\Delta_{n-1}$) displays participants’ subjective probabilities and disbelief rankings against a fixed objective probability [$33.\overline{3}$, $33.\overline{3}$, $33.\overline{3}$] marked in red. Blue dots, representing subjective probabilities, are widely dispersed, illustrating varied response densities from multiple participants. Yellow dots, indicating negative ranks, cluster at the center, reflecting a more consistent evaluation of objective probability, with each room ranked similarly. Larger dots indicate more common responses, highlighting that rankings provide a clearer picture of the uncertainty structure compared to subjective probabilities.}
  \label{fig:E4_simplexplot}
\end{figure}

Figure~\ref{fig:E4_simplexplot} shows the subjective probabilities as blue dots.
Note the dispersion across the simplex.
This reflects the varying probability estimates across participants.
In contrast, ranks (yellow dots) clustered centrally around the objective probability.

To quantify this impression we computed, for the uniform [$33.\overline{3}$, $33.\overline{3}$, $33.\overline{3}$] distribution, the percentage of participants whose reported negative ranks versus subjective probability estimates each fell within a small tolerance of the objective probability for every room.
In the hiding condition, 56.22\% of participants’ negative ranks agreed with the objective probabilities, compared to only 31.34\% for subjective probability estimates.
In the seeking condition, 58.21\% of ranks agreed with the objective probabilities, compared to 34.33\% for subjective probability.
Ranks thus matched the objective probability distribution more often than subjective probability estimates did in both conditions, consistent with the impression from Figure~\ref{fig:E4_simplexplot} that ranks provide a more coherent representation of participants’ assessments.

\subsection{Discussion}

Participants adapted their strategies based on opponents’ behaviors.
The strategy space of each distribution served as a benchmark for verifying participants’ understanding.
The [70,30,0] and [80,15,5] distributions were particularly relevant for observing a mix of optimal and non-optimal strategies.
In both distributions, participants employed a combination of matching and maximizing strategies when seeking, with a noticeable degree of non-optimal probability matching.

Experiment 4 provides a dynamic, learning environment.
Responses to the extremes (always look here; never look here) made sense for both two-sided ranking functions and subjective probability.
However, there were many instances where participants expressed finite ranking responses to these extremes.
These deviations suggest that participants’ belief systems may not strictly adhere to extreme probabilistic reasoning, but rather a more reserved approach to certainty.
Participants’ neutrality in belief systems — neither believed nor disbelieved — was clearly observed in the uniform condition

Deviations in subjective probability estimates were more pronounced for intermediate probabilities (e.g., 80\%, 70\%, 30\%, and 15\%).
This pattern is well-documented in the probabilistic reasoning literature, where individuals often struggle with estimating probabilities that fall in the middle of the spectrum \autocite{khaw2021individual, prelec1998probability, tversky1992advances}. 

The contrast between the structured and unstructured scenarios demonstrates the adaptability of negative ranks.
In everyday, unstructured situations, negative ranks accounted for the influence of external factors.
They permit exploring how cultural and personal biases shape belief formation.
This distinction deepens our understanding of cognitive processes and suggests new research avenues for examining how individuals adapt their beliefs in response to environmental cues.

The comparison between subjective probabilities, negative ranks, and objective probabilities highlights key differences in how individuals assess probabilistic information.
While subjective probabilities varied among participants, ranks aligned more closely with objective probabilities.
This alignment indicates that participants often struggled with accurate probability estimation.
Ranks offer a more straightforward and less error-prone tool for evaluating uncertainties.
This may be because ranking functions are simpler and less cognitively demanding.
Negative ranks arguably require fewer computational steps than traditional probability theory.
Negative ranks can function on a relative scale of disbelief, avoiding the precision required for probability estimates.

\section{Experiment 5: Rethinking the Ellsberg Paradox through Negative Ranking Functions}

\subsection{Introduction}

Our previous experiments demonstrated that individuals can use ranks to express belief.
Further, disbelief and belief ranks are not redundant to subjective probability.
Similar subjective probabilities can have very different ranks and vary with context.
All of this made us wonder if ranks would be useful in resolving any of the apparent paradoxes for uncertainty estimates interpreted probabilistically. 

The Ellsberg scenario is one such scenario.
The most frequently cited variation is the three-color one-urn problem \autocite{ghirardato2002ambiguity, slovic1974accepts, oechssler2015test}.
An urn is stated to be filled with 90 balls: 30 are red, while the remaining 60 are either black or yellow in an unknown proportion.
Decision-makers are presented with two pairs of bets based on the color of the drawn ball.
A bet on a ``red'' ball offers a definite \( \frac{1}{3} \) chance of winning a \$100 prize, and is generally preferred over a bet on a ``black'' ball where the probability of winning varies ambiguously between 0 and \( \frac{2}{3} \) with an expectation of \( \frac{1}{3} \).
Similarly, a bet on a ``black or yellow'' ball guarantees a \( \frac{2}{3} \) chance of winning, and is preferred usually to a bet on a ``red or yellow'' ball, where the win probability could be anywhere from \( \frac{1}{3} \) to 1 \autocite{ellsberg1961risk}.
This is interpreted as individuals preferring bets with known probabilities over ambiguous ones \autocite{ellsberg1961risk, oechssler2015test}.

This preference for known probabilities, termed ambiguity aversion, challenges expected utility theory \autocite{ellsberg1961risk, machina2014ambiguity, savage1972foundations}.
Expected utility theory asserts that decision-makers maximize expected utility based on subjective probabilities derived from their preferences \autocite{halevy2007ellsberg, savage1972foundations}.
Expected utility theory endorses the Sure-Thing Principle \autocite{aerts2011contextual, savage1972foundations}, which states that if an individual prefers one option over another they should continue to prefer that choice even when a fixed quantity is added to both alternatives \autocite{aerts2011contextual, savage1972foundations}.
In the one urn problem if red is preferred over black, then red plus yellow should be preferred over black plus yellow. 

\textcite{ellsberg1961risk} demonstrated that in real-world decision-making, individuals often exhibit non-neutral attitudes toward ambiguity, differentiating between risk, where probabilities are known, and uncertainty, where probabilities are unknown.
This distinction has practical implications, particularly in economics, where decision-making models typically rely on subjective probabilities to define and solve problems \autocite{halevy2007ellsberg}.
Ellsberg’s findings show that decision-makers do not always exhibit probabilistic sophistication, revealing descriptive limitations within expected utility theory and suggesting the need for models that account for non-standard attitudes toward ambiguity \autocite{halevy2007ellsberg}.

Ellsberg’s experiments inferred people’s subjective beliefs from their betting choices \autocite{ellsberg1961risk}, an inference that can be insecure in scenarios where there is no opportunity to explicitly express indifference or uncertainty.
Additionally, the presence of monetary stakes (hypothetical or real) is an additional complication.
Decision-makers could prioritize potential gains or losses over their true beliefs about outcomes.
Choice preferences could represent strategic responses to the incentives of the experimental setup \autocite{kovavrik2016ellsberg}.

We incorporated ranking responses into the one-urn Ellsberg scenario.
We used a negative ranking response to measure individuals’ degrees of disbelief about the outcomes of drawing from an urn.
Participants could articulate their levels of disbelief regarding each proposition.
For this first of two experiments on the Ellsberg Paradox (see also Experiment 6) we simplified the measurement process by using negative ranking functions that focused solely on disbelief in a proposition, rather than both the proposition and its negation.
This aligned the experimental procedure more closely with the Ellsberg study’s betting scenarios.

\subsection{Methods}

We conducted an on-line experiment and recruited 209 undergraduate students from the University of Waterloo to participate in exchange for course credit.
The participants’ median age was 21 years.
There were 152 females, 51 males, and 6 preferred not to answer.
The study obtained ethics clearance from the University of Waterloo Research Ethics Board (REB 45165).
Participants were informed about the study’s purpose and their right to withdraw from the study at any time without penalty.
To protect the privacy of the participants, all data were stored securely with no identifying information.

Participants were presented with a hypothetical scenario involving an urn filled with colored balls.
The urn contained a total of 90 balls, which were of three different colors: red, black, and yellow.
The number of red balls in the urn was fixed at 30 and was known to the participants.
However, the distribution of the remaining 60 black and yellow balls was not disclosed, which created an element of uncertainty.

The participants were informed that one ball would be randomly drawn from this urn.
They were then asked to provide their degree of disbelief if the ball drawn was ``red'', ``black'', as well as the disjunctions ``red or yellow'' and ``black or yellow''.
Participants typed in numbers to give their disbelief responses (see Figure \ref{fig:E5_E6_rankQ}).
Finally, the participants were asked to rate the difficulty of quantifying their disbelief, providing insights into the perceived challenges posed by the task.

The study used a custom-built on-line platform developed using JavaScript to capture participant responses \autocite{de2023jspsych}.
The platform provided an interface that allowed participants to input their answers as whole numbers, starting from 0 and going up to any positive value, with an option to select ‘infinity’ via a checkbox.
Participants completed the study in less than 15 minutes.

Descriptive statistics were computed for the degrees of disbelief associated with each proposition (red, black, red or yellow, and black or yellow).
We analyzed the proportion of participants displaying ambiguity aversion, attraction, indifference, and consistent preferences, as these tendencies are crucial for understanding decision-making under uncertainty.
To identify significant differences in degrees of disbelief across propositions, a one-way analysis of variance (ANOVA) was conducted.

\subsection{Results}

When using ranks only 6.4\% of participants exhibited paradoxical behavior.
Most of our participants who rated a ``red'' ball less surprising than a ``black'' ball also rated a ``red or yellow'' ball as less surprising than a ``black or yellow'' ball.
This paradoxical proportion found with ranks is much less than is typically found for the betting version of the task (see Experiment 6 below). 

We also saw rare cases of ambiguity attraction. 3.45\% preferred ambiguity.
These participants were less surprised by drawing a ``black'' ball compared to a ``red'' ball, and similarly were less surprised when a ``red or yellow'' ball was drawn compared to ``black or yellow'' ball.
Many participants expressed neutrality.
These proportions are illustrated in Figure~\ref{fig:E5_ranks_barchart}.

41.38\% of participants showed no difference in their degrees of disbelief between the propositions, indicating a general tendency to remain neutral, regardless of whether the probabilities were known or uncertain.

\begin{knitrout}
\definecolor{shadecolor}{rgb}{0.969, 0.969, 0.969}\color{fgcolor}\begin{figure}[H]
\includegraphics[width=\maxwidth,]{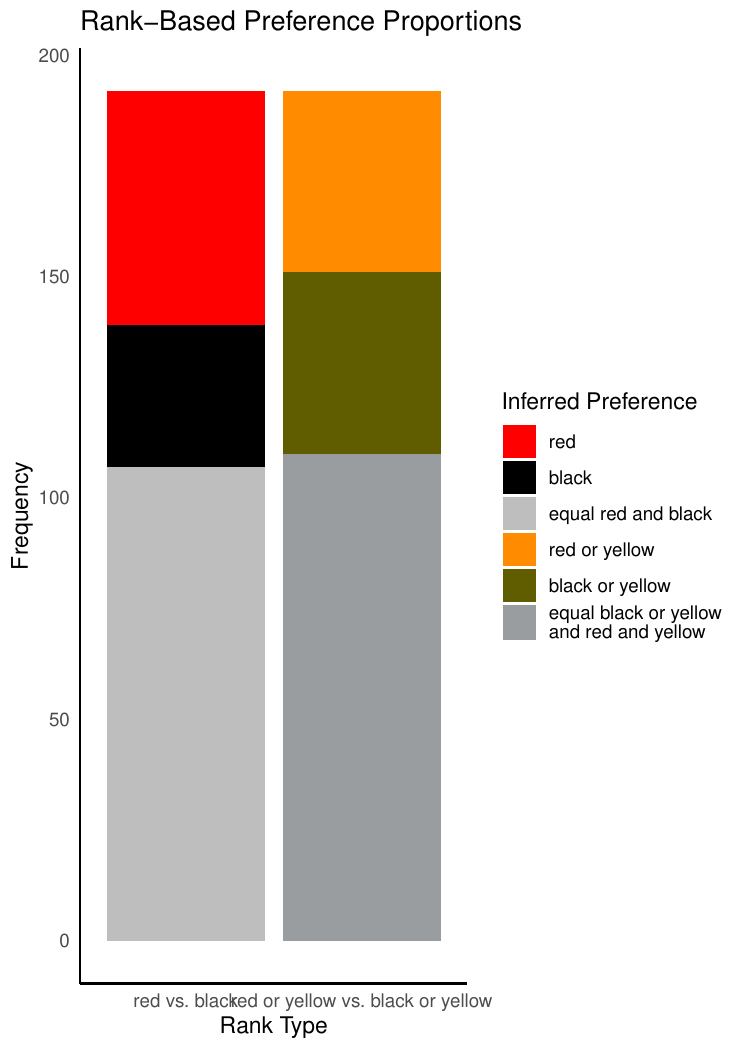} \caption[The figure displays the distribution of preferences based on degrees of disbelief (negative ranks) across participants]{The figure displays the distribution of preferences based on degrees of disbelief (negative ranks) across participants. The stacked bars in the figure show the distribution of preferences for (a) ``Red'' versus `Black', and (b) ``Red or Yellow'' versus ``Black or Yellow'', and situations of equal disbelief between the two. Preferences were determined by comparing the degrees of disbelief. For example, a preference for ``red'' over ``black'' was noted when disbelief in drawing a red ball was less than that for a ``black'' ball, and vice versa. Situations where disbelief in ``red'' and ``black'' were equivalent are represented as ``equal'' preferences. The results show that a higher proportion of participants had equal disbelief between the two propositions.}\label{fig:E5_ranks_barchart}
\end{figure}

\end{knitrout}

For the statistical analyses we excluded values of infinity as illogical.
The urn contained 90 red, black or yellow balls.
It was not possible for no ball to be drawn.
This exclusion accounted for 5.12\% of data.
Homogeneity of variances was confirmed via Levene’s test ( F (3, 793) = 0.1344, \(p\)  = .940).
Note that the Shapiro-Wilk test indicated a failure to meet the normality assumption \(W\) = 0.0405, \(p\)  < .001, and this was not corrected by a log transformation \(W\) = 0.831, \(p\)  < .001.
We decided to conduct the ANOVA anyway.
Readers should take note of this violation of statistical assumptions.
The reason for proceeding with ANOVA was because of its robustness to the normality assumption, particularly in cases of large sample sizes.
The impact on type I error rate is usually minimal \autocite{schmider2010really}.
The results of the ANOVA showed no significant difference in disbelief scores among the various propositions, \(F\)(3, 793) = 0.13, \(MSE\) = 139410403623.81, \(p\)  = .940, \(\eta_{p}^{2}\) < .001.
We also ran a non-parametric Kruskal-Wallis test.
This test also indicated no significant differences across groups, \(H\)(3) = 3.03, \(p\)  = .387.

\subsection{Discussion}

Participants in previous studies of the Ellsberg Urn Problem have shown varying degrees of ambiguity aversion \autocite{binmore2012much, slovic1974accepts}.
A comprehensive review indicated that slightly more than 50\% of subjects in 40 experimental studies of the Ellsberg urn problem exhibited ambiguity aversion \autocite{oechssler2015test}.
In contrast, our study revealed that only 6.4\% of participants displayed ambiguity aversion when utilizing negative ranking functions to assess degrees of disbelief.

Attitudes toward ambiguity can be categorized into three behavioral responses: aversion, attraction, and neutrality \autocite{dominiak2012dynamic}.
We found that neutrality was the most prevalent response, with a smaller proportion of participants displaying ambiguity aversion or attraction.
Participants demonstrated consistent neutrality across both simple (e.g., ‘red vs. black’) and compound probabilities (e.g., ‘red or yellow vs. black or yellow’), indicating logical consistency regardless of ambiguity.

Contrary to traditional findings, where ambiguity aversion is typically dominant, our results indicated that most participants showed logical consistency.
One possible explanation is that this was because our experimental design allowed participants to express neutrality through negative ranks.
These results challenge the assumption that the Ellsberg Paradox is inherently reflective of human cognitive processes.
Ambiguity aversion may be an artifact of specific experimental setups.
While our experiment was designed to reduce complexity and focus purely on measuring disbelief, it may also be that this simplification influenced the outcomes.
Experiment 6 more closely replicated the standard Ellsberg One Urn scenario.

\section{Experiment 6: Comparing Traditional Betting Choices and Negative Ranks in the Ellsberg Paradox}

\subsection{Introduction}

In Experiment 6 participants responded both to belief questions \emph{and} betting scenarios.
Disbelief judgments were made first.
This allowed a direct comparison to the results of Experiment 5.
In addition, this order prevented betting choices from influencing disbelief responses.
To expand the range of responses we followed the standard task with a version where outcomes were much less likely.
To allow for the expression of neutrality when betting we had four rounds. 

\subsection{Methods}
We recruited 301 undergraduate students from the University of Waterloo for an on-line experiment in exchange for a course credit.
We excluded 11 who failed attention checks. 290 participants were included in the analysis (median age: 19 years; 228 females; 52 males; 1 other; 9 preferred not to answer).
The study obtained ethics clearance from the University of Waterloo Research Ethics Board (REB 45629).
Participants were informed about the study’s purpose and their right to withdraw from the study at any time without any penalty.
To protect the privacy of the participants, all data were stored securely with no identifying information.

The three-colored one-urn scenario replicated that of Experiment 5 except as follows.
They faced two new betting scenarios: choose between receiving \$100 for (a) red vs. black, and (b) red or yellow vs. black or yellow.
These two bets were repeated for four rounds to capture higher-order confidence in their choices.
All four rounds were displayed together on the screen.
Participants did not have to remember their prior bets.
Participants could now express neutral betting by choosing the same gamble twice across the four rounds (see Figure~\ref{fig:E6_Q2}).

The rarity scenario asked participants to imagine an urn with 1000 balls of four colors: red, black, yellow, and green.
There was one red ball, two balls of either black or yellow, and 997 green balls.
Participants assigned degrees of disbelief to this scenario, maintaining the red-to-black/yellow ratio while increasing rarity to elicit stronger degrees of disbelief or surprise (see Figure \ref{fig:E6_Q3}).

To ensure attentiveness, participants were required to answer questions about the colors and total number of balls in the urn

The differences in negative ranks for the propositions were calculated and compared against participants’ betting preferences.
Preferences were quantified by the number of times participants chose red in scenario (a) and black or yellow in scenario (b), with counts ranging from 0 to 4.
A count of 0 indicated the participant never chose that color across all rounds, while a count of 4 indicated a consistent preference.

For scenario (a), the difference in negative ranks was computed as \(\kappa(black) - \kappa(red)\).
A negative value suggested a higher disbelief in red relative to black, while a positive value suggested lower disbelief in red.
A similar computation was used for scenario (b) with \(\kappa(\text{red or yellow}) - \kappa(\text{black or yellow})\).

Pearson’s correlation analyses were conducted to assess the relationship between betting choices and degrees of disbelief.
If a relationship existed, positive correlations were expected between the participants’ disbelief and their betting choices.
A neutral count of 2 would indicate no preference, aligning with a disbelief difference value of zero, indicating equal disbelief in both options.

Box plots were used to visualize the relationship between disbelief differences and betting choices.
For the black vs. red scenario, the y-axis represented disbelief differences, and the x-axis indicated the frequency of choosing red.
The same approach was applied to scenario (b).
A positive linear trend would indicate alignment between participants’ beliefs and their betting choices.

\subsection{Results}
8.62\% exhibited ambiguity aversion calculated from ranks in the standard condition and 5.17\% in the rarity condition.
There was no significant difference in the level of ambiguity aversion between the one-urn and rarity-urn scenarios ( \(\chi^2(1)\) = 1.85, \(p\)  = .174 ).
The lack of difference in participants’ degrees of disbelief between the one-urn and rarity-urn scenarios — both of which maintained the same proportion of red to black or yellow balls — makes a floor effect less likely.

Results were different when betting.
Round 1 bets, standard condition, found paradoxical choices for 48.97\% of participants.
Combined for the four rounds it was 1.38\%.
A chi-squared test for independence revealed a significant decrease in participants exhibiting ambiguity aversion over four betting rounds, ( \(\chi^2(1)\) = 17.93, \(p\)  < .001 ).
The four betting rounds allowed participants to express neutrality in their choice preferences.
Figure~\ref{fig:E6_flipped_plot} illustrates how participants’ betting choices changed or remained the same across the four betting rounds for both scenarios.
The y-axis indicates whether participants switched their choices (True) or retained their previous choices (False).
This methodology enabled participants to modify their preferences across rounds, while others maintained consistent choices throughout.

The chi-squared test for independence comparing ranks and bets revealed a significant reduction in the number of participants exhibiting ambiguity aversion when calculated from ranks ( \(\chi^2(1)\) = 45.53, \(p\)  < .001). 

\begin{knitrout}
\definecolor{shadecolor}{rgb}{0.969, 0.969, 0.969}\color{fgcolor}\begin{figure}[H]
\includegraphics[width=\maxwidth,]{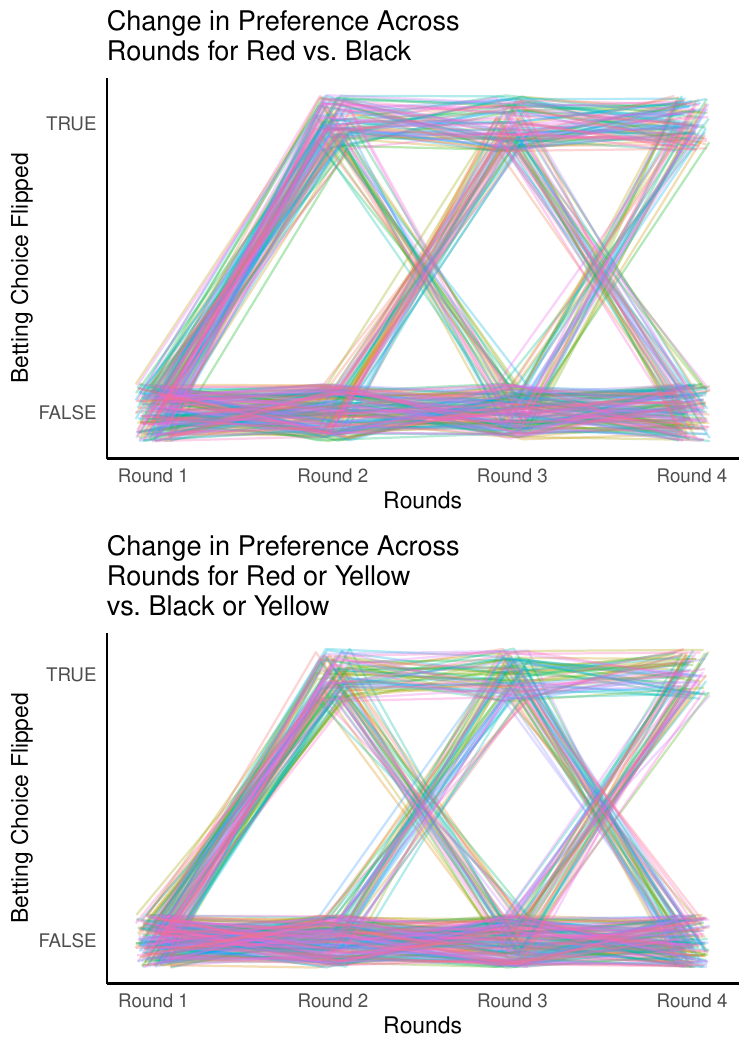} \caption[Change in choice preference across rounds for Red vs]{Change in choice preference across rounds for Red vs. Black and Red or Yellow vs. Black or Yellow. The plots show the percentage of participants who switched their choices for each round (2, 3, 4). A large proportion of participants did not change their choices across four rounds, while some switched their choices to show neutrality or some degree of confidence in their choices. The y-axis represents whether participants switched their choices (True) or kept the same choices across rounds (False).}\label{fig:E6_flipped_plot}
\end{figure}

\end{knitrout}

A two-way ANOVA of disbelief scores across propositions and urn-scenarios (one-urn and rarity urn) revealed no difference in disbelief scores among different propositions (F(3, 1996) = 0.003, p = 1), but did show a significant effect of urn-scenario ( F(1, 1996) = 5.206, p = 0.023).
Overall degrees of disbelief were greater for the rarity-urn scenario (\(\mbox{Median } =\)
98) than for the one-urn scenario (\(\mbox{Median } =\)
3).
Additionally, the interaction between proposition and urn-scenario was examined but was not significant, (F(3, 1996) = 0.008, p = 0.999 ).

\begin{knitrout}
\definecolor{shadecolor}{rgb}{0.969, 0.969, 0.969}\color{fgcolor}\begin{figure}[H]
\includegraphics[width=\maxwidth,]{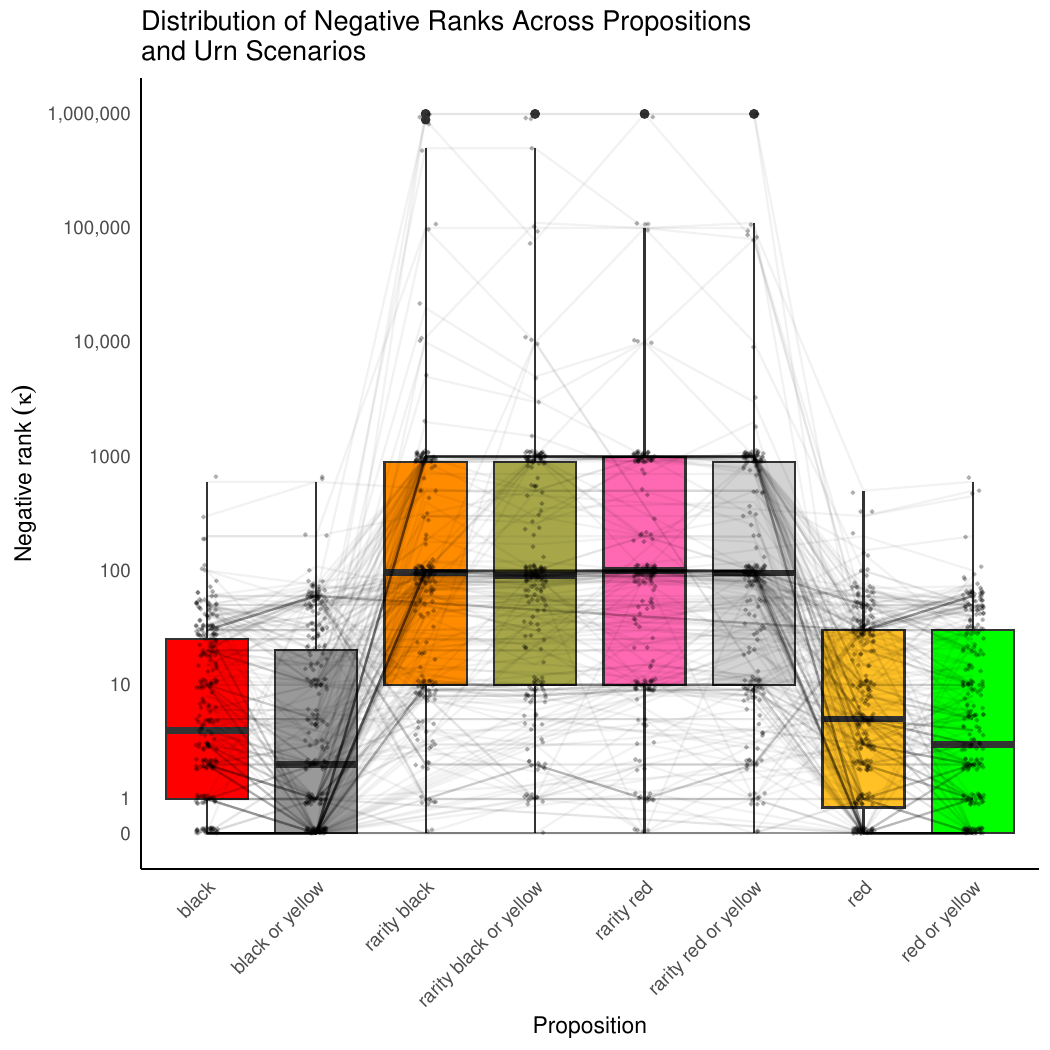} \caption[Visual Representation of Disbelief Scores Across Propositions and Urn Scenarios]{Visual Representation of Disbelief Scores Across Propositions and Urn Scenarios: This plot demonstrates the distribution of disbelief ranks on a logarithmic scale, utilizing boxplots to depict central tendencies and variability, jittered points to indicate individual responses, and connecting lines to trace changes in disbelief for each participant across different propositions and urn scenarios. The boxplots show a clear distinction in disbelief levels between one-urn and rarity-urn scenarios, with increased disbelief in the rarity scenario compare to the one-urn scenario. Importantly, within each scenario, no differences in disbelief scores were observed across the propositions.}\label{fig:E6_disbelief}
\end{figure}

\end{knitrout}

For both betting scenarios (red vs. black and red or yellow vs. black or yellow) we computed the Pearson’s correlation.
Betting choices and disbelief were not significantly correlated, (r = 0.07 ), \(p\)  = .264, \(95\%\) CI [-0.05, 0.18] and \(r(277) =\) .07, \(p\) 0.28, \(95\%\) CI [-0.05, 0.18].

A randomization analysis assessed whether the observed levels of ambiguity aversion in negative ranking responses could be due to chance.
We shuffled participants’ negative ranks within each of the four propositions (``Red,'' ``Black,'' ``Red or Yellow,'' ``Black or Yellow'') while preserving the grouping by proposition 1000 times.
That is the negative ranks for red were permuted, then the ranks for black, and so forth.
Then for each randomization, the percentage of ambiguity-averse responses was computed under two experimental conditions: the one-urn scenario and the rarity one-urn scenario.
The empirical data showed ambiguity aversion percentages for negative ranks of 8.62\% for the one-urn condition and 5.17\% for the rarity one-urn condition.
These percentages were compared with the randomized distributions, which yielded mean percentages of 21.15\% for the one-urn condition and 19.98\% for the rarity one-urn condition.
The observed empirical percentages were distinct from the central tendencies of the randomized distributions (see Figure~\ref{fig:E6_randomization_plot}).
The higher percentages of ambiguity aversion in the randomized distributions further suggest that the observed reduction in ambiguity aversion with negative ranks was not merely due to some quirk in how people assign numbers to the particular propositions. There is a within-participants assignment that leads to a lower degree of paradoxical behavior. 

\begin{knitrout}
\definecolor{shadecolor}{rgb}{0.969, 0.969, 0.969}\color{fgcolor}\begin{figure}[H]
\includegraphics[width=\maxwidth,]{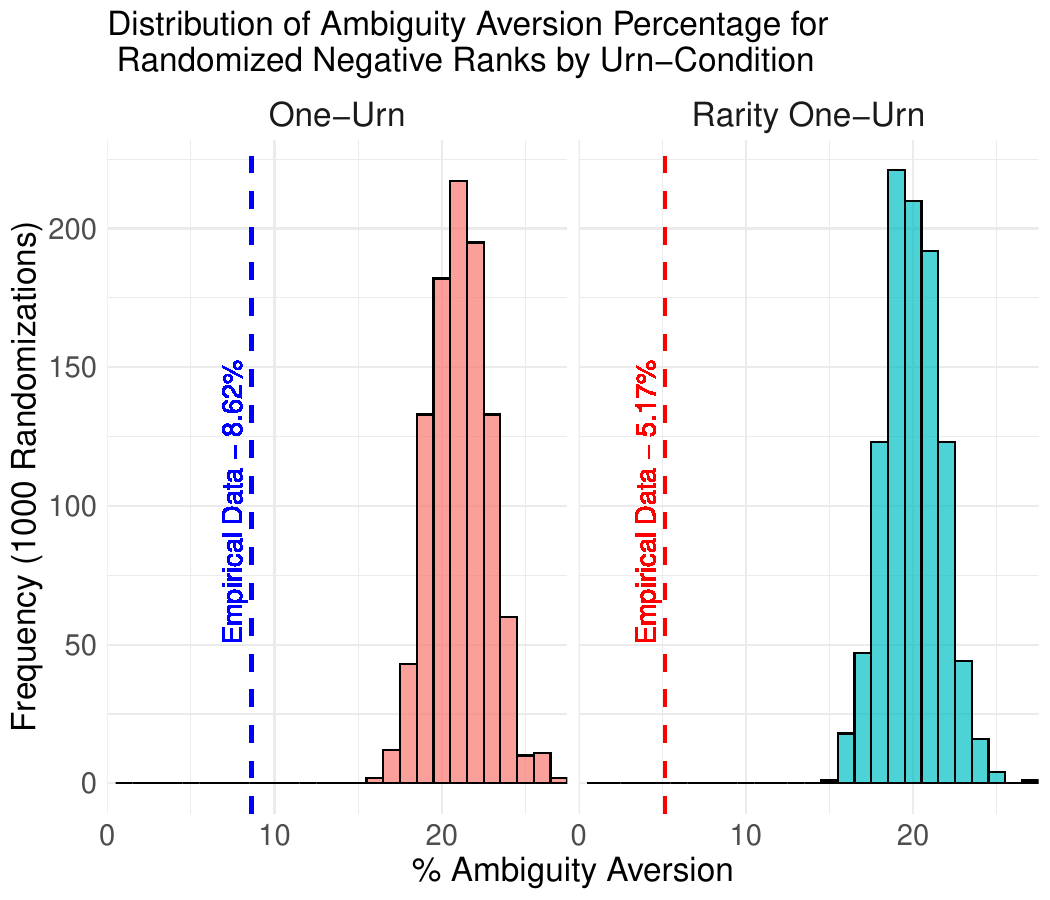} \caption[The distribution of ambiguity aversion percentages for randomized negative ranks is presented by scenario condition (one-urn and rarity one-urn)]{The distribution of ambiguity aversion percentages for randomized negative ranks is presented by scenario condition (one-urn and rarity one-urn). The dotted vertical lines represent the observed empirical data for ambiguity aversion: 8.62\% for the one-urn condition (blue line) and 5.17\% for the rarity one-urn condition (red line). The histograms show the distribution of ambiguity aversion percentages across 1000 randomizations, where negative ranks were shuffled within each proposition (``Red,'' ``Black,'' ``Red or Yellow,'' ``Black or Yellow'') across participants. The observed empirical percentages varied from the central tendencies of the randomized distribution.}\label{fig:E6_randomization_plot}
\end{figure}

\end{knitrout}

\subsection{Discussion}

Experiment 6 replicated the key findings of Experiment 5.
When participants expressed their degrees of disbelief using negative ranking functions ambiguity aversion was minimal.
What was new in Experiment 6 is that we were able to show for the same participants at the same time betting decisions did show ambiguity aversion to a degree common in the literature \autocite{oechssler2015test}.

Contrary to Ellsberg’s proposition that gambling choices are influenced by, or ‘reflect,’ varying degrees of belief \autocite{ellsberg1961risk}, no significant relationship emerged between participants’ disbelief ranks and their betting preferences.
This suggests that betting choices are more susceptible to the disjunction effect, which emphasizes decision avoidance or alteration when faced with uncertainty \autocite{aerts2014identifying, tversky1992disjunction}.
Ambiguity aversion, alongside the disjunction effect, highlights a broader discomfort with uncertainty in decision-making.
While ambiguity aversion demonstrates a preference for known risks over ambiguous ones, the disjunction effect reflects how the absence of complete information can lead to changes in decision-making, even when the missing information should not logically influence the choice \autocite{tversky1992disjunction}.

When using ranks inconsistent choices — those often associated with ambiguity aversion — were significantly reduced.
This suggests to us that ambiguity aversion may be driven more by the betting context and considerations of costs and benefits than epistemic estimates.
Ranks may allow for a more objective evaluation of uncertainty \autocite{chow2001comparative, fox1995ambiguity}.

A limitation of our study is that hypothetical monetary stakes were not included when participants provided negative ranking responses.
We intentionally excluded monetary incentives in the ranking tasks to evaluate the coherence of belief systems without the influence of financial factors.
Further exploration is needed to determine how incorporating hypothetical monetary incentives might affect belief assessments in the ranking theory framework.

\section{General Discussion}

We explored the normative model proposed by Wolfgang Spohn \autocite{spohn2012laws} for quantifying belief.
We show that people can reliably provide the negative ranks the theory is based on.
People can provide such ranks in scenarios where event probabilities are unknown, ambiguous, or even unknowable.
For no other reason than this, negative ranks are worth knowing about: they provide a valuable practical tool for psychological studies on belief and belief updating.
Seeing that negative ranks are less variable than subjective probabilities, reflect contextual influences of uncertainty independent of fixed probabilities and are less prone to the paradoxes suggests they may also be a superior framework for thinking about how we represent uncertainty \autocite{kern20215}.
We consider these points in turn.

Negative ranks effectively captured disbelief across various contexts.
Participants found generating negative ranks moderately difficult, but they clearly understood what was asked, and the results of Experiments 1 and 2, which compared different cohorts in on-line and lab environments, showed substantial concordance.
These are not particularly noisy measures.
Certainly less so than asking for explicit reports of subjective probability.
We found the subjective probability estimates noisy even when trying to use an OSF \autocite{tiede2022revisiting}, which we did because it was supposed to make it easier for people to give consistent subjective reports.

Negative ranks range from 0 to infinity.
Using negative ranks has an advantage common to many scaling metrics: it allows the individual to choose the response range that make most sense to them.
It does not require all people to convert their intuitions to the same fixed 0 to 1 scale.
It also seems intuitively simpler.
For each new judgment the participant has to mainly be concerned with ordering: this is less believable than this, but more believable than that, and then chooses an intermediate number.
If judgments of belief were based on subjective probability the participant would need to consider the entire space of possible events to make sure their answer was internally consistent.
If a participant were to learn of a new option the subjective probability perspective imagines them renormalizing all their prior estimates.
For negative ranks it is enough to find the right slot to place the new option in.
All these informal considerations lead us to conjecture that computing with ranks and orders is computationally easier and thus negative ranking theory deserving consideration as a candidate for how humans actually quantify uncertainty and belief.

Common sense leads us to expect a relationship between negative ranks and subjective probability, and we consistently find one.
It is non-linear, contextually variable, and noisy.
This is most notable when the actual probabilities for an event are obscure or impossible.
When we ask about the scenario of your cowboy hat wearing mechanic friend Tom what is supposed to be your probability space? What are you supposed to be sampling from \autocite{Sanborn_2021}? 

The low agreement between negative ranks and subjective probabilities implies they capture different aspects of uncertainty estimation.
Negative ranks showed a broad dispersion of disbelief levels for lower subjective probabilities.
Context also influenced disbelief levels.
An interesting case is the lottery scenario where many participants felt free to disbelieve, to some degree, that they could win or that they would lose.
Yet the gedanken lottery scenarios popular in philosophy rely on just such an assumption: that we are deductively coherent \autocite{nelkin2000lottery}. 

Not all our scenarios dealt with unknown or unknowable probabilities.
In Experiment 4, where participants played a computerized version of hide-and-seek, participants could learn the objective probabilities, and the data showed that they did.
This is inferred not just because they won more games than chance (when they could), but also because using the matched distribution gave a good fit to participant data in most cases.
It is in the hide-and-seek experiment that we found the best evidence for the theoretically predicted logarithmic relation between negative ranks and probability.

Our results with negative ranks should not be interpreted to say that people do not represent environmental statistics, but rather that is not the only, and maybe not even the primary, way they construct and update their degrees of belief.
While ranks can be computed from a transformation of probability, it is a rather arcane one, involving as it does a logarithm with an infinitesimal base.
If people actually do ever undertake some mental transformation like this, they are probably doing it in situations where the objective environmental probabilities are easy to detect. 

While negative ranks may be the theoretically preferred scale when talking about belief it often feels more natural to consider belief and disbelief as opposite ends of a scale with 0 representing true neutrality of a proposition.
We found that practically a two sided belief measure worked just fine.
Two sided belief can be calculated with a single ranking function and having participants report ranks for propositions and their negation.
The point of 0 on this two-sided ranking function provides another interesting comparison to probabilist perspectives such as Bayesianism.
A rank of zero means you have no belief in either of two outcomes, say whether a flipped coin will or won’t be heads.
However, the Bayesian neutral position, a prior of 0.5, is generally very strongly held.
Now neither the Rank-believer nor the Bayesian-believer would give odds on a coin flip, but that betting behavior would reflect very different belief structures.
Negative ranking theory is not the only alternative to Bayesianism when it comes to representing ignorance.
Dempster-Shafer theory also represents the maximally ignorant, no belief, position differently than does the Bayesian’s strong conviction of uniformity \autocite{haenni2003ignoring}. 

While these experiments show much to like about negative ranking theory as an empirical approach to belief and uncertainty representation in people, there are also problems.
Negative ranking theory, in its formal form, brings with it certain ``laws.'' One is the law of negation: one cannot disbelieve both a proposition and its negation.
We generally saw good alignment with this principle when we had people rate propositions and their negation, but their obedience was far from the normative standard.
Again, the lottery scenario provides the most stark example.
In a setting where only winning and losing were options many people had non-zero disbelief for both.
An even more challenging law is the law of disjunction.
This law requires that when considering a joint proposition of two disjoint propositions then the rank of the disjunction should be the rank of the minimum of the two.
This is not what we usually observed.
People tended to give intermediate ranks to such disjunctive propositions.
They did usually show a greater influence from the smaller of their two ranks, but that is not the same as reporting the minimum.
Of course, we have yet to really discover any true laws of thought.
The general adherence of human disbelief ranks to Spohn’s normative theory is enough to make it worth a closer look. 

Beliefs are shaped by many factors, including cognition, emotions, social interactions, and motivations.
Emotionally, beliefs are influenced by personal experiences and affective states \autocite{grayling2011psychology}.
Socially, cultural norms and interpersonal interactions also shape beliefs \autocite{grayling2011psychology}.
Additionally, motivational factors, such as the desire to maintain consistency in one’s opinions and beliefs and the need to uphold a favorable conception of oneself and the world, play a critical role \autocite{uso2016belief}.
Together, these diverse influences collectively form and reinforce personal and societal belief systems \autocite{halpern1992two, grayling2011psychology, uso2016belief, wyer2005belief}.
We did not explore this wide space of belief.
We were more interested in situations where probability is usually given as the answer to what is graded belief.
Thus, we stayed away from politics and religion.
We do not know if when asking questions about belief in God for instance one would see any other ranks than zero and infinity.

Subjective probabilities are important for rational decision-making, but they may not be equivalent to one’s degree of belief.
Ranking theory offers a flexible alternative.
As revealed in the Ellsberg Urn experiments (5 \& 6), there are practical advantages to asking for ranks instead of probabilities or bets.
Given the difficulty that people have when asked to express coherent subjective probabilities it may also be practically simpler to ask them for a number grading their degree of disbelief.
A researcher could still believe deeply that the core of belief was some cognitive estimate of subjective probability while benefiting from the superior consistency and absence of paradoxes when soliciting responses as ranks experimentally. 

One of the most interesting avenues for further research using this approach is to explore the dynamics of belief revision \autocite{Rott_2008}.
Belief revision theory examines how an ideal doxastic agent should revise their beliefs upon receiving new information \autocite{huber2013belief}.
Ranking theory offers a robust framework for analyzing the dynamics of belief by providing insights into how people should adjust their beliefs in response to contradicting or supporting information \autocite{spohn2009survey}.
This approach may align with the way humans naturally process information, where adjustments to beliefs do not necessarily require a ``recalibration'' of the entire belief system but rather an intuitive updating of ranks relative to new evidence \autocite{kern20215, spohn2009survey}.
Consider the scenario described by \textcite{eichhorn2018rational}.
They present two conditional beliefs: ``If the car had been out of gas, then (usually) it would have stalled'' and ``If the car had not been out of gas, then (usually) it would not have stalled''.
Suppose new evidence reveals that ``The car was out of gas.'' In this case, we only need to adjust the ranks directly affected by this new evidence \autocite{huber2013belief}.
A nice feature of negative ranking theory is that updates of specific beliefs do not require a complete overhaul of the entire belief system.

In conclusion, this research suggests that ranking theory is an effective and useful framework for interrogating human belief systems.
The experiments conducted throughout this paper show that ranking theory addresses some of the limitations of subjective probabilities, offering an intuitive and computationally simpler framework for understanding human beliefs.
By enhancing our understanding of how beliefs are formed, represented, and acted upon in an uncertain world, ranking-based approaches provide a distinct method for modeling subjective beliefs and have the potential to offer new insights into human cognition and behavior across various fields.

\textbf{Comment: This is a reproducible manuscript.}
Full source code for this manuscript can be found at \nolinkurl{https://codeberg.org/brittAnderson/neg-rank-belief}. With the material in this repository readers can see all the code for the statistical analyses and figures in the ~goAnderson.Rnw~ file. Compilation instructions are available in the ~Readme~ of the manuscript repository. 

\printbibliography

@article{aerts2011contextual,
	title={{A contextual risk model for the Ellsberg paradox}},
	author={Aerts, Diederik and Sozzo, Sandro},
	journal={arXiv preprint arXiv:1105.1814},
	year={2011}
}

@article{aerts2014identifying,
	title={{Identifying quantum structures in the Ellsberg paradox}},
	author={Aerts, Diederik and Sozzo, Sandro and Tapia, Jocelyn},
	journal={International Journal of Theoretical Physics},
	volume={53},
	pages={3666--3682},
	year={2014},
	publisher={Springer}
}

@article{binmore2012much,
	title={{How much ambiguity aversion? Finding indifferences between Ellsberg’s risky and ambiguous bets}},
	author={Binmore, Ken and Stewart, Lisa and Voorhoeve, Alex},
	journal={{Journal of Risk and Uncertainty}},
	volume={45},
	pages={215--238},
	year={2012},
	publisher={Springer}
}

@article{bland2012different,
	title={Different varieties of uncertainty in human decision-making},
	author={Bland, Amy R and Schaefer, Alexandre},
	journal={{Frontiers in Neuroscience}},
	volume={6},
	pages={85},
	year={2012},
	publisher={Frontiers Research Foundation}
}

@book{carnap1962logical,
	title={{Logical Foundations of Probability}},
	author={Carnap, Rudolf},
	volume={2},
	year={1962},
	publisher={Citeseer}
}

@article{chow2001comparative,
	title={Comparative ignorance and the Ellsberg paradox},
	author={Chow, Clare Chua and Sarin, Rakesh K},
	journal={{Journal of Risk and Uncertainty}},
	volume={22},
	pages={129--139},
	year={2001},
	publisher={Springer}
}

@book{cohen1977probable,
	title={{The Probable and the Provable}},
	author={Cohen, L Jonathan},
	year={1977},
	publisher = {{Oxford University Press}},
}

@article{cosmides1996humans,
	title={Are humans good intuitive statisticians after all? {R}ethinking some conclusions from the literature on judgment under uncertainty},
	author={Cosmides, Leda and Tooby, John},
	journal={{Cognition}},
	volume={58},
	number={1},
	pages={1--73},
	year={1996},
	publisher={Elsevier}
}

@incollection{cross2013judgments,
  title={On judgments of magnitude},
  author={Cross, David V},
  booktitle={{Social Attitudes and Psychophysical Measurement}},
  pages={73--88},
  year={2013},
  publisher={Psychology Press}
}

@book{de_finetti_theory_1975,
	title     = {{Theory of Probability}},
	author    = {de Finetti, Bruno},
	publisher = {Wiley},
	year      = {1975},
	volume    = {1 \& 2},
	address   = {New York}
}

@article{de2023jspsych,
	title={jsPsych: Enabling an open-source collaborative ecosystem of behavioral experiments},
	author={de Leeuw, Joshua R and Gilbert, Rebecca A and Luchterhandt, Bj{\"o}rn},
	journal={Journal of Open Source Software},
	volume={8},
	number={85},
	pages={5351},
	year={2023}
}

@mastersthesis{diberardino2023probability,
	title={{Probability Antimatching}},
	author={DiBerardino, Peter},
	year={2023},
	school={University of Waterloo}
}

@article{dominiak2012dynamic,
	title={{A dynamic Ellsberg urn experiment}},
	author={Dominiak, Adam and Duersch, Peter and Lefort, Jean-Philippe},
	journal={Games and Economic Behavior},
	volume={75},
	number={2},
	pages={625--638},
	year={2012},
	publisher={Elsevier}
}

@incollection{dubois2023possibility,
	title={Possibility theory},
	author={Dubois, Didier and Prade, Henri},
	booktitle={{Granular, Fuzzy, and Soft Computing}},
	pages={859--876},
	year={2023},
	publisher={Springer}
}

@inproceedings{eichhorn2018rational,
	title={Rational inference patterns based on conditional logic},
	author={Eichhorn, Christian and Kern-Isberner, Gabriele and Ragni, Marco},
	booktitle={Proceedings of the AAAI Conference on Artificial Intelligence},
	volume={32},
	year={2018}
}

@InBook{einstein2015,
  author       = {Einstein, A},               
  year	       = 2015,
  chapter      = {Credo: What I believe},
  booktitle    = {{An Einstein Encyclopedia}},
  publisher    = {Princeton University Press},
  pages	       = {xxi–xxiv},
}

@article{ellsberg1961risk,
	title={Risk, ambiguity, and the Savage axioms},
	author={Ellsberg, Daniel},
	journal={The Quarterly Journal of Economics},
	volume={75},
	number={4},
	pages={643--669},
	year={1961},
	publisher={MIT Press}
}

@book{feller1991introduction,
	title={{An Introduction to Probability Theory and its Applications, Volume 2}},
	author={Feller, William},
	volume={81},
	year={1991},
	publisher={John Wiley \& Sons}
}

@article{Fiore_2022,
title = {Similar network compositions, but distinct neural dynamics underlying belief updating in environments with and without explicit outcomes},
journal = {NeuroImage},
volume = {247},
pages = {118821},
year = {2022},
issn = {1053-8119},
author = {Vincenzo G. Fiore and Xiaosi Gu},
}

@article{Fleming_2024,
  author       = {Fleming, Stephen M.},
  title	       = {Metacognition and Confidence: A Review and
                  Synthesis},
  journal      = {Annual Review of Psychology},
  year	       = 2024,
  volume       = 75,
  pages	       = {241–268},
  issn	       = {1545-2085},
  publisher    = {Annual Reviews}
}

@article{forster1938what,
  author = {Forster, E. M.},
  title = {What {{I}} Believe},
  journal = {The Nation},
  year = {1938},
  pages = {},
  note = {Originally titled "Two Cheers for Democracy"; republished by Hogarth Press in 1939 and later collected in Two Cheers for Democracy (1951)}
}

@article{fox1995ambiguity,
	title={Ambiguity aversion and comparative ignorance},
	author={Fox, Craig R and Tversky, Amos},
	journal={The Quarterly Journal of Economics},
	volume={110},
	number={3},
	pages={585--603},
	year={1995},
	publisher={MIT Press}
}

@article{galavotti_interpretation_2017,
	title = {The {Interpretation} of {Probability}: {Still} an {Open} {Issue}? 1},
	author = {Galavotti, Maria Carla},
	journal = {Philosophies},
	volume = {2},
	number = {4},
	urldate = {2024-05-02},
	year = {2017},
	pages = {20}
}

@article{ghirardato2002ambiguity,
	title={Ambiguity made precise: A comparative foundation},
	author={Ghirardato, Paolo and Marinacci, Massimo},
	journal={Journal of Economic Theory},
	volume={102},
	number={2},
	pages={251--289},
	year={2002},
	publisher={Elsevier}
}

@article{go2022saccadic,
	title={Saccadic eye movement metrics reflect surprise and mental model updating},
	author={Go, Hanbin and Danckert, James and Anderson, Britt},
	journal={Attention, Perception, \& Psychophysics},
	volume={84},
	number={5},
	pages={1553--1565},
	year={2022},
	publisher={Springer}
}

@article{grayling2011psychology,
  author       = {Grayling, A. C.},
  title	       = {Psychology: How we form beliefs},
  journal      = {Nature},
  year	       = 2011,
  volume       = 474,
  number       = 7352,
  pages	       = {446–447},
  issn	       = {1476-4687},
  publisher    = {Springer Science and Business Media LLC}
}

@article{griffiths2006optimal,
	title={Optimal predictions in everyday cognition},
	author={Griffiths, Thomas L and Tenenbaum, Joshua B},
	journal={Psychological Science},
	volume={17},
	number={9},
	pages={767--773},
	year={2006},
	publisher={SAGE Publications Sage CA: Los Angeles, CA}
}

@article{haenni2003ignoring,
  title={Ignoring ignorance is ignorant},
  author={Haenni, Rolf},
  journal={Philosophy and Probability Working Paper},
  volume={6},
  year={2003}
}

@article{halevy2007ellsberg,
	title={Ellsberg revisited: An experimental study},
	author={Halevy, Yoram},
	journal={Econometrica},
	volume={75},
	number={2},
	pages={503--536},
	year={2007},
	publisher={Wiley Online Library}
}

@article{halpern1992two,
	title={Two views of belief: belief as generalized probability and belief as evidence},
	author={Halpern, Joseph Y and Fagin, Ronald},
	journal={Artificial Intelligence},
	volume={54},
	number={3},
	pages={275--317},
	year={1992},
	publisher={Elsevier}
}

@article{henckel2022belief,
	title={Belief adjustment: A double hurdle model and experimental evidence},
	author={Henckel, Timo and Menzies, Gordon D and Moffatt, Peter G and Zizzo, Daniel J},
	journal={Experimental Economics},
	volume={25},
	number={1},
	pages={26--67},
	year={2022},
	publisher={Springer}
}

@incollection{huber_ed_2009,
	title={{Belief and Degrees of Belief}},
	author={Huber, Franz},
	booktitle={{Degrees of Belief}},
	pages={1--33},
	year={2009},
	publisher={Springer}
}

@article{huber2013belief,
	title={Belief revision II: Ranking theory},
	author={Huber, Franz},
	journal={Philosophy Compass},
	volume={8},
	number={7},
	pages={613--621},
	year={2013},
	publisher={Wiley Online Library}
}

@InBook{jackson_2024,
  author = {Jackson, Elizabeth},
  chapter = {The cognitive science of credence},
  title = {{The Oxford Handbook of the Cognitive Science of Belief}},
  editor = {Neil Van Leeuwen and Tania Lombrozo},
  publisher = {Oxford University Press},
  year = {forthcoming}
  }

@article{kern20215,
	title={5.3 Ranking Theory},
	author={Kern-Isberner, Gabriele and Skovgaard-Olsen, Niels and Spohn, Wolfgang},
	journal={{The Handbook of Rationality}},
	pages={337},
	year={2021},
	publisher={MIT Press}
}

@book{keynes2013treatise,
	title={{A Treatise on Probability}},
	author={Keynes, John Maynard},
	year={2013},
	publisher={Courier Corporation}
}

@article{khaw2021individual,
	title={Individual differences in the perception of probability},
	author={Khaw, Mel W and Stevens, Luminita and Woodford, Michael},
	journal={PLoS Computational Biology},
	volume={17},
	number={4},
	pages={e1008871},
	year={2021},
	publisher={Public Library of Science San Francisco, CA USA}
}

@Book{kolmogorov2018foundations,
	title={{Foundations of the Theory of Probability}},
	author={Kolmogorov, Andrei Nikolaevich},
	year={2018},
	edition={Second English Edition},
	publisher={Courier Dover Publications},
}

@article{kovavrik2016ellsberg,
	title={Ellsberg paradox: Ambiguity and complexity aversions compared},
	author={Kov{\'a}{\v{r}}{\'\i}k, Jarom{\'\i}r and Levin, Dan and Wang, Tao},
	journal={Journal of Risk and Uncertainty},
	volume={52},
	pages={47--64},
	year={2016},
	publisher={Springer}
}

@book{Kyburg1961,
  author    = {Kyburg, Henry E.},
  title     = {{Probability and the Logic of Rational Belief}},
  publisher = {Wesleyan University Press},
  year      = {1961},
  address   = {Middletown, Connecticut}
}

@incollection{machina2014ambiguity,
	title={Ambiguity and ambiguity aversion},
	author={Machina, Mark J and Siniscalchi, Marciano},
	booktitle={{Handbook of the Economics of Risk and Uncertainty}},
	volume={1},
	pages={729--807},
	year={2014},
	publisher={Elsevier}
}

@inproceedings{maddalena2015judging,
  title={Judging relevance using magnitude estimation},
  author={Maddalena, Eddy and Mizzaro, Stefano and Scholer, Falk and Turpin, Andrew},
  booktitle={European Conference on Information Retrieval},
  pages={215--220},
  year={2015},
  organization={Springer}
}

@article{nelkin2000lottery,
	title={The lottery paradox, knowledge, and rationality},
	author={Nelkin, Dana K},
	journal={The Philosophical Review},
	volume={109},
	number={3},
	pages={373--409},
	year={2000},
	publisher={JSTOR}
}

@Article{Nickerson_2002,
  author       = {Nickerson, Raymond S.},
  title	       = {The production and perception of randomness.},
  journal      = {Psychological Review},
  year	       = 2002,
  volume       = 109,
  number       = 2,
  pages	       = {330–357},
  issn	       = {0033-295X},
  publisher    = {American Psychological Association (APA)}
}

@article{oechssler2015test,
	title={A test of mechanical ambiguity},
	author={Oechssler, J{\"o}rg and Roomets, Alex},
	journal={Journal of Economic Behavior \& Organization},
	volume={119},
	pages={153--162},
	year={2015},
	publisher={Elsevier}
}

@book{pearl1988probabilistic,
	author = {Judea Pearl},
	title = {{Probabilistic Reasoning in Intelligent Systems: Networks of Plausible Inference}},
	year = {1988},
	isbn = {1558604790},
	publisher = {Morgan Kaufmann Publishers Inc.}
}

@article{prelec1998probability,
	title={The probability weighting function},
	author={Prelec, Drazen},
	journal={Econometrica},
	pages={497--527},
	year={1998},
	publisher={JSTOR}
}

@article{raidl2017,
	title={Bridging ranking theory and the stability theory of belief},
	author={Raidl, Eric and Skovgaard-Olsen, Niels},
	journal={Journal of Philosophical Logic},
	volume={46},
	number={6},
	pages={577--609},
	year={2017},
	publisher={Springer}
}

@InBook{Rott_2008,
  title	       = {Belief Revision},
  year	       = 2008,
  author       = {Rott, Hans},
  booktitle    = {Reasoning},
  publisher    = {Cambridge University Press},
  isbn	       = 9780511814273,
  pages	       = {514–534},
}

@Book{Russell_2013,
  author       = {Russell, Bertrand},
  title	       = {{What I Believe}},
  year	       = 2013,
  publisher    = {Routledge},
  isbn	       = 9781134517398,
}

@article{sanborn2016bayesian,
	title={Bayesian brains without probabilities},
	author={Sanborn, Adam and Chater, Nick},
	journal={Trends in Cognitive Sciences},
	volume={20},
	number={12},
	pages={883--893},
	year={2016},
	publisher={Elsevier}
}

@InBook{Sanborn_2021,
  title	       = {Sampling as the Human Approximation to Probabilistic
                  Inference},
  year	       = 2021,
  author       = {Sanborn, Adam and Zhu, Jian-Qiao and Spicer, Jake
                  and Sundh, Joakim and León-Villagrá, Pablo and
                  Chater, Nick},
  booktitle    = {{Human-Like Machine Intelligence}},
  publisher    = {Oxford University Press},
  isbn	       = 9780191895333,
  pages	       = {430–448},
}

@book{savage1972foundations,
	title={{The Foundations of Statistics}},
	author={Savage, Leonard J},
	year={1972},
	publisher={Courier Corporation}
}

@article{schmider2010really,
	title={Is it really robust?},
	author={Schmider, Emanuel and Ziegler, Matthias and Danay, Erik and Beyer, Luzi and B{\"u}hner, Markus},
	journal={Methodology},
	year={2010},
	publisher={Hogrefe Publishing}
}

@book{shafer1976mathematical,
	title={{A Mathematical Theory of Evidence}},
	author={Shafer, G},
	year={1976},
	publisher={Princeton university press}
}

@article{slovic1974accepts,
	title={Who accepts {S}avage's axiom?},
	author={Slovic, Paul and Tversky, Amos},
	journal={Behavioral science},
	volume={19},
	number={6},
	pages={368--373},
	year={1974},
	publisher={Wiley Online Library}
}

@article{skovgaard2016ranking,
	title={Ranking theory and conditional reasoning},
	author={Skovgaard-Olsen, Niels},
	journal={Cognitive Science},
	volume={40},
	number={4},
	pages={848--880},
	year={2016},
	publisher={Wiley Online Library}
}

@incollection{spohn2009survey,
	title={A survey of ranking theory},
	author={Spohn, Wolfgang},
	booktitle={{Degrees of belief}},
	pages={185--228},
	year={2009},
	publisher={Springer}
}

@book{spohn2012laws,
	title={{The laws of belief: Ranking theory and its philosophical applications}},
	author={Spohn, Wolfgang},
	year={2012},
	publisher={Oxford University Press}
}

@article{spohn2013ranking,
	title={A ranking-theoretic approach to conditionals},
	author={Spohn, Wolfgang},
	journal={Cognitive Science},
	volume={37},
	number={6},
	pages={1074--1106},
	year={2013},
	publisher={Wiley Online Library}
}

@article{stalnaker1976possible,
  title={Possible worlds},
  author={Stalnaker, Robert C},
  journal={No{\^u}s},
  pages={65--75},
  year={1976},
  publisher={JSTOR}
}

@Article{Szollosi_2023,
  author       = {Szollosi, Aba and Donkin, Chris and Newell, Ben R.},
  title	       = {Toward nonprobabilistic explanations of learning and
                  decision-making.},
  journal      = {Psychological Review},
  year	       = 2023,
  volume       = 130,
  number       = 2,
  pages	       = {546–568},
  issn	       = {0033-295X},
  publisher    = {American Psychological Association (APA)}
}

@article{tiede2022revisiting,
	title={Revisiting the Open Sampling format: Improving risky choices through a novel graphical representation},
	author={Tiede, Kevin E and Henninger, Felix and Kieslich, Pascal J},
	journal={Psychonomic Bulletin \& Review},
	pages={1--12},
	year={2022},
	publisher={Springer}
}

@article{tversky1983extensional,
	title={Extensional versus intuitive reasoning: The conjunction fallacy in probability judgment.},
	author={Tversky, Amos and Kahneman, Daniel},
	journal={Psychological Review},
	volume={90},
	number={4},
	pages={293},
	year={1983},
	publisher={American Psychological Association}
}

@article{tversky1992advances,
	title={Advances in prospect theory: Cumulative representation of uncertainty},
	author={Tversky, Amos and Kahneman, Daniel},
	journal={Journal of Risk and Uncertainty},
	volume={5},
	pages={297--323},
	year={1992},
	publisher={Springer}
}

@article{tversky1992disjunction,
	title={The disjunction effect in choice under uncertainty},
	author={Tversky, Amos and Shafir, Eldar},
	journal={Psychological Science},
	volume={3},
	number={5},
	pages={305--310},
	year={1992},
	publisher={SAGE Publications Sage CA: Los Angeles, CA}
}

@article{uso2016belief,
	title={What are belief systems?},
	author={Uso-Domenech, Josep-Lluis and Nescolarde-Selva, Josue},
	journal={Foundations of Science},
	volume={21},
	pages={147--152},
	year={2016},
	publisher={Springer}
}

@book{von_mises_probability_1957,
	title     = {{Probability, Statistics and Truth}},
	author    = {von Mises, Richard},
	edition   = {Second},
	publisher = {Dover Publications},
	year      = {1957},
	address   = {New York}
}

@Article{Walker_2023,
  author       = {Walker, Edgar Y. and Pohl, Stephan and Denison,
                  Rachel N. and Barack, David L. and Lee, Jennifer and
                  Block, Ned and Ma, Wei Ji and Meyniel, Florent},
  title	       = {Studying the neural representations of uncertainty},
  journal      = {Nature Neuroscience},
  year	       = 2023,
  volume       = 26,
  number       = 11,
  pages	       = {1857–1867},
  issn	       = {1546-1726},
  publisher    = {Springer Science and Business Media LLC}
}

@article{wedgwood2002aim,
	title={The aim of belief},
	author={Wedgwood, Ralph},
	journal={Philosophical Perspectives},
	volume={16},
	pages={267--297},
	year={2002},
	publisher={JSTOR}
}

@Book{ggplot2016,
	author = {Hadley Wickham},
	title = {ggplot2: Elegant Graphics for Data Analysis},
	publisher = {Springer-Verlag New York},
	year = {2016},
	isbn = {978-3-319-24277-4},
	url = {https://ggplot2.tidyverse.org}
}

@article{wyer2005belief,
	title={Belief formation, organization, and change: Cognitive and motivational influences},
	author={Wyer, Robert S and Albarracin, Dolores},
	journal={The Handbook of Attitudes},
	volume={273},
	pages={322},
	year={2005}
}

@article{yu2012influence,
	title={The influence of initial beliefs on judgments of probability},
	author={Yu, Erica C and Lagnado, David A},
	journal={Frontiers in psychology},
	volume={3},
	pages={25632},
	year={2012},
	publisher={Frontiers}
}

\appendix
\section*{Appendix}
\subsection*{Experiments 1 \& 2 Questionnaires}
\label{AppendixE1}

Participants completed questionnaires to express disbelief toward various propositions in Experiments 1 \& 2.

\begin{figure}[H]
    \centering
    \includegraphics[width=1\textwidth]{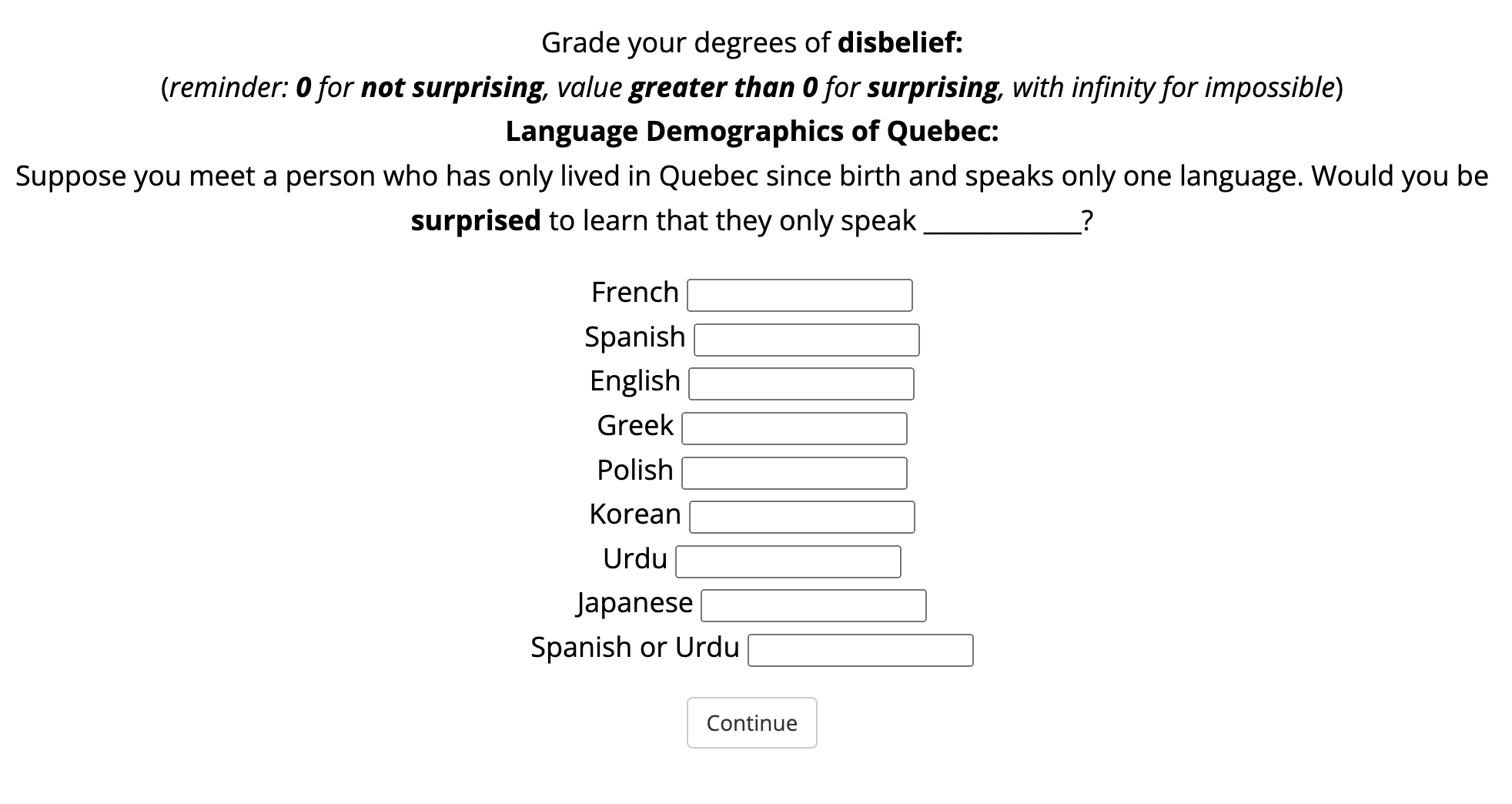}
  
    \caption{Ranking questionnaire for language demographics of Quebec.}
    \label{fig:E1_rankQ1}
\end{figure}

\begin{figure}[H]
    \centering
    \includegraphics[width=1\textwidth]{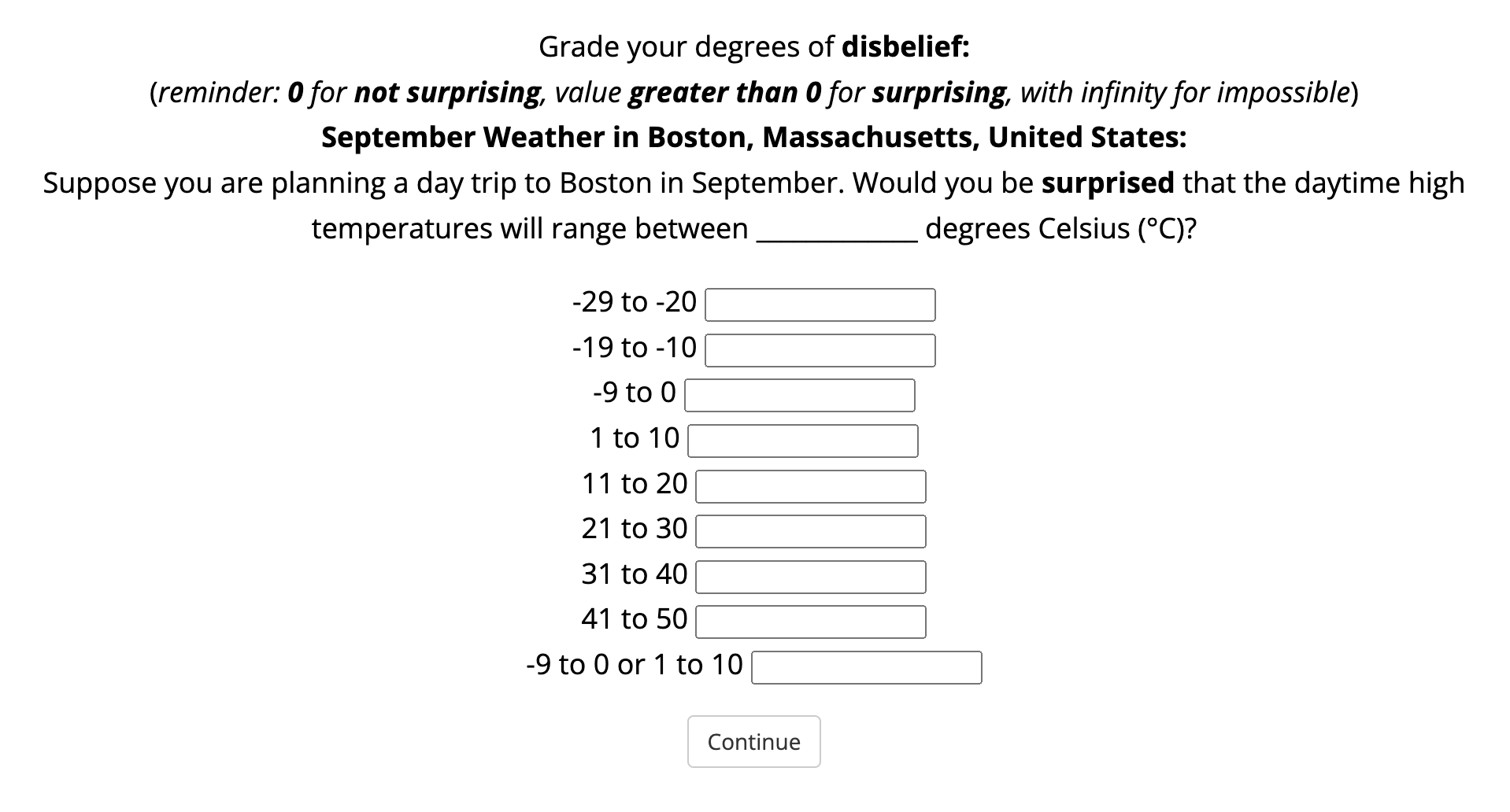}
  
    \caption{Ranking questionnaire for September weather in Boston, Massachusetts, United States.}
    \label{fig:E1_rankQ2}
\end{figure}

\begin{figure}[H]
    \centering
    \includegraphics[width=1\textwidth]{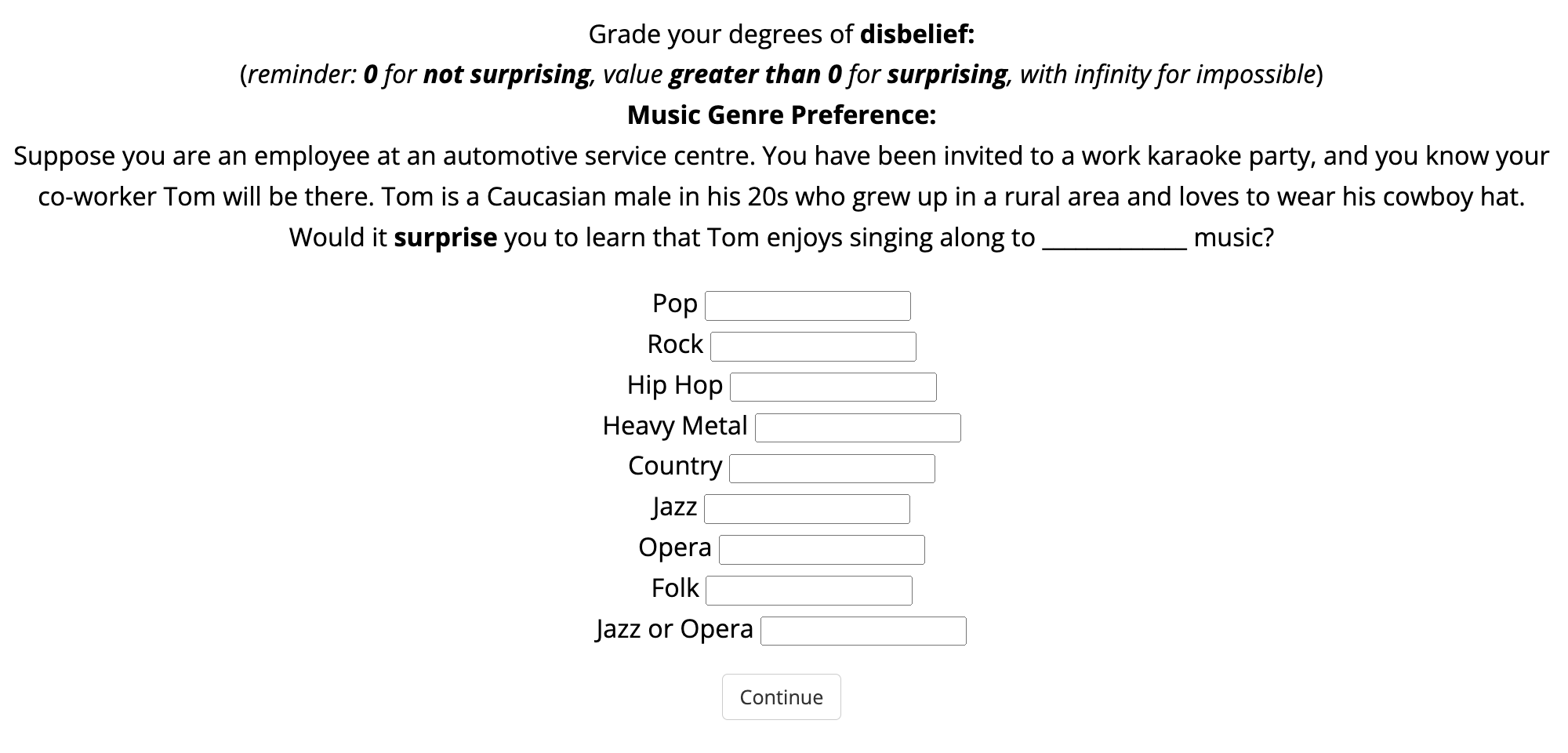}
  
    \caption{Ranking questionnaire for co-worker’s music genre preferences.}
    \label{fig:E1_rankQ3}
\end{figure}

\begin{figure}[H]
    \centering
    \includegraphics[width=1\textwidth]{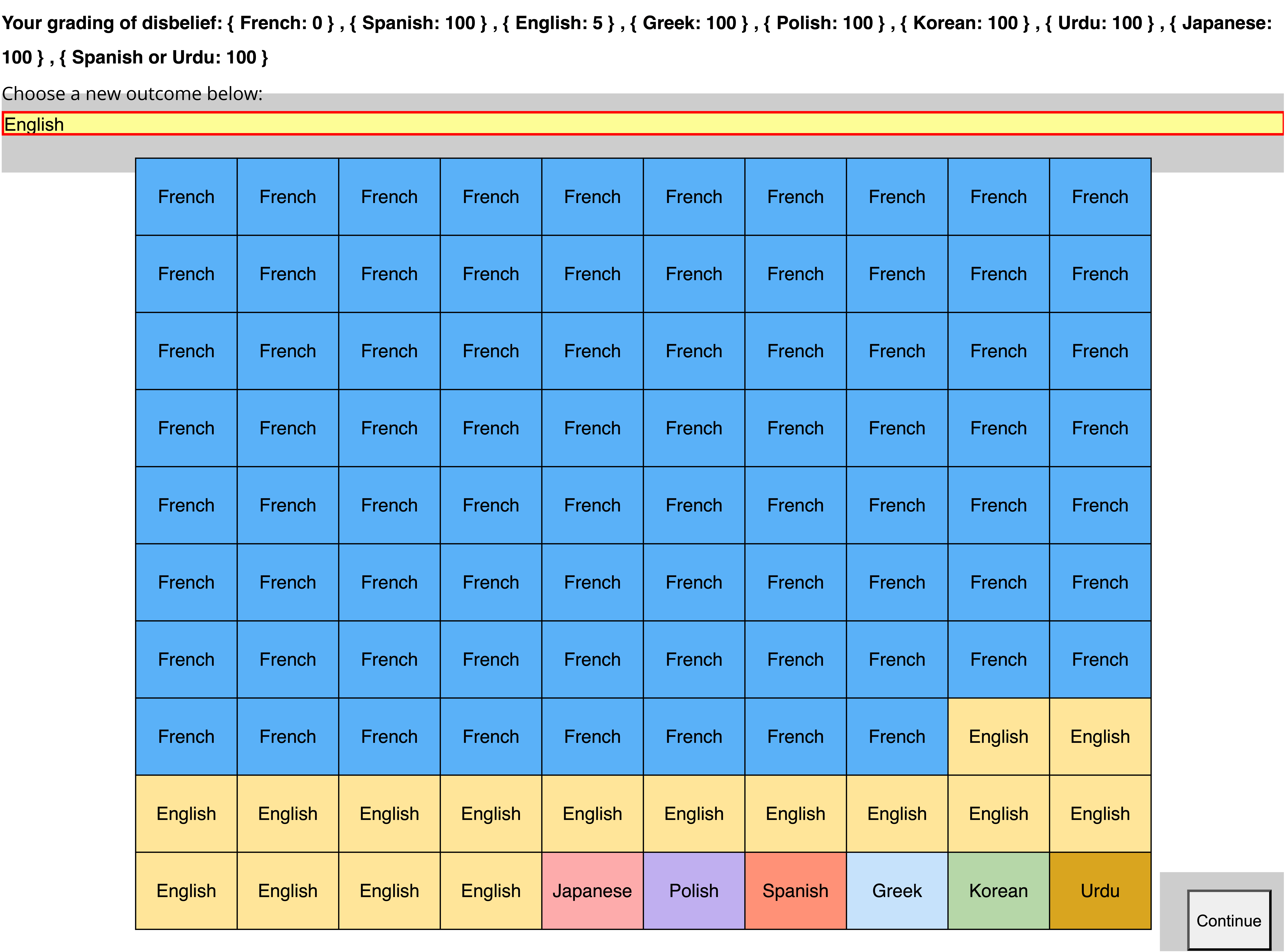}
  
    \caption{OSF used to measure subjective probability, represented by a 10 by 10 grid of possible outcomes. In this example, the language demographics questionnaire shows French as the most expected, followed by English, with equal expectancy for the remaining languages.}
    \label{fig:E1_E3_OSF}
\end{figure}

\subsection*{Experiment 3 Questionnaires}
\addcontentsline{toc}{section}{Experiment 3 Materials}
\label{AppendixE3}

Two new questionnaires:
\begin{figure}[H]
    \centering
    \includegraphics[width=1\textwidth]{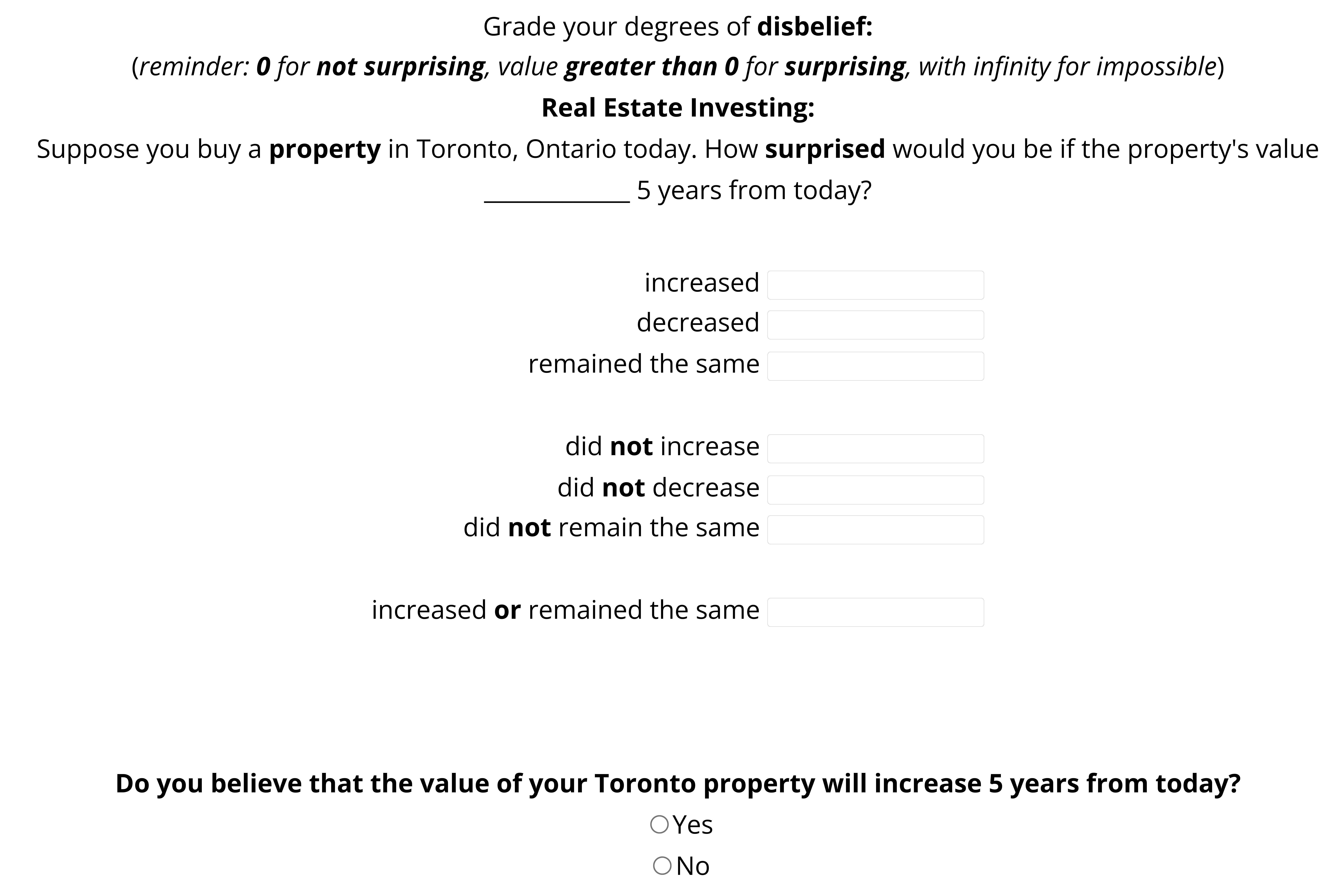}
    \caption{Ranking questionnaire for real estate investment outcomes.}
    \label{fig:E3_rankQ3}
\end{figure}

\begin{figure}[H]
    \centering
    \includegraphics[width=1\textwidth]{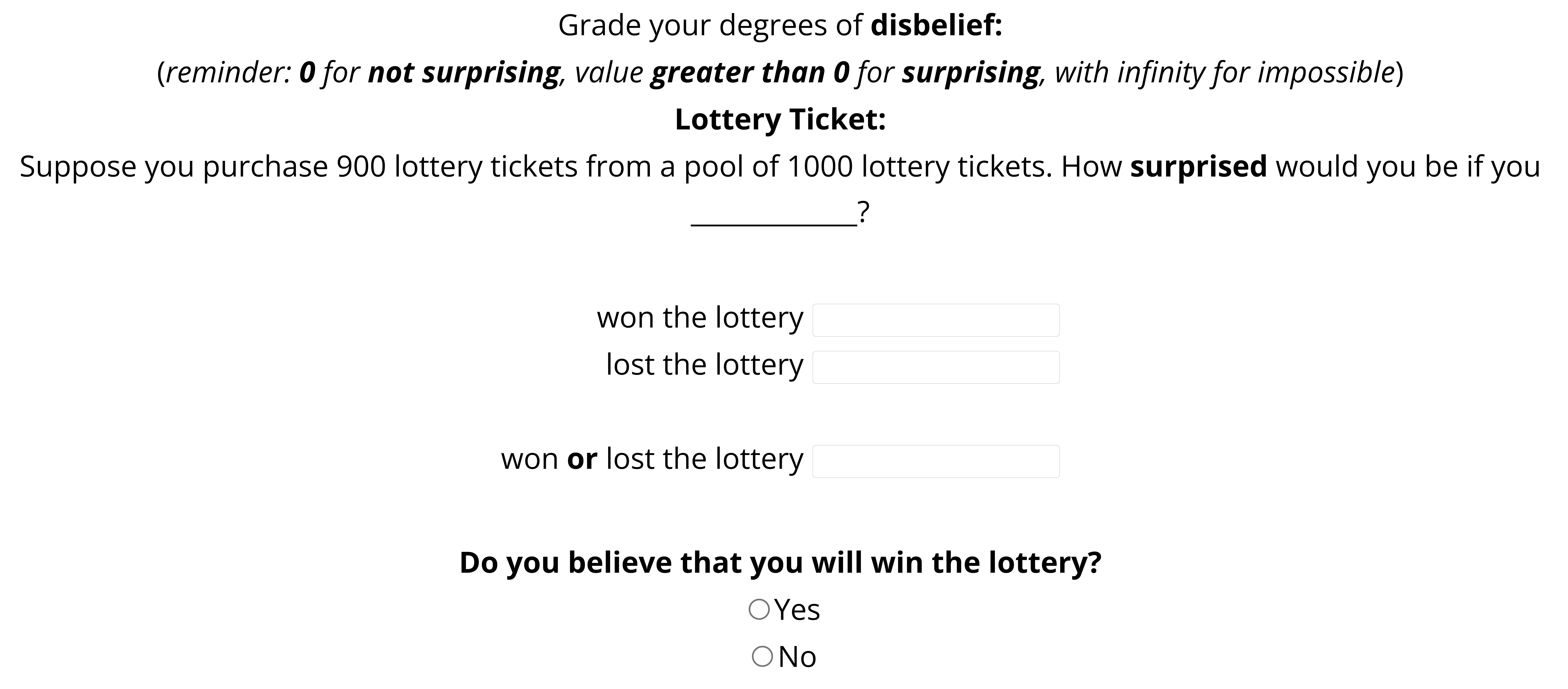}
  
    \caption{Ranking questionnaire for lottery outcomes.}
    \label{fig:E3_rankQ4}
\end{figure}

\subsection*{Experiments 5 and 6 Materials}
\addcontentsline{toc}{section}{Experiments 5 and 6 Materials}
\label{AppendixE6}
This section provides the materials for Experiments 5 and 6. In both experiments, participants graded their disbelief toward the one-urn Ellsberg scenario. Specifically for Experiment 6, participants also completed betting scenarios and the rarity-urn Ellsberg scenario.

\begin{figure}[H]
    \centering
    \includegraphics[width=1\textwidth]{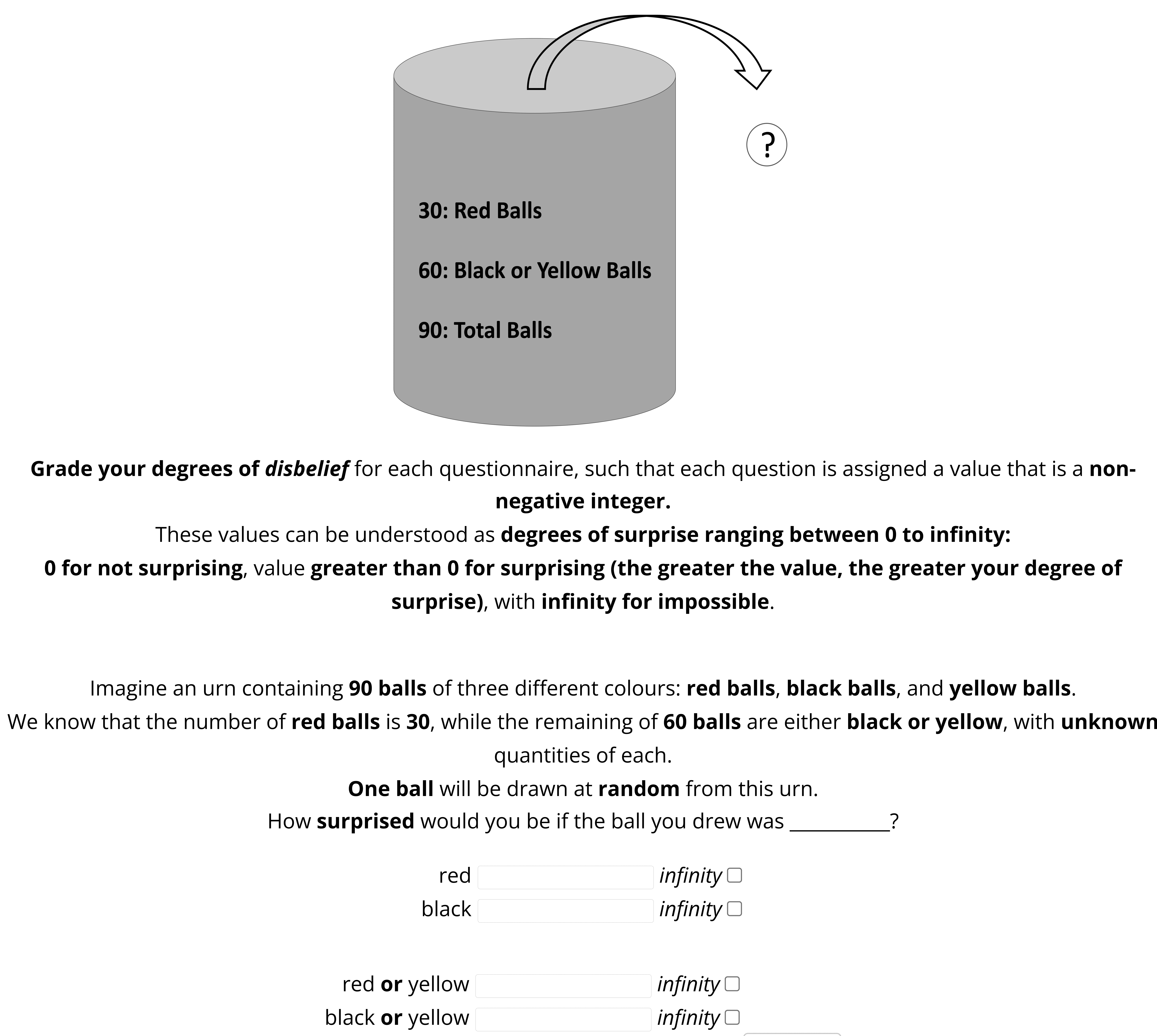}
  
    \caption{Ranking questionnaire for the one-urn scenario.}
    \label{fig:E5_E6_rankQ}
\end{figure}

\begin{figure}[H]
    \centering
    \includegraphics[width=1\textwidth]{./E6_Q2.png}
  
    \caption{Participants chose between different gambles across four rounds, in Experiment 6.}
    \label{fig:E6_Q2}
\end{figure}

\begin{figure}[H]
    \centering
    \includegraphics[width=1\textwidth]{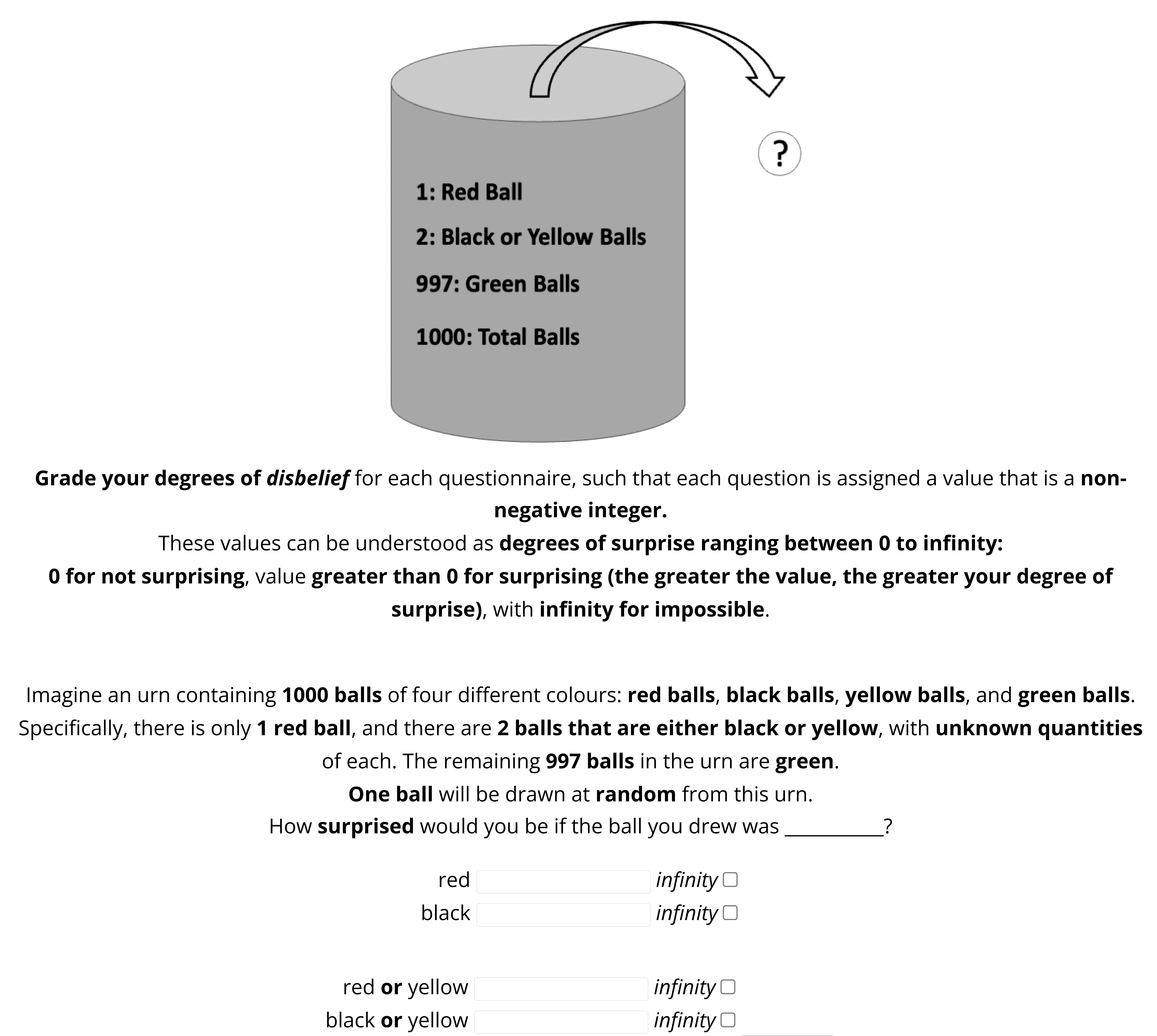}
  
    \caption{Ranking questionnaire for the rarity one-urn scenario in Experiment 6.}
    \label{fig:E6_Q3}
\end{figure}

\end{document}